\documentclass[onecolumn,authoryear]{els-mrw} 

\usepackage{amsmath,amssymb,amsfonts,amsthm,makeidx,graphicx}
\usepackage{txfonts}
\usepackage{helvet}
\usepackage{slashed}
\usepackage[dvipsnames,svgnames]{xcolor}

\usepackage{soul}
\usepackage{wrapfig}

\usepackage[%
pdfsubject={Sign},
pdfkeywords={Sign},
colorlinks=true,    %color the links, not boxes
citecolor=blue,     %Change colors of citations
linkcolor=black,    %All internal links. contents page, etc %Eqn/fig set in main.tex hypersetup
urlcolor=blue,      %Color of URLs in bib
backref=section
]{hyperref}

\newcommand{\be}{\begin{equation}}
\newcommand{\ee}{\end{equation}}

\newcommand{\half}{\frac{1}{2}}
\newcommand{\ket}{\rangle}
\newcommand{\bra}{\langle}
\newcommand{\qqquad}{\qquad\quad}

\newcommand{\im}{{\mathrm{Im}}}
\newcommand{\re}{{\mathrm{Re}}}
\newcommand{\Tr}{{\mathrm{Tr}\,}}

\newcommand{\xv}{{\mathbf x}}
\newcommand{\yv}{{\mathbf y}}

\renewcommand{\Re}{\mbox{Re}\,}
\renewcommand{\Im}{\mbox{Im}\,}
\newcommand{\Var}{\mbox{Var}}
\newcommand{\bean}{\begin{eqnarray*}}
\newcommand{\eean}{\end{eqnarray*}}
\newcommand{\id}{1\!\!1}
\newcommand{\rmR}{{\rm R}}
\newcommand{\rmI}{{\rm I}}

\newcommand{\phiR}{\phi_{\rm R}}
\newcommand{\phiI}{\phi_{\rm I}}

\newcommand{\bea}{\begin{eqnarray}}
\newcommand{\eea}{\end{eqnarray}}
\newcommand{\bit}{\begin{itemize}}
\newcommand{\eit}{\end{itemize}}

\begin{document}

\chapter{Lattice field theories with a sign problem}
\label{chap1}

\author[1]{Gert Aarts}%

\address[1]{\orgdiv{Centre for Quantum Fields and Gravity, Department of Physics}, \orgname{Swansea University}, \orgaddress{\\ \;\;Swansea, SA2 8PP, United Kingdom}}

\author[2]{D\'enes Sexty}%

\address[2]{\orgdiv{Institute of Physics, NAWI Graz}, \orgname{University of Graz},  \orgaddress{Universit\"atsplatz 5, 8010 Graz, Austria}}

\articletag{Chapter Article tagline: update of previous edition,, reprint..}

\maketitle

\begin{glossary}[Glossary]
\term{Sign problem} 
failure of Monte Carlo importance sampling caused by the Boltzmann weight being complex-valued, leading to exponentially severe cancellations and a vanishing signal-to-noise ratio, e.g., in QCD at non-zero baryon chemical potential.

\term{Overlap problem} 
inefficiency of Monte Carlo sampling when the sampling distribution has poor overlap with the regions that dominate the target distribution, leading to exponentially rare relevant configurations and large statistical errors.

\term{Holomorphic extensions} 
analytic continuation of fields and the action into complexified variables, allowing the original complex-valued path integral to be evaluated over a complexified manifold, with the aim of ameliorating the sign problem.

\end{glossary}

\begin{abstract}[Abstract]
The sign problem obstructs the determination of the QCD phase diagram in the temperature–baryon chemical potential plane using lattice QCD. We review the sign problem in QCD and related field theories, including applications to real-time dynamics. We focus on approaches where the sign problem can potentially be solved or controlled, irrespective of its severeness. These include holomorphic extensions -- Lefschetz thimbles, holomorphic flow, contour deformations, and complex Langevin dynamics --, and the introduction of new degrees of freedom -- dual variables and the tensor renormalisation group. We also highlight directions in which machine learning approaches have shown promise. Since many methods are first tested in simpler models, we provide an outlook on their feasibility for lattice systems.
\end{abstract}

\section{Introduction}
\label{sec:intro}

Properties of strongly interacting matter at non-zero temperature can be computed from first principles using non-perturbative simulations of Quantum Chromodynamics (QCD), rotated to Euclidean time and discretised on a four-dimensional grid, i.e., lattice QCD [\cite{PhysRevD.10.2445}]. This has produced a wealth of information on the thermal transition between the confined and the deconfined phase and the behaviour of hadrons as the temperature increases
[\cite{Aarts:2023vsf,Borsanyi:2025ttb}].
Extending these studies to finite baryon density, to map out the QCD phase diagram in the temperature–chemical potential plane, is, however, not straightforward. The famous (or rather infamous) {\em sign problem} is a fundamental stumbling block when the standard formulation of lattice QCD is used, rendering importance-sampling based numerical methods infeasible [\cite{deForcrand:2009zkb}]. Due to the presence of a complex phase in the Boltzmann weight, fluctuating from one field configuration to the next, importance sampling loses its efficiency. This problem gets more severe as the thermodynamic limit is taken at low temperature. 
Such a sign problem is not unique to QCD, but affects many theories in which the Boltzmann weight is not real and semi-positive, including bosonic theories with a chemical potential, and is hence of general interest. One could argue that the ultimate sign problem appears in real-time (as opposed to imaginary-time) dynamics, in which the path integral weight is a pure phase, $\exp(iS)$, with $S$ the Minkowski action. 

At a basic level, the sign problem can be seen as a computational issue and hence {\em all one needs to do} is to find an algorithm that gets around it. How hard can this be? After all, there has been tremendous progress in computational physics, lattice QCD, and hardware and software development over the past decades. 
As it turns out, the sign problem is indeed very hard -- some may say NP-hard [\cite{Troyer:2004ge}] -- and attempts to solve it has motivated a rich research programme in quantum and lattice field theory, going beyond purely algorithmic developments. It turns out that in the most ambitious approaches to solve the sign problem the path integral is reconsidered, deformed, analytically extended, or fully taken apart and reassembled, leading to new formulations of (lattice) field theories, with sometimes deep connections to mathematical physics, at first sight unrelated to lattice QCD.

In this overview, we summarise the status of the sign problem in lattice field theory, focussing on QCD and methods which may (eventually) allow exploration of the full phase diagram. 
The adoption of machine learning methods in (lattice) QCD has increased dramatically in recent years, see, e.g., the reviews [\cite{Cranmer:2023xbe,Kanwar:2024ujc,Aarts:2025gyp,Lawrence:2025rnk,Tomiya:2025quf}]. In the context of the sign problem, a number of directions is being explored, in particular in relation to contour deformations and complex Langevin dynamics. A discussion on progress in these directions is included.
We will not address the sign problem in quantum many-body systems nor focus on sign problems specific to fermions and their anticommuting nature [\cite{Troyer:2004ge}], which are less relevant for QCD with two light flavours. 
We also do not mention quantum algorithms, except in the outlook.

This contribution is organised as follows. In Sec.~\ref{sec:general} we summarise some basic facts about lattice QCD at non-zero chemical potential and the resulting sign problem. 
A brief outline on importance-sampling based methods is given in Sec.~\ref{sec:MC}. Two widely used field theories with a sign problem used to test novel approaches are introduced as well.
Sec.~\ref{sec:holo} starts the main focus of this review by  introducing holomorphic extensions to explore the complexified field space. We discuss Lefschetz thimbles, holomorphic flow, and contour deformations, as well as machine-learning inspired approaches. Contour deformations for variance reduction in the case of real actions are mentioned as well. Progress in complex Langevin dynamics, the other main approach to analytical continuation, is presented in Sec.~\ref{sec:CL}, again including recent machine learning contributions.
Real-time path integrals are reviewed in Sec.~\ref{sec:realtime}. 
A radical rewriting of the path integral is examined in Sec.~\ref{sec:dof}, via dual variables and tensor networks. 
The final section concludes. 
There are a number of excellent reviews on the sign problem in the context of lattice field theory [\cite{deForcrand:2009zkb,Aarts:2015tyj,Gattringer:2016kco,Berger:2019odf,Alexandru:2020wrj,Nagata:2021ugx}]; this review follows a similar pattern as the ones above, paying attention to advances over the past five years or so.

\section{QCD at non-zero baryon chemical potential and the sign problem}
\label{sec:general}

In the standard formulation of lattice QCD, see e.g.~the textbooks [\cite{Smit:2002ug,Gattringer:2010zz}], fermions are integrated out and the path integral over the remaining gluon fields, the links $U_{\nu}(x)$, is given by
\be
Z = \int DU\, \det M(U;\mu) \, e^{-S_{\rm W}[U]},
\ee
where $S_{\rm W}$ is the Wilson action (or an improvement thereof) and $\det M$ is the determinant of the fermion matrix, including $N_f$ quark flavours. Here $\mu$ denotes the quark chemical potential, with the baryon chemical potential given by $\mu_B=3\mu$. It is well known how to include the chemical potential in the lattice fermion action [\cite{Hasenfratz:1983ba}]; see also the lecture notes [\cite{Aarts:2015tyj}]. Observables are defined in the usual way,
\be
\bra O\ket = \frac{1}{Z} \int DU\, \det M(U;\mu) \, e^{-S_{\rm W}[U]}\, O(U).
\ee
The fermion determinant is real in the absence of a chemical potential, 
\be
\left[\det M(U;0)\right]^* = \det M(U;0) \in \mathbb R,
\ee
and positive for an even number of (light) quarks, enabling importance-sampling based simulations of lattice QCD in vacuum and at non-zero temperature. In the presence of a real chemical potential, however, the determinant is complex-valued,
\be
\label{eq:det}
\left[\det M(U;\mu)\right]^* = \det M(U;-\mu^*) \in \mathbb C.
\ee
The derivation, based on $\gamma_5$ hermiticity, takes only a few lines, see, e.g., the reviews [\cite{deForcrand:2009zkb,Aarts:2015tyj}]. 
For a purely imaginary (but unphysical) chemical potential, the determinant is again real [\cite{DElia:2002tig}].

The usual argument that the {\em sign problem} is (exponentially) hard follows from an analysis of the phase of the determinant. 
Since we are dealing with complex-valued observables $O$, we remind the reader that the variance is defined as 
\be
\label{eq:var}
\Var(O) = \big\bra | O -\big\bra O\big\ket |^2\big\ket = \big\bra |O|^2\big\ket - |\big\bra O\big\ket |^2.
\ee
It is real and non-negative, when evaluated with respect to a real and positive measure. It is also non-holomorphic, which will become relevant further down. 
This is in contrast to a susceptibility,
\be
\chi \sim \big\bra (O -\big\bra O\big\ket)^2\big\ket  =\big\bra O^2 \big\ket - \big\bra O\big\ket^2,
\ee 
which is holomorphic but can be complex-valued. 

Writing 
\be
\det M(U;\mu) = \left| \det M(U;\mu) \right| e^{i\varphi}
\ee
the expectation value of the phase can be estimated in the {\em phase-quenched} theory, i.e., the theory in which the phase is ignored in the measure, 
\be
\label{eq:sign}
\big\bra e^{i\varphi} \big\ket_{\rm pq}  
\equiv \frac{ \int DU\, e^{i\varphi} |\det M(U;\mu) | \, e^{-S_{\rm W}[U]}}{ \int DU\, |\det M(U;\mu)| \, e^{-S_{\rm W}[U]}} = \frac{Z}{Z_{\rm pq}} = \exp{[-(F-F_{\rm pq})/T]}.
\ee
This expectation value is equal to the ratio of the full and phase-quenched partition functions, by definition. Writing these as free energies, $Z_{\rm (pq)}=\exp[-F_{\rm (pq)}/T]$, the final expression shows that the expectation value of the phase depends exponentially on the difference between the free energies. 
Since these are extensive, and the two theories generally differ, the average phase goes to zero exponentially as the volume increases and the temperature decreases. The signal-to-noise ratio (SNR) deteriorates similarly,
\be
\mbox{SNR}\big(e^{i\varphi}\big) = \frac{|\big\bra e^{i\varphi} \big\ket_{\rm pq} | }{\sqrt{\Var(e^{i\varphi})} }
= \frac{|\big\bra e^{i\varphi} \big\ket_{\rm pq} |}{\sqrt{1- |\big\bra e^{i\varphi} \big\ket_{\rm pq} |^2}}
\to \exp{[-(F-F_{\rm pq})/T]}.
\ee
This exponential behaviour rules out naive implementations of {\em reweighting}, in which the phase is combined with an observable, as
\be
\label{eq:rew}
\bra O\ket = \frac{\bra e^{i\varphi} O\ket_{\rm pq}}{\bra e^{i\varphi} \ket_{\rm pq}}.
\ee
While formally correct, the poor SNR of the phase makes this approach feasible only for small systems. 
Numerically, it implies that the number of configurations should grow exponentially to overcome the sign problem: one observes that as the phase-quenched and the full theories become more distant, the averages in Eq.~(\ref{eq:rew}) are dominated by very few configurations.
Physically, the interpretation is that the target theory, say QCD at non-zero chemical potential, and the simulated theory, say phase-quenched  QCD, have poor overlap. Hence this manifestation of the sign problem is usually known as the {\em overlap problem}.

As an illustrative example, consider QCD with two light degenerate quarks. Writing the determinant separately for the up and down flavour, we find that under phase quenching  
\be
\left[ \det M(U;\mu) \right]^2
\to
\left|\det M(U;\mu) \right|^2 = \left[\det M(U;\mu)\right]^* \det M(U;\mu) = \det M(U;-\mu) \det M(U;\mu),
\ee
where we used relation (\ref{eq:det}). Hence for two degenerate flavours phase-quenched QCD corresponds to QCD at non-zero isospin chemical potential $\mu_{\rm iso}$ [\cite{Son:2000xc}]. 
This theory and QCD behave very differently at vanishing and low temperature. At strictly zero temperature, QCD at non-zero $\mu_{\rm iso}$ develops a pion condensate when $\mu_{\rm iso}$ exceeds $m_\pi/2$. On the other hand, in QCD at non-zero quark chemical potential, a transition occurs when $\mu$ exceeds (approximately) one third of nucleon mass, $m_N/3$, the onset to nuclear matter. Since $m_\pi/2<m_N/3$, the two theories differ radically physically when $m_\pi/2<\mu<m_N/3$. The absence of the formation of a condensate in QCD in this intermediate region has been dubbed the {\em Silver Blaze problem} and is another manifestation of the sign problem [\cite{Cohen:2003kd}]; for a recent review, see [\cite{Cohen:2026pzh}]. Despite the explicit dependence of the Boltzmann weight on $\mu$, exponentially delicate phase cancellations prevent a condensate from forming. 
Since the Silver Blaze problem is in essence a physical manifestation of the sign problem, the ability to reproduce Silver Blaze behaviour is a useful benchmark for numerical methods aimed at solving the sign problem. Of course, QCD at non-zero isospin chemical potential can also be studied in its own right [\cite{Brandt:2017oyy}].

The sign, overlap and Silver Blaze problems are all intricately connected, 
although their manifestation is specific to the approach followed to resolve them.
If the sign problem can be eliminated altogether, it is expected that all three problems will disappear. It should be noted, however, that the sign problem may re-emerge in unexpected ways.

\section{Importance-sampling based methods and the QCD phase diagram}
\label{sec:MC}

Understanding the phase structure of QCD at non-zero temperature and baryon chemical potential is a central objective of lattice field theory. While simulations at vanishing chemical potential are well established, the extension to finite density is hindered by the sign problem, as discussed above.
Current lattice studies of the QCD phase diagram rely therefore on a set of indirect methods, all of which are controlled only in the regime of small baryon chemical potential over temperature, and for which importance sampling is applicable.
Here we briefly mention three approaches. Comprehensive reviews of these and their status can be found elsewhere, see, e.g., 
[\cite{deForcrand:2009zkb,DElia:2018fjp,Aarts:2023vsf,Schmidt:2025ppy,Borsanyi:2025ttb,Ding:2026gao}].

In the Taylor series method, thermodynamic observables are expanded in powers of $\mu_B/T$. For instance,  the pressure is written as 
\be
\frac{p(T, \mu_B)}{T^4} = \sum_{n=0}^\infty c_n(T) \left(\mu_B/T\right)^n.
\ee
The coefficients $c_n(T)$ can be computed, up to some order, at $\mu_B=0$, where simulations are sign-problem free. 
This method has been widely used to determine the equation of state, fluctuations and higher moments of conserved charges, and the curvature of the crossover line in the $(T, \mu_B)$ plane separating the confined phase with broken chiral symmetry from the deconfined, chirally symmetric phase.
The applicability of this method is limited by the radius of convergence of the series and the rapidly increasing noise in higher-order coefficients, where the sign problem strikes back. In practice, it provides reliable results for small $\mu_B/T$, relevant for the region probed in high-energy heavy-ion collisions.
An alternative strategy is to simulate at imaginary chemical potential, $\mu=i\mu_I$, where the fermion determinant remains real, see Eq.~(\ref{eq:det}). Observables are then analytically continued to real $\mu$.
This approach has been widely used to extract the phase boundary and thermodynamic observables, often in combination with Taylor expansions, yielding consistent results.
Finally, reweighting techniques attempt to evaluate observables at non-zero $\mu$ by reweighting configurations generated at $\mu=0$, via
\be
\bra O \ket_{\mu\neq 0} = \frac{\bra O w(\mu)\ket_{\mu=0}}{\bra w(\mu)\ket_{\mu=0}},
\ee
where there is some freedom in choosing the optimal reweighting factor $w(\mu)$ [\cite{ deForcrand:2009zkb}], improving on the naive phase reweighting (\ref{eq:rew}).  
While formally exact, reweighting suffers from the overlap problem and is restricted to very small $\mu_B/T$ and moderate volumes.

Taken together, these methods provide a consistent picture of QCD thermodynamics at small baryon density in the chemical potential range $\mu_B/T \lesssim 3$, indicating that the finite-temperature transition at $\mu_B=0$ is a crossover. The curvature of the transition line as well as fluctuations and higher-order moments of conserved charges can be computed reliably.
However, the region of large baryon density and low temperature, where a critical endpoint or first-order transition may occur, remains inaccessible to these techniques.
See [\cite{Ding:2015ona,DElia:2018fjp,Aarts:2023vsf,Schmidt:2025ppy,Borsanyi:2025ttb,Ding:2026gao}] for details and a quantitative comparison.

The limitations of the Taylor expansion, imaginary chemical potential, and reweighting methods highlight the need for approaches that can address the sign problem directly. This has motivated the development of alternative techniques, which aim to enable first-principles calculations at any chemical potential. It should be noted, however, that these methods are not yet in a position to determine the QCD phase diagram, and hence are of a more exploratory nature. Research in this area is therefore often classified as high-risk, high-gain.

Since the implementation of a new approach directly in QCD is usually not feasible (or recommended, for that matter), it is common to first try a simpler field theory. 
A particularly well studied bosonic theory with a severe sign problem in $d=4$ dimensions is the one with a self-interacting relativistic complex scalar field at non-zero $\mu$ (often called the Bose gas), with the continuum action
 \be
 S = \int d^4x\,\left[ |\partial_\nu\phi|^2
 + (m^2-\mu^2)|\phi|^2 
 + \mu\left(\phi^*\partial_4\phi - \partial_4\phi^* \phi \right)
 + \lambda|\phi|^4 \right].
\label{eq11}
\ee
The euclidean action is complex and satisfies $S^*(\mu) = S(-\mu^*)$. On the lattice, with lattice spacing $a_{\rm  lat}\equiv 1$, the action reads
\be
 S = \sum_x \bigg[ \left(2d+m^2\right) \left|\phi(x)\right|^2 
 + \lambda \left|\phi(x)\right|^4 
- \sum_{\nu=1}^4\left(  \phi^*(x) e^{-\mu\delta_{\nu,4}} \phi(x+\hat\nu)
+ \phi^*(x+\hat\nu) e^{\mu\delta_{\nu,4}} \phi(x)\right)
\bigg].
\ee
This theory has some features which makes it a useful benchmark [\cite{Aarts:2008wh}]: it is a four-dimensional theory with a severe sign problem, according to the definition (\ref{eq:sign}); it is easier to implement approaches to solve the sign problem in this bosonic theory than in QCD; the theory has a Silver Blaze problem, with $\mu_{\rm onset}\approx m_R$, where $m_R$ is the renormalised mass at $T=0$.
Indeed, this theory has been solved with many of the methods we will discuss below. 

Popular fermionic models are theories with a four-fermion interaction term in $d=1,2,3$ spacetime dimensions. An example is the Thirring model at non-zero chemical potential, with the continuum action
\be
S = \int d^dx\, \left[ \bar\psi \left( \slashed{\partial} +\mu\gamma_0 +m\right)\psi +\frac{g^2}{2} \left(\bar\psi\gamma_\nu\psi\right)^2\right]. 
\ee
The fermions can be integrated out by introducing an auxiliary vector field, making this an effective bosonic model with a sign problem. There are various lattice formulations available, see, e.g., [\cite{Alexandru:2015sua,Alexandru:2017czx}] in the context of the sign problem.
The verification of results, ideally using the same theory and choice of parameters, allows for important cross-checking.

\section{Holomorphic extensions}
\label{sec:holo}

For elementary integrals, a common way to evaluate integrals with a complex-valued integrand is to deform the integration contour into the complex plane, relying on analytic continuation. We assume below that the integrand is holomorphic, such that Cauchy’s theorem applies.
Variations of this general idea have been explored  in lattice field theory since the 1980s [\cite{Parisi:1983mgm}], with variable success. 
Since many methods can be best explained using simple integrals, we use here the integral
\be
\label{eq:Z}
Z = \int_{\mathbb R} dx\, \rho(x), \qqquad\qqquad \rho(x) = e^{-S(x)} \in {\mathbb C},
\ee
as an illustration, but include only methods which have been implemented in field theory.
Qualitatively, there are two approaches [\cite{Sexty:2014dxa}]:
\begin{enumerate}
\item deform the integration contour, i.e., ${\mathbb R} \to {\cal M} \in {\mathbb C}$. For one degree of freedom, this entails
\be
\label{eq:Zth}
\int_{\mathbb R} dx\, \rho(x) \to \int_{\cal M} dz\, \rho(z) = \int_{\mathbb R} dt\, J(t)\rho(z(t)),
\ee
where ${\cal M}$ is the deformed contour and $z(t)=x(t)+iy(t)$ is some parametrisation with Jacobian $J(t) = dz(t)/dt$. Observables computed in the analytically extended theory should be equal to the target ones, i.e.,
\be
\left\bra O(x)\right\ket \equiv \frac{1}{Z}\int_{\mathbb R} dx\, \rho(x)O(x) \stackrel{?}{=} \frac{1}{Z} \int_{\cal M} dz\, \rho(z) O(z).
\ee
Proposals that fall in this category are Lefschetz thimbles, holomorphic flow, and contour deformations.
\item extend the integral into the complex manifold completely, i.e., ${\mathbb R} \to {\mathbb C}$, such that one can require the measure on the complexified manifold to be real and semi-positive. Again for one degree of freedom one writes
\be
\label{eq:ZC}
\int_{\mathbb R} dx\, \rho(x) \to \int_{{\mathbb R}^2} dxdy\, P(x,y),
\ee
with
\be
\left\bra O(x)\right\ket \equiv \frac{1}{Z}\int_{\mathbb R} dx\, \rho(x)O(x) \stackrel{?}{=} \frac{1}{Z} \int_{{\mathbb R}^2} dxdy\, P(x,y)O(x+iy).
\ee
Here the prime example is complex Langevin dynamics, but the problem of relating averages over complex-valued Boltzmann weights to statistical averages over real, semi-positive probability distributions is an open problem worth studying more generally [\cite{Weingarten:2002xs,Seiler:2017vwj,Salcedo:2018fvt}].

\end{enumerate}

\subsection{Holomorphic extensions in field theory}

We now describe how the analytical continuation or complexification is implemented in field theory. 

Let us start with the path integral for a real scalar field,
\be
Z = \int D\phi\, e^{-S[\phi]}, \qqquad\qqquad  \phi(x)\in\mathbb{R}.
\ee
The field is analytically extended as 
\be
\widetilde\phi(x) = \phiR(x) +i\phiI(x) \in \mathbb{C}.
\ee
Here and below we denote analytically continued fields as $\widetilde\phi(x)$.
Using the original field $\phi(x)$ to parametrise the contour, the path integral includes a Jacobian and is written as
\be
Z = \int_{\cal M} D\widetilde\phi\, e^{-S[\widetilde\phi]} = 
\int D\phi\, \det J(\phi) \, e^{-S[\widetilde\phi(\phi)]}, 
\qqquad\qqquad
J(\phi) = \left(\frac{\delta\widetilde\phi(x)}{\delta\phi(y)}\right).
\ee
Sampling along the contour should now be done with a (most likely complex-valued) effective action, which includes the Jacobian,  
\be
S_{\rm eff}[\phi] = S[\widetilde\phi(\phi)] - \log\det J(\phi).
\ee
Note that on a lattice with $\Omega= N_s^3N_\tau$ lattice sites, the Jacobian is an $\Omega\times \Omega$ matrix.

A complex scalar field is first written in terms of two real components $\phi_a$ ($a=1,2$), which are each continued analytically,
\be
\phi(x) = \frac{1}{\sqrt{2}} \left( \phi_1(x) + i\phi_2(x)\right) \in \mathbb{C}, 
\qquad \phi_a \in\mathbb{R} 
\qquad \to 
\qqquad \phi_a(x)\to \widetilde \phi_a(x) = \phi_a^\rmR(x) +i\phi_a^\rmI(x), 
\qquad \phi_a^{\rmR,\rmI}(x) \in \mathbb{R}.
\ee
The Jacobian is now a $2\Omega\times 2\Omega$ matrix.

Finally, for non-abelian gauge theories, SU($N$) matrices $U_{\nu}(x)$ are analytically continued to SL($N,\mathbb{C}$). This can be seen by writing a gauge link in terms of real vector potentials and analytically continue those,
\be
U_{\nu}(x) = \exp\left[i\sum_a T_a A_{a,\nu}(x)\right], \qqquad\qqquad A_{a,\nu}(x) \in\mathbb{R} \to A_{a,\nu}(x) \in\mathbb{C},
\ee
where $T_a$ are the traceless, hermitian generators of SU($N$). After the analytical extension, $\det U_\nu(x)=1$ but $U_{\nu}^\dagger(x)U_{\nu}(x)\neq \id$. 
Note that to use holomorphicity, it is necessary to first replace every appearance of $U_{\nu}^\dagger(x)$ by $U_{\nu}^{-1}(x)$ and only then analytically continue. 

When the integral is not deformed but properly extended into the complex manifold, the path integral measure is doubled. For instance, for a real scalar field one writes $\phi(x)=\phiR(x)+i\phiI(x)$ and
\be
\left\bra O(\phi)\right\ket = \frac{\int D\phi\, e^{-S[\phi]} O(\phi)}{\int D\phi\, e^{-S[\phi]} }
\qqquad \to 
\qqquad 
\left\bra O(\phi)\right\ket_P = \frac{\int D\phiR D\phi_I\, P[\phiR, \phiI] O(\phiR+i\phiI)}{\int D\phiR D\phi_I\, P[\phiR, \phiI]}.
\ee
Validity of the extended theory depends crucially on the properties of $P[\phiR, \phiI]$. Note that in the case of gauge theories, the original Haar measure is compact, while the complex extension is non-compact.

We continue below with a review of approaches that fall in the first category. Complex Langevin dynamics is discussed in Sec.~\ref{sec:CL}.

\subsection{Lefschetz thimbles}

The method of Lefschetz thimbles, based on Picard-Lefschetz theory, is in principle a mathematically rigorous approach to the sign problem. While the formulation is a classic topic in complex analysis, it was brought to the attention of lattice field theorists [\cite{Cristoforetti:2012su}] through a paper by [\cite{Witten:2010cx}].
The core idea is the first one mentioned above, namely to move the integration of the path integral from the original real manifold (on which the Boltzmann weight oscillates) to a specific union of curved manifolds in the complex space.
Thimbles can be defined as contours passing through critical points $\widetilde\phi^{(k)}_0(x)$, labelled by $k$, which are extrema of the action in the complexified space.
Along these thimbles the imaginary part of the Boltzmann weight is constant, see below. 
If Cauchy's theorem applies, the original integration contour can be deformed into a union of some of these contours without changing the value of the integral, but with a potentially significant improvement of the sign problem. 

We use a real scalar field $\phi(x)$ to demonstrate the approach and denote its analytic extension with $\widetilde\phi(x)$. We also introduce a fictitious or auxiliary time dependence, $\widetilde\phi(x,t)$.
Critical points and stable thimbles are determined by 
\be
\label{eq:thimblephi}
\frac{\partial S[\widetilde\phi]}{\partial\widetilde\phi(x)}\Bigg|_{\widetilde\phi=\widetilde\phi^{(k)}_0(x)}=0,
\qqquad\qqquad 
\frac{\partial \widetilde\phi(x,t)}{\partial t} = - \overline{\frac{\partial S[\widetilde\phi]}{\partial\widetilde\phi(x,t)}}.
\ee
The derivative is taken with respect to a parameter $t$ parametrising the thimble. A configuration on a stable thimble ends up at its associated critical point as $t\to\infty$. Unstable thimbles are obtained by reversing the sign of $t$. 
Using these equations, it is easy to see that the imaginary part of the action is indeed constant along a thimble, using the chain rule, since [\cite{Cristoforetti:2012su}]
\be
\frac{d}{dt}S[\widetilde\phi] =  - \left| \frac{\partial S[\widetilde\phi]}{\partial \widetilde\phi} \right|^2
\qqquad 
\mbox{or}
\qqquad
\frac{d}{dt}\Im S[\widetilde\phi] =0, 
\qqquad 
\frac{d}{dt}\Re S[\widetilde\phi] <0.
\ee
It also follows that along the stable (unstable) thimble the real part of the action is minimal (maximal) at the critical point.

We denote the stable thimbles as ${\cal J}_k$. The crucial mathematical statement [\cite{Witten:2010cx}] is that the integral along the union of all stable contributing thimbles, ${\cal M} = \cup\, {\cal J}_k$, is equivalent to the original path integral, i.e.,
\be
Z = \int D\phi\, e^{-S[\phi]}  
\to 
\int_{\cal M} D\widetilde\phi\, e^{ -S[\widetilde\phi]} 
= \sum_k n_k \int_{{\cal J}_k} D\widetilde\phi\, e^{ -S[\widetilde\phi]}
= \sum_k n_k e^{-i\im\, S[\widetilde\phi^{(k)}_0]} \int_{{\cal J}_k} D\widetilde\phi\, e^{ -\re\, S[\widetilde\phi]}.
\ee
Here the sum goes over all stable thimbles ${\cal J}_k$, $n_k$ is the intersection number of corresponding unstable thimble with the original contour, and it is noted that the imaginary part of the weight is constant along each thimble and hence can be taken out of each integral. 

\begin{figure}[t]
    \centering
    \includegraphics[width=0.6\linewidth]{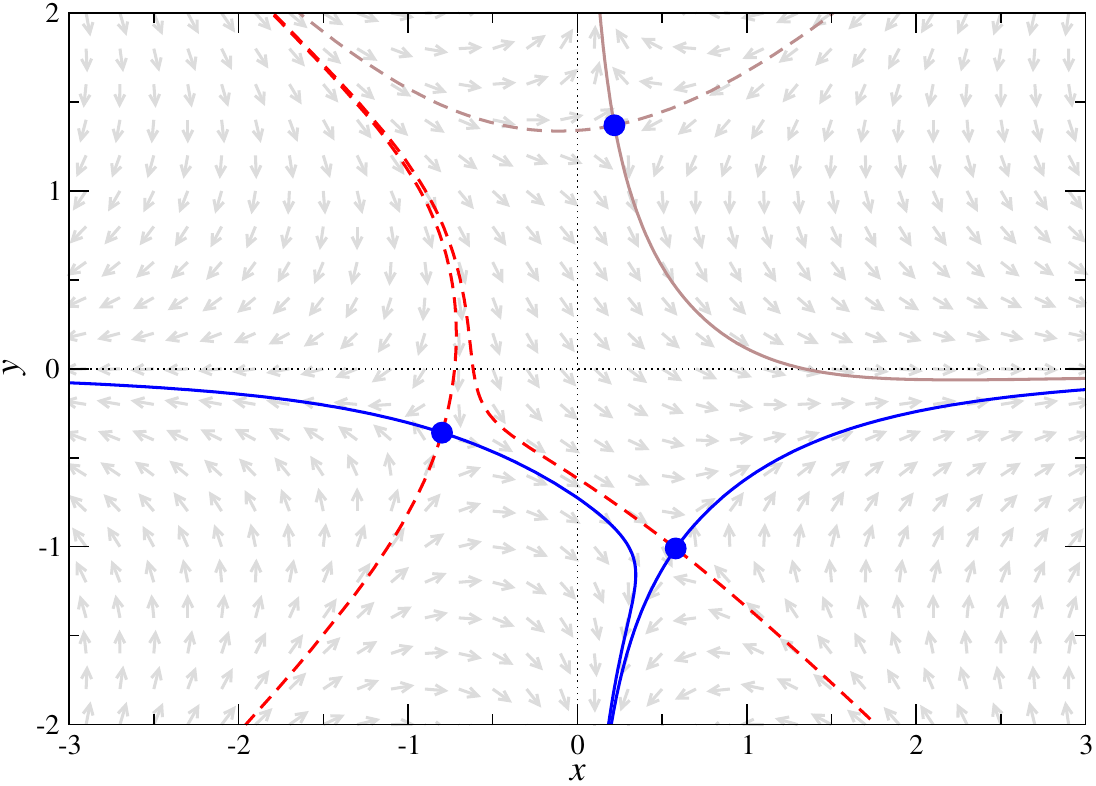}
\caption{Thimbles and holomorphic flow in the quartic model (\ref{eq:Sthimble}) with $\sigma=1$ and $h=1+i$: the blue circles denote the fixed points, the (normalised) arrows the holomorphic flow (\ref{eq:holo}), and full (dashed) lines the stable (unstable) thimbles. The two blue thimbles contribute. The third fixed point does not contribute. 
The flow {\em defining} a thimble, see Eq.~(\ref{eq:thimble}), is obtained by reversing the direction of the holomorphic flow. 
The complex Langevin drift (see below) is obtained by reversing the sign of the flow in the real direction.
Adapted from [\cite{Aarts:2014nxa}].
}
 \label{fig:thimble}
\end{figure}

Before going into more detail in field theory, it is useful to illustrate the approach with a simple example. We consider the partition function and action [\cite{Aarts:2014nxa}]
\be
\label{eq:Sthimble}
Z = \int dx\, e^{-S(x)}, 
\qqquad\qqquad
S(x) = \half\sigma x^2 + \frac{1}{4}x^4 +hx,
\qqquad\qqquad  h\in \mathbb{C}.
\ee
Thimbles are defined by (the dot denotes the time derivative)
\be
\label{eq:thimble}
 \partial_z S(z) \big|_{z=z^{(k)}_0}=0, 
 \qqquad\qqquad  \dot z = -\overline{\partial_z S(z)},  
\ee
and are given in Fig.~\ref{fig:thimble} for a specific choice of parameters, $\sigma=1$, $h=1+i$. 
Here the critical points $z^{(k)}_0$ ($k=1,2,3$) are indicated with blue circles, and stable (unstable) thimbles with full (dashed) lines. Only the two blue thimbles contribute. 
The critical point in the upper-half plane does not contribute, since the corresponding unstable thimble does not intersect the real axis. 
The (normalised) arrows indicate the holomorphic flow, see Eq.~(\ref{eq:holo}). 
The flow {\em defining} a thimble is obtained by reversing the sign of this flow. 
All critical points are saddle points under the flow. 
The original complex integral along the real axis can be replaced by the integral along the two curved (blue) paths. 

It is important to note that the sign problem has not been eliminated completely. First, two thimbles contribute and the imaginary parts of the action along these will in general differ. Hence there is a {\em global sign problem}, associated with the relative phase between thimbles. And second, along each thimble the Jacobian, $J(t) = dz/dt$ in this simple example, is complex-valued, resulting in a {\em residual sign problem}. Both of these sign problems also appear in field theory, to which we turn now.

The application of Lefschetz thimbles to lattice field theories with a sign problem was proposed in [\cite{Cristoforetti:2012su}] and quickly applied to the Bose gas at non-zero $\mu$ [\cite{Cristoforetti:2013wha,Fujii:2013sra}]. In the numerical implementation, some simplifying assumptions were made, which have already been alluded to in the context of Fig.~\ref{fig:thimble}. 
In a field theory, one expects a great number of critical points and thimbles, possibly increasing exponentially with the volume. An approximation is to include only one, potentially the dominant or main thimble, associated with the global minimum of the action, i.e., $\phi(x)=0$ for a scalar field theory in its symmetric phase. Including only one thimble bypasses the global sign problem. 
Concerning the complex-valued Jacobian, it has been argued that the residual sign problem along a thimble is possibly very mild, as the (real) weight along the thimble drops quickly and only the region close to the critical point contributes [\cite{Cristoforetti:2013wha}]. It can then be taken into account using reweighting. The method remains, however, numerically expensive, due to the need to compute the thimble(s) and the Jacobian, carefully mapping the deformation from the real manifold to the thimble(s). Subsequently one has to sample along the thimble(s), using, e.g., hybrid Monte Carlo or Langevin algorithms.  
Nevertheless, promising results have been obtained [\cite{Cristoforetti:2012su,Cristoforetti:2013wha,Fujii:2013sra}].

Further work has focussed on the inclusion of multiple thimbles. Here it has been observed that since the weight on each thimble is localised near the critical point, the integration domain effectively separates into disjoint domains for each thimble. This can cause issues with ergodicity. 
Sampling from the resulting multi-modal distribution can be improved using parallel tempering to connect the different thimbles [\cite{Fukuma:2017fjq,Alexandru:2017oyw}].
Many of these assumptions can be analysed in toy models, where it is often possible to determine all thimbles [\cite{Tanizaki:2015rda,DiRenzo:2017igr}], and in lower-dimensional (fermionic) models, where substantial progress can be made, see, e.g., [\cite{Ulybyshev:2019fte,DiRenzo:2021kcw}].
Some of these complications can be addressed using holomorphic flow as a natural extension of thimbles, to which we turn next.

\subsection{Holomorphic flow}

A key insight [\cite{Alexandru:2015sua}] is the realisation that any smooth manifold between the original one and the union of thimbles will have an improved sign problem, but suffer less from the issues raised above. A family of manifolds, parametrised by a flow time $t$, can be constructed by solving the (anti-)holomorphic flow equation,
\be
\label{eq:holo}
\frac{\partial \widetilde\phi(x,t)}{\partial t} =  \overline{\frac{\partial S[\widetilde\phi]}{\partial\widetilde\phi(x,t)}},  
\qqquad\qqquad 
\widetilde\phi(x,0) = \phi(x) \in \mathbb{R},
\ee
as an initial value problem, interpolating between the original manifold at flow time $t=0$ and a flowed manifold ${\cal M}_t$ at time $t$. The thimbles are recovered as $t\to\infty$. 
We note the opposite signs comparing Eqs.~(\ref{eq:thimble}) and (\ref{eq:holo}); the former determines the flow on a thimble, whereas the latter yields the flow towards a thimble. 
This formulation is known as {\em holomorphic flow} or the {\em generalised thimble approach}, referring to the holomorphic deformation of the integration manifold, not the anti-holomorphic flow equation itself. 
In Fig.~\ref{fig:thimble}, the (normalised) arrows indicate the holomorphic flow (\ref{eq:holo}), which also makes clear that all critical points are saddles. In this particular example, holomorphic flow yields a family of deformed contours in the lower half plane, lying between the real axis and the blue thimbles. One noticeable advantage is that the contributions from the two thimbles in the lower right quadrant effectively cancel as $y\to-\infty$, and a simpler manifold connecting the thimbles, obtained through holomorphic flow, might suffice. 
The Jacobian of holomorphic flow satisfies a first-order differential equation governed by the Hessian of the action, obtained by differentiating the flow equation with respect to the initial fields [\cite{Alexandru:2015sua}]. The Jacobian remains expensive to compute.

This method has been successfully applied to two-dimensional models, notably the Thirring model at finite density and temperature [\cite{Alexandru:2016ejd}], QED [\cite{Alexandru:2018ngw}], and U(1) gauge theory with a complex coupling [\cite{Pawlowski:2021bbu}], as well as the four-dimensional Bose gas [\cite{Alexandru:2016san}]. A detailed review is given in [\cite{Alexandru:2020wrj}].
Real-time dynamics will be discussed in Section \ref{sec:realtime}.

An important question is how to apply holomorphic flow to theories beyond the lower-dimensional ones mentioned above, without running into problems related to ergodicity, multi-modality, and numerical cost. [\cite{Fukuma:2020fez}] suggest to consider not only one manifold at flow time $t$, but instead a continuous union of all flowed manifolds with $t_0<t<t_1$. This smears the flowed manifold to non-zero thickness, creating a worldvolume of one dimension higher. It is argued that the extra flow-time direction allows a Markov chain to move around barriers that would be impenetrable on a single flowed manifold. 
This approach is similar in spirit to parallel tempering or annealed importance sampling, except that the tempering parameter is the flow time. An advantage is that the Jacobian only has to be computed when a measurement is performed. 
This {\em worldvolume Hybrid Monte Carlo} (WV-HMC) approach [\cite{Fukuma:2020fez}] has been applied to a chiral random matrix model [\cite{Fukuma:2020fez}] and the four-dimensional Bose gas [\cite{Namekawa:2024ert}]. Recent attention has shifted towards lattice gauge theory [\cite{Fukuma:2021aoo,Fukuma:2025gya,Fukuma:2025cxg}].
[\cite{Fujisawa:2021hxh}] contains an HMC algorithm evaluating the dynamics for generalised thimbles on the original contour, avoiding again issues with multi-modality.

\subsection{Contour deformations}

Thimbles and holomorphic flow provide specific deformations of the original manifold, determined by the gradient of the action. While these approaches improve the severity of the sign problem, in general they do not yield the optimal manifold for Monte Carlo sampling. As mentioned, large flow times are computationally expensive and the probability distributions typically become strongly multi-modal, with wide barriers of low probability separating isolated regions of high probability. Such distributions are difficult to sample through local updates. 
These observations have motivated a broader class of contour deformations, in which the integration manifold is treated as a variational object optimised to minimise phase fluctuations. 

We write again the partition function for a real scalar field along a manifold ${\cal M}$ in the complex plane, parametrised as $\widetilde\phi(\phi)$, 
\be
Z = \int D\phi\, e^{-S[\phi]} 
\to \int_{\cal M} D\widetilde\phi\, e^{-S[\widetilde\phi]} 
= \int D\phi\, \det J(\phi) \,  e^{-S[\widetilde\phi(\phi)]} 
= \int D\phi\, e^{-S_{\rm eff}[\widetilde\phi(\phi)]},  
\ee
with the effective action and phase angle,  
\be
S_{\rm eff}[\widetilde\phi(\phi)] = S[\widetilde\phi(\phi)] - \log\det J(\phi),
\qqquad\qqquad
\varphi(\phi) = -\im \, S_{\rm eff}[\widetilde\phi(\phi)].
\ee
The severity of the sign problem is determined by 
\be
\big\bra e^{i\varphi} \big\ket_{\cal M} 
= \frac{ \int D\phi\, e^{i\varphi(\phi)}  \, e^{-{\rm Re}\, S_{\rm eff}[\widetilde\phi(\phi)]}}{ \int D\phi\, e^{-{\rm Re}\, S_{\rm eff}[\widetilde\phi(\phi)] }},
\ee
which suggests to choose the manifold ${\cal M}$ such that phase fluctuations are minimised. Hence, contour deformations can be seen as a variational problem over manifolds in the complex extension of the original path integral domain.

This idea has been applied in the three-dimensional Thirring model by considering a parametrised deformation of one component of the auxiliary field into the complex plane [\cite{Alexandru:2018ddf}].
Similar parametrised deformations have been considered in a chiral random matrix model at non-zero $\mu$ [\cite{Giordano:2023ppk}]. A detailed analysis of integrals appearing in fermionic models is given in [\cite{Pasztor:2025wsj}], comparing Lefschetz thimbles, contours obtained by holomorphic flow at different flow times, and numerically optimised contours.
In some cases it turns out that even a constant shift into the complex plane is already sufficient to improve the sign problem substantially [\cite{Gantgen:2023byf}], something that has also been observed in complex Langevin dynamics [\cite{Aarts:2010gr}].
Considering complexified contours may also help in understanding real-time path integral representations of quantum tunnelling 
[\cite{Nishimura:2023dky}].
More formal aspects of contour deformations are discussed in [\cite{Lawrence:2023sfc}], where deformations are related to certain convex optimisation problems. In particular, it is shown that for two-dimensional U(1) gauge theory with a complex coupling it is not possible to remove the sign problem for many values of the coupling via contour deformation.
Finally, a method where the path integral is modified by a {\em subtraction}, which integrates to zero but improves the severity of the sign problem, can be found in [\cite{Lawrence:2022dba}]. 
Since contour deformations can be seen as optimisation problems, it is natural to consider them in the context of machine learning, to which we return below.

\subsection{Contour deformations for variance reduction}

While contour deformations as discussed above are developed to address sign problems in theories with a complex action, similar ideas can be applied in a different context: the reduction of signal-to-noise ratios in conventional lattice QCD calculations, where the path integral measure is real and positive. Since this approach relies on holomorphicity and Cauchy's theorem, as above, it is worth to consider this problem as well. We follow the presentation of [\cite{Detmold:2020ncp}].

In standard lattice QCD simulations, expectation values (with or without dynamical quarks at zero chemical potential)
\be
\label{eq:O}
\left\bra O\right\ket = \frac{\int DU\, e^{-S[U]} O(U)}{\int DU\, e^{-S[U]}}
\ee
can be evaluated using importance sampling. However, many observables of physical interest, including Wilson loops and hadronic correlation functions, exhibit exponentially degrading signal-to-noise ratios. This phenomenon is closely related to the presence of large phase fluctuations in the observable itself, even when the measure is positive.

This observation suggests a reinterpretation of signal-to-noise problems as a form of effective sign problem in the observable. Building on this idea, contour deformations can be applied not to the sampling measure, but to the observable itself. It relies on the fact that holomorphic observables are unchanged when the contour is deformed, but the variance, which is non-holomorphic, is not [\cite{Detmold:2020ncp}].
Because the method does not require modifications to existing sampling algorithms, it can in principle be combined with standard techniques such as hybrid Monte Carlo and multi-level algorithms.

Let us define a new integration manifold ${\cal M}$ with links $\widetilde U$, parametrised as $\widetilde U(U)$. Assuming holomorphicity, we can write
\be
\left\bra O\right\ket_{\cal M}
= \frac{1}{Z} \int_{\cal M} D\widetilde U\, e^{-S[\widetilde U]} O(\widetilde U) 
= \frac{1}{Z} \int DU\, \det J(U) \, e^{-S[\widetilde U(U)]} O(\widetilde U(U)) 
= \frac{1}{Z} \int DU\,  e^{-S[U]} \widetilde O(U) = \bra \widetilde O\ket, 
\ee
where the {\em deformed observable} now includes the Jacobian and the action difference,
\be
\widetilde O(U) = \det J(U) \, e^{S[U]- S[\widetilde U(U)]} O(\widetilde U(U))
\qqquad\qqquad
J(U) = \left(\frac{\delta\widetilde U}{\delta U} \right).
\ee
The expectation value of $\widetilde O(U)$ with respect to the original path integral weight should be identical to the desired one in Eq.~(\ref{eq:O}), for any deformation that satisfies Cauchy's theorem. As stated, configurations are generated here with respect to the real-valued weight $\sim\exp(-S[U])$ and only the manner in which observables are computed is changed.  

To determine the optimal manifold, the aim is to minimise the variance of an observable. Consider an observable $O$ with expectation value $\bra O\ket = \bra \widetilde O\ket$. The non-holomorphic variances (\ref{eq:var}) of the original and the deformed observable will in general differ, $\Var(\widetilde O) \neq \Var(O)$, since $\widetilde O$ is complex-valued after the deformation, even when $O$ is real. The task is therefore to find a manifold on which the variance is minimal, beating the signal-to-noise problem. As above, this can be formulated as an optimisation problem, for instance by introducing parametrised families of maps, $U\to \widetilde U(U;\alpha)$,
and optimising the parameters $\alpha$, e.g., using machine-learning techniques. 
Recent applications have focused on Wilson loops in two-dimensional SU(2) and SU(3) and three-dimensional SU(2) gauge theory; it has been shown that even constant imaginary shifts can improve the signal-to-noise ratio considerably [\cite{Detmold:2021ulb,Lin:2023svo}].

\subsection{Machine learning thimbles, flows and contour deformations}

Generalised thimbles, holomorphic flow, and contour deformations can be formulated in terms of maps in complexified field space, which are then amenable to parametrisation and optimization using machine learning methods.
From this perspective, the sign problem becomes an optimisation problem over a space of transformations,
\be
\label{eq:ftheta}
\phi(x) \to \widetilde\phi(x)= f_\theta[\phi](x)
\qqquad \mbox{or} \qqquad
\phi(x) \to \widetilde\phi(x)= \phi(x) + i f_\theta[\phi](x),
\ee
where $f_\theta[\phi]$ is a parametrised map, implemented for instance as a neural network (here and below, $\theta$ denotes the collection of trainable parameters). As above, the goal is to construct a transformation that reduces phase fluctuations and improves the signal-to-noise ratio.

The deformed path integral takes the same form as before, 
\be
Z = \int D\phi\, e^{-S[\phi]} 
\to  
\int_{{\cal M}_\theta} D\widetilde\phi\, e^{-S[\widetilde\phi]} 
= \int D\phi \, \det J_\theta(\phi) \,  
e^{-S[\widetilde\phi_\theta(\phi)]} 
= \int D\phi\, e^{-S_{\rm eff}[\widetilde\phi_\theta(\phi)] },
\ee
with the Jacobian, effective action and phase angle
\be
J_\theta(\phi) = \left(\frac{\delta\widetilde\phi_\theta(x)}{\delta\phi(y)}\right),
\qqquad\qqquad
S_{\rm eff}[\widetilde\phi_\theta(\phi)] 
= S[\widetilde\phi_\theta(\phi)] -\log\det J_\theta(\phi),
\qqquad\qqquad
\varphi = -\Im S_{\rm eff}[\widetilde\phi_\theta(\phi)].
\ee
The learning problem is then to determine $f_\theta[\phi]$ such that phase fluctuations are minimised.

Various objective functions or loss functions have been proposed, all related to the optimisation of the average phase $|\bra e^{i\varphi} \ket |$ on the manifold ${\cal M}_\theta$. We note that maximising the average phase factor is identical to minimising its variance, since
\be
\Var(e^{i\varphi}) = 1- |\bra e^{i\varphi} \ket |^2.
\ee
Minimising the variance of the imaginary part of the effective action leads to comparable results, since $\Var(\varphi) = \Var(e^{i\varphi})$ when $\varphi\sim 0$.

We now discuss several implementations of these ideas.
[\cite{Alexandru:2017czx}] introduced the concept of a {\em learnifold}, an approximation to a flowed manifold ${\cal M}_t$. A feed-forward network encodes the imaginary part of the field, see Eq.~(\ref{eq:ftheta}), and is trained using a small number of field configurations obtained via holomorphic flow (supervised learning).
The trained network then defines a new manifold on which the sign problem is reduced and sampling is more efficient. Symmetries, such as lattice translations, can be implemented in the network. 
This method has been applied to the two-dimensional Thirring model at non-zero $\mu$ and it is found that problems related to multi-modality, a stumbling block in (generalised) thimbles, are reduced. 

In the {\em path optimisation method} [\cite{Mori:2017pne}], contour deformation is treated as an unsupervised learning problem, in which the imaginary part of the field, see Eq.~(\ref{eq:ftheta}), is learned by maximising the average phase.
The method was first applied to a simple model using an explicit parametrisation of the contour [\cite{Mori:2017pne}] and subsequently to the two-dimensional Bose gas at non-zero $\mu$ using a feed-forward neural network [\cite{Mori:2017nwj}], the Polyakov-loop extended Nambu--Jona-Lasinio model [\cite{Kashiwa:2019lkv}], as well as to a single plaquette in $U(1)$ gauge theory with a complex coupling [\cite{Kashiwa:2020brj}]. 
The method has recently been extended to the one-dimensional massive lattice Thirring model [\cite{Hisayoshi:2025adi}].

Normalising flows are invertible, parametrised maps with a tractable Jacobian, usually encoded in neural networks, which transform field configurations from a simple prior, such as the normal distribution, to configurations representative of the target distribution, $\sim\exp(-S[\phi])$. This approach is a popular generative method in lattice field theory [\cite{Albergo:2019eim,Kanwar:2020xzo,
Cranmer:2023xbe}].
[\cite{Lawrence:2021izu}] show how normalising flows may be generalised to path integrals with a sign problem, including real-time dynamics. Moreover, it is argued that {\em complex normalising flows} only exist when an integration contour is available which solves the sign problem exactly and that this is always possible locally, by modifying holomorphic flow. Flow-based methods have also been considered in the context of the density of states [\cite{Fodor:2007vv,Langfeld:2012ah}] for complex actions [\cite{Pawlowski:2022rdn}] and for the Hubbard model away from half-filling using complex-valued neural networks [\cite{Rodekamp:2022xpf}].

Finally, [\cite{Ihssen:2026njd}] recently proposed a new generative framework based on so-called {\em physics-informed kernels} (PIKs). As before, the idea is to construct a generalised contour deformation or transport map, which is now constructed via physics-informed kernels and with analytical control of the mapping. These kernels provide explicit access to the information transport between complex distributions and (nearly) real and positive ones. First results in toy models and for real-time dynamics are promising.

\section{Complex Langevin dynamics}
\label{sec:CL}

Contour deformations do not resolve the sign problem completely. It would be desirable to find a representation of the theory on the complexified manifold in terms of a real and semi-positive probability density $P[\phiR, \phiI]$, such that 
\be
\label{eq:CLErealcomp}
 \int D\phiR D\phiI\, P[\phiR, \phiI] O(\phiR+i\phiI) 
=  \int D\phiR\, \rho[\phi_R] O(\phi_R),
\qqquad\qqquad \rho[\phiR] \sim e^{-S[\phiR]} \in \mathbb{C}, 
\qqquad P[\phiR, \phiI] \in \mathbb{R}, 
\ee
for all holomorphic observables $O(\phi)$. Here we assumed that both distributions are normalised. If we are allowed to shift integration variables and demand Eq.~(\ref{eq:CLErealcomp}) to hold for all observables, we find the relation
\be
 \label{eq:CL_ID}
 \int D\phiI\, P[\phiR-i\phi_I, \phiI] = \rho[\phi_R],
\ee
which indicates how the complex and real-valued weights are related.
Such representations can exist quite generally [\cite{Weingarten:2002xs,Seiler:2017vwj,Salcedo:2018fvt}], but constructing them is not necessarily easy. Complex Langevin dynamics is one approach to arrive at such a distribution. Complex Langevin dynamics explores the complexified manifold via a stochastic process [\cite{Parisi:1983mgm,klauder}], which is the complexification of the Langevin process used in stochastic quantisation [\cite{Parisi:1980ys,Damgaard:1987rr}].
For a real scalar field $\phi(x)$ with complex extension $\widetilde\phi(x,t)$, depending on an auxiliary {\em Langevin time} $t$, this process is given by 
\be
 \frac{\partial\widetilde\phi(x,t)}{\partial t} 
= -\frac{\partial S[\widetilde\phi]}{\partial \widetilde\phi(x,t)} +\eta(x,t)
\qqquad\qqquad
\bra \eta(x,t)\eta(x',t')\ket = 2\delta(t-t')\delta(x-x'),
\ee
 where $\eta(x,t) $ is uncorrelated random noise.
Comparing the dynamics with the one encountered in holomorphic flow and on Lefschetz thimbles, see Eqs.~(\ref{eq:thimblephi}, \ref{eq:holo}), we note that the complex conjugation in the drift term is absent. A consequence is that the saddle points encountered in thimbles and holomorphic flow turn into attractive and repulsive fixed points under the complex Langevin drift [\cite{Aarts:2013fpa,Aarts:2014nxa}], see Fig.~\ref{fig:thimble} in the context of the simple model discussed above. 
The effect of the noise is to sample the complexified field space properly, i.e., not to be confined to a manifold of the same dimension as the original one. We note that the distribution on the LHS of Eq.~(\ref{eq:CLErealcomp}) is not known explicitly (see below), but is effectively sampled by the stochastic process.

After complex Langevin dynamics was introduced in the early 1980s [\cite{Parisi:1983mgm,klauder}], it was explored to some extent but  both numerical and conceptual issues were encountered, see, e.g., [\cite{Ambjorn:1985iw,Ambjorn:1986fz}]. The method was revived in the 2000s, first for real-time dynamics [\cite{Berges:2005yt}] and subsequently at non-zero chemical potential [\cite{Aarts:2008rr}]. 
Considerable progress has been made since, but the method is not still not without its problems. 

It turns out that the equivalence (\ref{eq:CLErealcomp}) can fail in several ways. 
First of all, the numerical solution of the complex Langevin equation can be unstable due to {\it runaway trajectories}, paths of the fields where infinity is reached in a finite Langevin time in the absence of noise. In the presence of noise, using a discretised form of the Langevin equation, paths close to runaway trajectories can be sensitive to this, leading to numerical instabilities. This can be improved by using a very small or (preferably) adaptive stepsize  [\cite{Aarts:2009dg}] or by using implicit updates [\cite{Alvestad:2021hsi}].
Second, the complex Langevin process can give {\em ``wrong'' convergence}, i.e., the distribution effectively sampled by the process does not satisfy the correspondence (\ref{eq:CLErealcomp}, \ref{eq:CL_ID}). Since the expected results are not known in realistic cases, this problem is harder to diagnose.
Importantly, the failure is not linked to the severeness of the sign problem and can already occur in toy models 
[\cite{Ambjorn:1985iw,Aarts:2010aq,Aarts:2012ft}].
On the other hand, there exist numerous examples where complex Langevin dynamics gives correct and non-trivial results, including for lattice models with a severe sign problem in three [\cite{Aarts:2011zn}] and four [\cite{Aarts:2008wh}] Euclidean dimensions. Significantly, this includes QCD in four dimensions, see Sec.~\ref{sec:QCD}.

A careful derivation of the formal relation between the complex distribution on the real manifold and the real and semi-positive distribution on the complexified manifold has led to considerable insight and developments [\cite{Aarts:2009uq,Aarts:2011ax,Nishimura:2015pba,Aarts:2017vrv,Seiler:2020mkh,Seiler:2023kes,Mandl:2025mav}].
It is now understood that the failure of the complex Langevin approach can be attributed to {\em boundary terms}, which can in principle arise at infinity and near zeroes of the measure on the complexified manifold. This resulted in practical {\em criteria for correctness} involving boundary terms, which need to be verified {\em a posteriori} [\cite{Aarts:2009uq,Aarts:2011ax,Nishimura:2015pba,Scherzer:2018hid,Scherzer:2019lrh}], as well as the related {\it drift criterion}, which involves the norm of the drift terms [\cite{Nagata:2016vkn,Nagata:2018net}].
Even in the absence of boundary terms, one can arrive at incorrect results due to the possibility of complex Langevin dynamics yielding results on different {\it integration cycles}. This will be discussed below.
When the naive complex Langevin process fails, it can sometimes be cured by changing the stochastic process, e.g., by using a variable transformation [\cite{Mollgaard:2014mga}], reformulating the theory
[\cite{Berges:2007nr,Boguslavski:2024yto}], dampen the fluctuations in imaginary directions
[\cite{Attanasio:2018rtq,Hansen:2024lkn}], or by introducing kernels, as described below.

A recent, thorough comparison of many proposed criteria for correctness [\cite{Mandl:2026vdc}] studies also a recent addition: the so-called {\em configurational temperature}, which tests for a correct implementation and thermalisation of the stochastic process (and can be used for any Monte Carlo process) [\cite{Dhindsa:2025xfv,Joseph:2025fcd,Joseph:2025xbn}].

\subsection{Kernels}

The theoretical understanding of (complex) Langevin dynamics comes via the Fokker-Planck equation, describing the evolution of the probability distribution. For a real scalar field, with a real action and drift term (i.e., without a sign problem), it is given by 
\be
\frac{\partial P[\phi,t]}{\partial t} = L^T P[\phi,t],
\qqquad\qqquad
L^T = \int d^dx\, \frac{\partial}{\partial \phi(x)} \left(\frac{\partial}{\partial\phi(x)} + \frac{\partial S[\phi]}{\partial\phi(x)} \right).
\ee
Here $L^T$ is the (formally adjoint of the) Langevin operator, see, e.g., [\cite{Aarts:2009uq}]. It is easy to find the stationary distribution, $P[\phi]\sim \exp(-S[\phi])$, satisfying $L^TP[\phi]=0$.
Kernels allow for a generalisation of the Langevin process, without changing the stationary distribution (in the real case).
The stochastic equation is modified as (see, e.g., [\cite{Alvestad:2022abf}]),
\be
\label{eq:kernlang}
\frac{\partial\phi(x,t)}{\partial t} 
= - \int d^dx'\, K(x,x';t) \frac{\partial S[\phi]}{\partial\phi(x',t)}
 + \int d^dx'\, \frac{\partial K(x,x';t)}{\partial\phi(x',t)}
 + \int d^dx'\, H(x,x';t) \eta(x',t),
\ee
with
\be
K(x,x';t) = \int dx''\, H(x,x'';t)H(x'',x';t).
\ee
This structure is chosen such that the corresponding Fokker-Planck equation remains as above, with a modified adjoint Langevin operator, 
\be
\label{eq:LKT}
L^T_{K} = \int d^dx\, d^dx'\, \frac{\partial}{\partial \phi(x)}  K(x,x';t)\left(\frac{\partial}{\partial\phi(x')} + \frac{\partial S[\phi]}{\partial\phi(x')} \right).
\ee
Since the kernel $K$ appears as a factor up-front, the stationary solution is unchanged, as promised. 
In its most general form, the kernel can depend on Langevin time, spacetime coordinates, and the fields themselves. A simple example is to include only time dependence, $H(x,x';t)=g(t)\delta(x-x')$, $K(x,x';t)=g^2(t)\delta(x-x')$, which yields
\be
\frac{\partial\phi(x,t)}{\partial t} 
= - g^2(t)\frac{\partial S[\phi]}{\partial\phi(x,t)} + g(t)\eta(x,t),
\qqquad\qqquad
L^T_K = g^2(t) L^T_{K=\id},
\ee
i.e., a time-dependent diffusion coefficient. This diffusion coefficient can also be field dependent.
The kernel changes the time-dependent behaviour of the stochastic process, which can be used to improve  thermalisation rates or autocorrelations, e.g., by using Fourier acceleration [\cite{Batrouni:1985jn}].

Returning now to complex actions, here kernels can be complex-valued and change the equilibrium distribution on the complexified manifold. This follows from the study of the Fokker-Planck equation underpinning the complex Langevin process, whose adjoint Langevin operator differs from $L_K^T$.
Moreover, it has been known for some time that kernels can indeed change an incorrect result into a correct one [\cite{Okamoto:1988ru,Okano:1991tz,Hansen:2024kjm}]. In [\cite{Aarts:2012ft}] it was shown that a subset of kernels can be thought of as variable transformations, i.e., that kernels are more general modifications of the stochastic process. 
While the complex Langevin approach and contour deformations (including thimbles and holomorphic flow) both rely on analytical continuation, they differ in detail: the complex Langevin method extends the theory to a complexified manifold with twice the dimension of the real manifold, while contour deformations move the original real integration manifiold into the complex extension.
Nevertheless, the distributions effectively sampled by the complex Langevin process tend to concentrate around thimbles [\cite{Aarts:2013fpa,Aarts:2014nxa}].
Distributions on or around different thimbles in the theory can be changed using kernels [\cite{Hansen:2024kjm}].
The relation between both methods was also elucidated in [\cite{Nishimura:2017vav}]. 
A transformation similar to kernels can also be used in the case of generalised Lefschetz thimbles, to avoid instabilities related to the multi-modality in large systems  [\cite{Nishimura:2024bou}].
Kernels have mostly been explored for real-time dynamics, to which we turn in Sec.~\ref{sec:realtime}.

\subsection{Integration cycles}

Integration cycles are paths on the complexified field manifold, which connect zeroes of the complex measure $\rho$. Note that the Lefschetz thimbles above can also be thought of as integration cycles, or representatives thereof.
Normally theories are defined by integration over the real manifold (denoted as the ``real cycle''), connecting zeroes at infinity, but in general zeroes not at infinity are also possible.
According to Cauchy's theorem,  paths can be continuously modified without changing the results of the integral. Thus, in the complexified theory, one could define expectation values as integrals over some, properly normalised, integration cycles. Typically one considers a subset of cycles such that all zeroes of the measure are reachable with linear combinations of cycles from a minimal spanning set.

According to the Salcedo-Seiler theorem [\cite{Salcedo:2018fvt}], in the absence of boundary terms, the results of the complex Langevin approach should be given by a linear combination of the results on the possible integration cycles of the theory. 
This was proven for a single degree of freedom [\cite{Salcedo:2018fvt}, see also \cite{Pehlevan:2007eq}] and evidence was presented that it holds also more generally, for more degrees of freedom [\cite{Hansen:2024kjm}].
It was demonstrated for a toy model [\cite{Hansen:2024kjm}] that using different kernels in Eq.~(\ref{eq:kernlang}) can lead to different equilibrium states of the stochastic process, and that in some of those boundary terms are absent. Complex Langevin results are then given by different linear combinations of the results on integration cycles. This is of course only explicitly verifiable when exact results are known. A new necessary and sufficient criterion for correctness, which can help differentiate between integration cycles, was developed recently [\cite{Mandl:2025mav,Mandl:2025ins}].
In fact, the conditions for complex Langevin dynamics to give results which correspond to the integration over the cycle ${\cal I}$ are that all Schwinger-Dyson equations are fulfilled as well as certain bounds.
For one degree of freedom, $z=x+iy$, these bounds are based on a split of the original integration measure $\rho(z) = w(z) \rho_r(z) \in \mathbb{C}$ into a positive and real $w(z)$ and a complex remainder $\rho_r(z)$. The bounds are then
\bea
 \left| \int dx dy\, P(x,y) f(x+iy) \right| < C\,  || f ||_w,
\eea
with $P(x,y)$ the equilibrium probability of the complexified process and $f(x+iy)$ essentially any observable of interest. The norm on the RHS is given by the integral over a representation of the cycle ${\cal I}$, parametrised as $z(t)$ with Jacobian $J(t)=dz(t)/dt$,
\bea
|| f ||_ w = \left( \int dt\, J(t)\, \big| \, f(z(t)) \, w(z(t)) \, \big| ^{p/(p-1)} \right)^{(p-1)/p} 
\underset{{\cal I} = \mathbb{R}}{=}  \;
\left( \int dx\, \big| \, f(x) w(x) \, \big|^{p/(p-1)} \right)^{(p-1)/p},
\eea
 where $p\ge 1$ can be chosen at will and and $C$ is a constant given by
\bea
 C=\left( \int dt\, J(t)\, \big| \,\rho_r(z(t)) \, |^p \right)^{1/p}
\underset{{\cal I} = \mathbb{R}}{=} \;
 \left( \int dx\, \big| \,\rho_r(x) \, \big|^p \right) ^{1/p}.
\eea
Choosing ${\cal I} = \mathbb{R}$ leads to the simpler formulas on the RHS. Since the conditions must hold for all observables, this criterion is typically used by finding an observable $f$ (e.g., some polynomial of the fields) and a $p$ such that the bounds are violated.
This criterion was successfully used to differentiate complex Langevin results corresponding to different integration cycles as the kernel is changed [\cite{Mandl:2025mav}].

\subsection{Non-abelian gauge theories and QCD}
\label{sec:QCD}

The naive application of complex Langevin dynamics to non-abelian gauge theories with a complex drift term breaks down quickly, as the complexified gauge degrees of freedom increasingly fluctuate in the non-compact direction of SL($N, \mathbb{C}$) and boundary terms are rapidly induced. This can be monitored using, e.g., the unitary norm $\Tr (UU^\dagger)/N\geq 1$ [\cite{Aarts:2008rr}].
To counter this behaviour, a new procedure called gauge cooling was introduced [\cite{Seiler:2012wz}], which uses complexified gauge transformations to nudge the fields closer to the real manifold after every update. Further analysis of gauge cooling can be found in [\cite{Aarts:2013uxa,Nagata:2015uga,Cai:2019vmt,Cai:2020tgd,Dong:2020mtk}].
Using this procedure the whole phase diagram of HDQCD (heavy dense QCD, where spatial hopping terms of the quarks are neglected)
was mapped out using complex Langevin simulations [\cite{Aarts:2016qrv}].
Gauge cooling allows for simulations of full QCD as well [\cite{Sexty:2013ica,Kogut:2019qmi,Scherzer:2020kiu,Attanasio:2022ogk,Ito:2020mys,Tsutsui:2025jez}]. 
Dynamical stabilisation [\cite{Attanasio:2018rtq}] provides an alternative way to control large excursions on the complexified manifold.
As a side comment, we note that dynamical stabilisation has also been applied to block copolymer melts in chemical physics [\cite{Matsen2024,Matsen2026}].
Similar to lattice QCD simulations employing hybrid Monte Carlo, most of the computational cost is spent on calculating the fermionic drift or force terms, which are defined as the left derivative $D_{x\nu a}$ of the log of the determinant of Dirac matrix $M$ with respect to the link variable $U_{x\nu}$.
For large systems this is facilitated using noisy estimators,
\be
D_a {\ln \det M}=\Tr({ M^{-1} D_a M } )= \frac{1}{\Omega} \langle \eta^+ 
(D_a M )M^{-1} \eta \rangle,
\ee
with the spacetime volume $\Omega$ and a noise field $\eta$ satisfying $ \langle \eta_{x} \eta_y \rangle = \delta_{xy} $. For staggered fermions, rooting is particularly easy in this setup: using $N_f/4$ degenerate quark flavours just leads to a factor of $N_f/4$ in front of the drift term. 
Improved actions, such as the Symanzik action for the gauge fields, can also be implemented. For quarks, one has to 
stick to holomorphic improvement methods; while smearings using projections are 
not well-defined on the complexified manifold, stout smearing is usable [\cite{Sexty:2019vqx}].
Exploratory studies of the use of kernels for QCD at non-zero density can be found in [\cite{Mandl:2025ins}]. The main challenge applying kernels in non-abelian gauge theories is to find a sufficiently non-trivial kernel which keeps the gauge symmetry intact.
Thermodynamic quantities such as the pressure can be determined [\cite{Sexty:2019vqx}] using the integral formula 
\bea
\ln Z(\mu) - \ln Z(0) 
= \int_0^\mu d\mu'\, \frac{ \partial \ln Z}{\partial \mu'} 
=  \Omega \int_0^\mu d\mu'\, n(\mu'),  
\eea
where the quark number density $n(\mu)$ at non-zero $\mu$ is measured using complex Langevin simulations.

The method has been applied to QCD in four dimensions, in the heavy dense limit [\cite{Seiler:2012wz,Aarts:2016qrv}],
to all orders in the hopping expansion [\cite{Aarts:2014bwa}],
and with dynamical quarks 
[\cite{Sexty:2013ica,Fodor:2015doa,Attanasio:2018rtq,Sexty:2019vqx,Kogut:2019qmi,Scherzer:2020kiu,Ito:2020mys,Attanasio:2022mjd,Tsutsui:2025jez}].

The thermal transition line in QCD, $T_c(\mu)$, indicating the change from the confined to the deconfined phase, was calculated using Wilson fermions, with quark masses much larger than in the nature, up to large chemical potentials [\cite{Scherzer:2020kiu}]. It was observed that the line is well described by a quadratic function of $\mu$.

Very recently, the first QCD simulations at the physical point, i.e., using physical quark masses, have appeared, including a continuum extrapolation [\cite{Mandl:2026ngb}].
The equation of state was computed at temperatures above the thermal crossover and up to large chemical potentials, $0\leq \mu_B/T \lesssim 12$.
The values of chemical potential that are reachable are limited by a discretisation effect: at large chemical potentials all the available fermionic modes are filled, which is known as {\em saturation}. This lattice artefact goes away in the continuum limit.
In practice, for a given set of lattice spacings this limits the chemical potential up to which simulations are useful.
With regard to the temperature, simulations are bound by the cost of the fermion matrix inversion, which becomes more expensive as the temperature is lowered and the condition number of the Dirac matrix increases quickly. Hence simulations at the physical point are currently restricted to the high-temperature phase, where this is under control.

\subsection{Machine learning the distribution}

Diffusion models are a class of generative methods that learn directly from data [\cite{2510.21890}]. After training, they can be used to generate additional configurations and to investigate the (unknown) distribution underpinning the data set. The dynamics of diffusion models, a stochastic process in a fictitious time direction, is closely related to stochastic quantisation [\cite{Wang:2023exq}] and first applications to two-dimensional field theories are available for scalar [\cite{Wang:2023exq}], U(1) [\cite{Zhu:2025pmw}] and non-abelian [\cite{Aarts:2026zzr,Alharazin:2026lcb}] gauge theories. Diffusion models can learn a time-dependent drift or force, the so-called score, in score-based models and the distribution itself in energy-based models.  

The ability to learn distributions is relevant in the context of the sign problem, where a crucial role is played by a (typically unknown) distribution on some complexified manifold. For contour deformation and variants thereof, the dimension of the manifold is identical to the dimension of the original manifold. In the case of complex Langevin dynamics, its dimension is doubled for scalar fields and extended from SU($N$) to SL($N,\mathbb{C}$) for each non-abelian gauge link. The distribution on this extended manifold is real and non-negative, and is sampled by the complexified stochastic process. Hence, it should correspond to the stationary solution of the associated Fokker-Planck equation.
Unfortunately, the Fokker-Planck equation linked to this Langevin process cannot be solved in general and the distribution turns out to be elusive. A better characterisation of the distribution would be helpful in understanding success and failure of the complex Langevin approach [\cite{Aarts:2009uq,Aarts:2011ax,Mandl:2026vdc}]. If the distribution is known completely, one could in principle sample from it directly, providing a full solution to the sign problem, along the lines of [\cite{Weingarten:2002xs,Seiler:2017vwj,Salcedo:2018fvt}].

[\cite{Aarts:2025lpi}] propose to train diffusion models on data generated by the complex Langevin process, assuming correct convergence. Using a simple complex-valued quartic model, it is demonstrated that the learned score can subsequently be used to generate additional configurations or, in the energy-based approach, one can sample from the learned distribution. 
The application of a diffusion model is natural from the perspective of machine learning, with Langevin dynamics serving as a data generator.
Of course, when complex Langevin dynamics fails, the trained model will learn the wrong distribution, i.e., this approach does not solve the sign problem.
Conceptually, diffusion models or other AI methods that learn from data are applicable to any approach where an a priori unknown distribution is used; hence this setup is not limited to complex Langevin dynamics.

\section{Real-time dynamics}
\label{sec:realtime}

Lattice field theory is conventionally formulated in Euclidean time, where it is used for the computation of, e.g., hadron masses and thermodynamic quantities at zero and finite temperature (but vanishing chemical potential). Some physically very relevant questions, however, are inherently stated in real time, both in and out of thermal equilibrium. In or close to equilibrium, these include properties of (quasi-) particles in a medium, such as thermal masses and widths, and transport coefficients, such as viscosities, conductivities, and diffusion coefficients. 
Out of equilibrium, the interest is in the evolution after a sudden change, such as in a heavy-ion collision or during a first-order phase transition, and the subsequent path to equilibrium eventually, i.e., thermalisation.  The evolution contains multiple timescales, which in the sense of effective field theory are captured by far-from-equilibrium attractors, kinetic theory, and hydrodynamics, among others [\cite{Jeon:1995zm,Berges:2020fwq}].

\begin{wrapfigure}[9]{R}{0.42\textwidth}
    \centering
\includegraphics[width=0.6\linewidth]{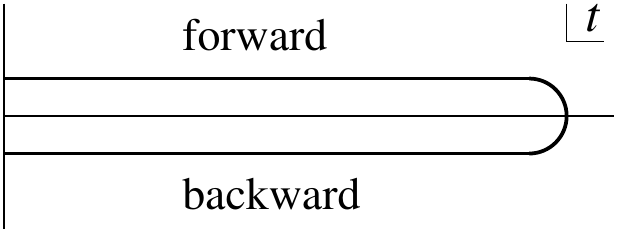}
\caption{Closed-time path or Schwinger-Keldysh contour.} 
 \label{fig:contour}
\end{wrapfigure}

Real-time dynamics in the path integral formulation is controlled by the phase $\exp(iS)$, where $S$ is the Minkowski action. Hence the sign problem is at its most severe. In thermal equilibrium, it is in principle possible to analytically continue from imaginary time back to real time, in what could be called an inverse Wick rotation. This is usually referred to as spectral function reconstruction and is classified as an ill-posed {\em inverse problem} [\cite{Asakawa:2000tr,Rothkopf:2019ipj}]. Out of equilibrium, this is by definition not possible.

The path integral formulation of initial-value problems has been known since the 1960s and goes under the name {\em Schwinger-Keldysh formalism} [\cite{Schwinger:1960qe,Keldysh:1964ud}]. Here we briefly summarise the main features as they will appear further down. 
For simplicity, we consider a real scalar field and focus on the two-point function. Starting from an arbitrary density matrix $\rho_0$ at time $t_0$, the propagator is defined as (we take $t_2>t_1$)
\be
\bra \phi(t_1, \xv) \phi(t_2, \yv) \,
= \frac{1}{Z} \Tr\left[\rho_0 \, \phi(t_1,\xv) \phi(t_2,\yv) \right]
= \frac{1}{Z} \Tr\left[\rho_0 \, U^\dagger(t_1,t_0) \phi(0,\xv) U(t_1,t_2) \phi(0,\yv) U(t_2,t_0) \right],
\qquad\qquad Z = \Tr \rho_0,
\ee
where $U(t,t_0)=\exp[-iH(t-t_0)]$ is the time evolution operator, 
$H$ is the Hamiltonian determining the time evolution, and we used the Heisenberg picture for time-dependent operators. The trace is over a complete set of states. 
This expression has a convenient interpretation along the {\em closed-time path}, shown in Fig.~\ref{fig:contour}: time starts at $t=t_0$ (initial density matrix), evolves to $t=t_1$ (insertion of $\phi(t_1,\xv)$), then to $t=t_2$ (insertion of $\phi(t_2,\yv)$), and finally returns to $t=t_0$, due to the trace being taken. The path integral hence features a doubling of degrees of freedom along the upper and lower part of the contour. Some comments are in order: different time orderings along the contour are possible, leading to, e.g., retarded and advanced Green functions; the contour can be extended to $t\to\infty$, since the forward and backward evolution operators with $t>t_2$ cancel; and the two-point function depends on $t_1-t_0$ and $t_2-t_0$ separately, since the density matrix defines an initial time. 
On the other hand, if the density matrix is thermal, $\rho_0\sim \exp(-H/T)$, and hence commutes with the time evolution operator, the propagator depends on $t_1-t_2$ only (time translation invariance in thermal equilibrium). The thermal density matrix can then be interpreted as an evolution operator in imaginary time, $\tau=it$, with $0<\tau<1/T$, adding a vertical part to the contour.

\subsection{Real-time path integrals and (machine learning) kernels}

Numerically calculating real-time observables in lattice field theory by brute force, e.g., using naive reweighting, is hopeless. One may, however, employ methods relying on holomorphicity [\cite{Callaway:1985vz,Berges:2005yt}].
As described above, defining the theory on a Schwinger-Keldysh contour allows for both a thermal and a non-equilibrium setting, starting from a general initial density matrix [\cite{Berges:2006xc}].
In the numerical implementation, the Schwinger-Keldysh contour is often {\em unfolded}, parametrising the forward and backward parts with one ``real-time'' parameter.
Initially it was noticed that the naive application of complex Langevin dynamics gives correct results if the real-time extent of the contour is sufficiently small; otherwise boundary terms build up and results are incorrect. This behaviour was observed in a $0+1$ dimensional scalar field theory, i.e., the anharmonic quantum oscillator, as well as for SU(2) gauge theory in $3+1$ dimensions [\cite{Berges:2006xc,Berges:2007nr}].

[\cite{Alexandru:2017lqr}] studied this problem using a thimble-inspired contour deformation, obtained by flowing the real manifold using holomorphic flow (\ref{eq:holo}) for some finite time. The resulting manifold is then sampled with importance sampling using the absolute value of the action measure on the contour. Since the contour is not exactly the Lefschetz thimble, a complex phase remains (as well as the complex Jacobian), which is taken into account with reweighting. The sign problem is much milder. However, when the real-time extent is increased, calculating the flowed contour becomes very costly, since large flow times are needed to have a contour with a mild enough sign problem, and thus large real times are inaccessible in practice.

Recently, it was noticed that one can optimise kernels in the kernelled complex Langevin equation (\ref{eq:kernlang}) to improve the results [\cite{Alvestad:2022abf,Lampl:2023xpb}].
This idea was first tested for the scalar quantum oscillator, with a field-independent matrix kernel for simplicity. Since ``correct'' results for real-time dynamics are generally unknown, one instead relies on {\em a priori} knowledge about the theory to determine suitable kernels, such as time translation invariance of $n$-point functions in a thermal setting and information on the boundary terms.
In practice, it turns out that there are cheap proxies, such as
\bea
L[\phi]= \int d^dx\, \left(\Im \phi(x) \right)^2,
\eea
which, when minimised, lead to a suitable distribution close to the real manifold.
The optimal matrix kernel can also be determined using machine learning methods, with $L[\phi]$ as a loss function.
Using the optimised kernel, the accessible real-time range roughly doubles compared to the one arising from using the naive Langevin equation, as shown for scalar field theories
[\cite{Alvestad:2022abf,Lampl:2023xpb,Alvestad:2023jgl}].

A slightly simpler kernel has been used in connection with $3+1$ dimensional SU(2) gauge theory [\cite{Boguslavski:2022dee}]. In this case the kernel is chosen to be diagonal, with a different rescaling factor for temporal and spatial link variables. This enables complex Langevin simulations without boundary terms for large real-time extents. Subsequently it was shown [\cite{Boguslavski:2023unu}] that all real-time two-point functions can be recovered when also using the backward part of the Schwinger-Keldysh contour, and that they satisfy the consistency relations between Feynman and Wightman two-point functions.

Overall, the time extent accessible by these methods is limited to a few elementary oscillations at most, and hence studies of transport and hydrodynamics is out of reach by a considerable amount.
The introduction of field-dependent kernels into the complex Langevin equation is, however, largely unexplored and it is worthwhile pursuing this further.

\section{Changing degrees of freedom}
\label{sec:dof}

The sign problem is present in the usual formulation of lattice QCD, in which the fermions are integrated out and the remaining gluonic path integral includes the fermion determinant in the Boltzmann weight. Hence a possible solution for the sign problem is to reconsider the integration order and arrive at a partition function written in terms of different degrees of freedom. Here we consider two examples, dual variables and tensor networks, in particular the Tensor Renormalisation Group.

\subsection{Dual variables}
\label{sec:dual}

Dual variables, or worldline and flux representations, are a physically intuitive way to evade the sign problem altogether. By rewriting the theory in terms of its dual variables, the complex phases often cancel out or recombine into strictly positive weights, see the reviews [\cite{Gattringer:2014nxa,Gattringer:2016kco}] and [\cite{Chandrasekharan:2013rpa}] with a focus on fermion models.  

In this approach, the Boltzmann weight is expanded via a high-temperature expansion to all orders or a character expansion, and the integrals over the original field variables are carried out analytically. What remains are sums over new integer-valued variables, which represent fluxes or closed loops of particle worldlines. For many systems, even those with a complex action in the original formulation, the weights for worldline configurations become real and non-negative.
Chemical potential no longer leads to a complex phase, but instead enhances worldlines wrapping around the temporal direction in the positive direction (as $e^{\mu/T}$) and suppresses those going in the opposite direction (as $e^{-\mu/T}$).
Since the weights therefore only depend on the winding number, they remain strictly positive [\cite{Endres:2006xu,Gattringer:2011gq,Gattringer:2012df,DelgadoMercado:2012tte,DelgadoMercado:2013ybm}].
In the case of the $\theta$ term, the topological charge is often linked to the total flux and dual variables represent surfaces or flux tubes. In the dual language, the $\theta$ term shifts the allowed values of the flux, again keeping the weights real [\cite{Gattringer:2015baa}].
Dual formulations benefit significantly from so-called {\em worm algorithms} [\cite{Prokof_ev_2001}], now a standard tool for efficient sampling in these representations.

It should be noted, however, that this approach is notoriously hard to extend to non-abelian gauge theories. For example, when using a character expansion for SU(3) gauge theory, the group integration yields weights that are not strictly positive. For the non-abelian case, progress has therefore slowed down [\cite{Gattringer:2016kco}].

\subsection{Tensor Renormalisation Group}
\label{sec:TRG}

The Tensor Renormalisation Group (TRG) method also takes a very different approach, again in the sense that the original field degrees of freedom are replaced by different ones. This approach is free from the sign problem, see [\cite{Akiyama:2026yhn}] for a recent review.

In TRG, originally proposed for spin systems [\cite{Levin:2006jai}], the partition function is represented as a network of tensors, which are contracted over the indices. Tensor indices are related to the original degrees of freedom, with the bond dimension indicating the dimension of each tensor index. For continuous field theories, the bond dimension is formally infinite and a truncation has to be introduced. An RG step is formulated by contracting indices and introducing the next level of ``coarse-grained'' tensors, in the spirit of block spinning. A convenient way to do this is by performing a singular-value decomposition of the tensors and reducing the degrees of freedom by keeping only the larger singular components. 

Returning to the sign problem, in TRG the partition sum is evaluated directly through matrix multiplication and partial traces, and not through importance sampling. Hence, even though tensor elements can be complex-valued, tensors are contracted analytically and cancellations are accounted for, also in the case of Grassmann integrals [\cite{Shimizu:2014uva}]. Instead, the difficulty of the sign problem is moved from sampling to computational memory, which scales with the bond dimension.

Traditionally it is assumed that tensor networks work well in two Euclidean dimensions, but there has been significant progress in three and four dimensions, including for non-abelian theories.
Several well-known theories with a sign problem have been considered in this approach. Examples in two dimensions with a non-zero chemical potential include the Bose gas [\cite{Kadoh:2019ube}], the O(3) non-linear $\sigma$ model [\cite{Luo:2024lbh}], and the SU(2) principal chiral model [\cite{Luo:2022eje}], and, with a non-zero $\theta$ term, U(1) gauge theory [\cite{Kuramashi:2019cgs}], the Schwinger model [\cite{Shimizu:2014fsa}], and the U(1) gauge Higgs model [\cite{Akiyama:2024qer}].
In four dimensions, TRG has been applied to the Bose gas [\cite{Akiyama:2020ntf}], the $Z_3$ gauge-Higgs model [\cite{Akiyama:2023hvt}], the cold dense phase of the Nambu--Jona-Lasinio model [\cite{Akiyama:2020soe}] and strong-coupling QCD [\cite{Sugimoto:2026wnw}], all at non-zero $\mu$. A tensor-network formulation of QCD at non-zero $\mu$ in the strong-coupling expansion is also given in [\cite{Samberger:2025hsr}].

It is noted that worldlines and tensor networks are to some extent related; the starting point for the tensor network formulation is a reorganisation of the path integral, obtained by expressing the Boltzmann weight as a product of local factors and carrying out the integrals in a non-standard fashion.
Depending on the way this is implemented, this is closely related to the high-temperature, hopping, or character expansions used in dual formulations [\cite{Kadoh:2019ube}].
In fact, for SU($N$) theories, it is possible to use a character expansion to formulate a tensor network [\cite{Yosprakob:2024sfd}]. It is stimulating to see that progress is being made by the tensor network approach also for non-abelian theories.

\section{Outlook}
\label{sec:outlook}

In this review we have surveyed a range of approaches which address the sign problem in lattice field theory, with a focus on recent developments in holomorphic extensions, complex Langevin dynamics, and tensor networks. We have also discussed the adoption of machine learning methods. 

Contour deformation methods, which include (generalised) Lefschetz thimbles and holomorphic flow, provide an appealing framework in which phase fluctuations can be reduced by deforming the integration manifold into complexified field space. While the formulations are firmly rooted in mathematical physics, their practical implementation is hindered by the computational cost of evaluating the Jacobian, the complexity of multi-modal thimble structures, and challenges in extending these methods to four-dimensional systems.
The worldvolume approach offers an alternative perspective, in which the flowed manifold is smeared, creating a worldvolume of one dimension higher, which ameliorates these obstructions.
Complex Langevin dynamics can solve the sign problem in principle completely, but it may suffer from incorrect convergence. 
By now, the failure modes of the complex Langevin approach are well understood and good diagnostic methods are available. Ways to adapt the process, e.g., by introducing kernels, are being explored.
Complex Langevin dynamics has solved the sign problem in several lattice field theories and as of now, it is the only method which can solve the sign problem in QCD at high baryon density. At present, for physical quark masses this
solution is limited to large temperature, i.e., in the deconfined phase.

Dual formulations and tensor-network based methods demonstrate that in certain cases the sign problem can be eliminated entirely by reformulating the theory, although such transformations are not known in general. Tensor networks offer promise for the future, with recent applications in non-abelian and in four-dimensional theories.

We note that many theories have been treated by now using a number of very different approaches. One example is the three-dimensional SU(3) spin model at non-zero $\mu$, which has been solved numerically using complex Langevin dynamics [\cite{Karsch:1985cb,Aarts:2011zn}], a dual formulation [\cite{DelgadoMercado:2012gta}] and also with TRG [\cite{Luo:2025qtv}].
The four-dimensional Bose gas has been solved with essentially all the methods we discussed above: complex Langevin dynamics [\cite{Aarts:2008wh,Aarts:2009hn}], dual methods
[\cite{Endres:2006xu,Gattringer:2012df}], Lefschetz thimbles [\cite{Cristoforetti:2013wha,Fujii:2013sra}], holomorphic flow [\cite{Alexandru:2016san}], worldvolume hybrid Monte Carlo [\cite{Namekawa:2024ert}], and tensor networks [\cite{Akiyama:2020ntf}].
As stated earlier, a verification of results obtained with entirely independent methods is crucial in building confidence and determining the best path forward.

A particularly encouraging direction is the incorporation of machine learning methods. It is applicable to contour deformations when framed as optimisation problems over parametrised maps in complexified field space. Neural networks, normalising flows, and physics-informed kernels provide flexible Ans\"atze for these maps, allowing one to directly minimise phase fluctuations or improve sampling efficiency. At the same time, machine learning has begun to play a role in the analysis and stabilisation of complex Langevin dynamics, as well as in learning the distribution effectively sampled. 

Despite this progress, significant challenges remain. In particular, the extension of these methods to full QCD at finite baryon density at low temperatures continues to be an open problem. Issues such as the scaling of computational cost with volume, sampling efficiency, and the control of systematic errors are not yet fully resolved. Moreover, it is not clear whether a single universal method exists or whether different regimes of parameter space will require fundamentally different techniques. As stated, of the methods discussed, complex Langevin dynamics is so far the only one which has been applied to QCD in four dimensions with dynamical quarks. 

Looking forward, several directions appear especially promising. Hybrid approaches that combine analytical insight with machine learning may offer improved scalability. The further development of generative models and diffusion-based methods could provide new ways to sample complex distributions or diagnose failures of existing algorithms. 
An active area of research not discussed here concerns quantum algorithms. These are fundamentally different from classical Monte Carlo methods because they operate in the Hamiltonian formulation rather than the Lagrangian path integral and quantum states are evolved directly in real time, without a sign problem. Instead, challenges lie in scaling the number of qubits and quantum gates in a fault-tolerant manner, while maintaining entanglement and coherence; for further detail we refer to the review articles [\cite{Funcke:2023jbq}], specifically in the context of lattice field theory, and [\cite{Halimeh:2025vvp}], with an emphasis on real-time dynamics.

\begin{ack}[Acknowledgments]

We thank our collaborators and colleagues for stimulating discussions on the sign problem over the past 20 years.
We acknowledge the use of ChatGPT (OpenAI) for assistance with literature review and conceptual explanations. All content has been written by the authors.
GA is supported by STFC grant ST/X000648/1 and by a Royal Society Leverhulme Trust Senior Research Fellowship.
DS gratefully acknowledges funding by the Austrian Science Fund (FWF) [\href{https://doi.org/10.55776/P36875}{10.55776/P36875}].

\end{ack}

\begin{thebibliography*}{191}
\providecommand{\bibtype}[1]{}
\providecommand{\natexlab}[1]{#1}
{\catcode`\|=0\catcode`\#=12\catcode`\@=11\catcode`\\=12
|immediate|write|@auxout{\expandafter\ifx\csname
  natexlab\endcsname\relax\gdef\natexlab#1{#1}\fi}}
\renewcommand{\url}[1]{{\tt #1}}
\providecommand{\urlprefix}{URL }
\expandafter\ifx\csname urlstyle\endcsname\relax
  \providecommand{\doi}[1]{doi:\discretionary{}{}{}#1}\else
  \providecommand{\doi}{doi:\discretionary{}{}{}\begingroup
  \urlstyle{rm}\Url}\fi
\providecommand{\bibinfo}[2]{#2}
\providecommand{\eprint}[2][]{\url{#2}}

\bibtype{Article}%
\bibitem[Aarts(2009{\natexlab{a}})]{Aarts:2008wh}
\bibinfo{author}{Aarts G} (\bibinfo{year}{2009}{\natexlab{a}}).
\bibinfo{title}{{Can stochastic quantization evade the sign problem? The
  relativistic Bose gas at finite chemical potential}}.
\bibinfo{journal}{{\em Phys. Rev. Lett.}} \bibinfo{volume}{102}:
  \bibinfo{pages}{131601}. \bibinfo{doi}{\doi{10.1103/PhysRevLett.102.131601}}.
\href{http://arxiv.org/abs/0810.2089}{{\tt arXiv:0810.2089}}.

\bibtype{Article}%
\bibitem[Aarts(2009{\natexlab{b}})]{Aarts:2009hn}
\bibinfo{author}{Aarts G} (\bibinfo{year}{2009}{\natexlab{b}}).
\bibinfo{title}{{Complex Langevin dynamics at finite chemical potential: Mean
  field analysis in the relativistic Bose gas}}.
\bibinfo{journal}{{\em JHEP}} \bibinfo{volume}{05}: \bibinfo{pages}{052}.
  \bibinfo{doi}{\doi{10.1088/1126-6708/2009/05/052}}.
\href{http://arxiv.org/abs/0902.4686}{{\tt arXiv:0902.4686}}.

\bibtype{Article}%
\bibitem[Aarts(2013)]{Aarts:2013fpa}
\bibinfo{author}{Aarts G} (\bibinfo{year}{2013}).
\bibinfo{title}{{Lefschetz thimbles and stochastic quantization: Complex
  actions in the complex plane}}.
\bibinfo{journal}{{\em Phys. Rev. D}} \bibinfo{volume}{88}
  (\bibinfo{number}{9}): \bibinfo{pages}{094501}.
  \bibinfo{doi}{\doi{10.1103/PhysRevD.88.094501}}.
\href{http://arxiv.org/abs/1308.4811}{{\tt arXiv:1308.4811}}.

\bibtype{Article}%
\bibitem[Aarts(2016)]{Aarts:2015tyj}
\bibinfo{author}{Aarts G} (\bibinfo{year}{2016}).
\bibinfo{title}{{Introductory lectures on lattice QCD at nonzero baryon
  number}}.
\bibinfo{journal}{{\em J. Phys. Conf. Ser.}} \bibinfo{volume}{706}
  (\bibinfo{number}{2}): \bibinfo{pages}{022004}.
  \bibinfo{doi}{\doi{10.1088/1742-6596/706/2/022004}}.
\href{http://arxiv.org/abs/1512.05145}{{\tt arXiv:1512.05145}}.

\bibtype{Article}%
\bibitem[Aarts and James(2010)]{Aarts:2010aq}
\bibinfo{author}{Aarts G} and  \bibinfo{author}{James FA}
  (\bibinfo{year}{2010}).
\bibinfo{title}{{On the convergence of complex Langevin dynamics: The
  Three-dimensional XY model at finite chemical potential}}.
\bibinfo{journal}{{\em JHEP}} \bibinfo{volume}{08}: \bibinfo{pages}{020}.
  \bibinfo{doi}{\doi{10.1007/JHEP08(2010)020}}.
\href{http://arxiv.org/abs/1005.3468}{{\tt arXiv:1005.3468}}.

\bibtype{Article}%
\bibitem[Aarts and James(2012)]{Aarts:2011zn}
\bibinfo{author}{Aarts G} and  \bibinfo{author}{James FA}
  (\bibinfo{year}{2012}).
\bibinfo{title}{{Complex Langevin dynamics in the SU(3) spin model at nonzero
  chemical potential revisited}}.
\bibinfo{journal}{{\em JHEP}} \bibinfo{volume}{01}: \bibinfo{pages}{118}.
  \bibinfo{doi}{\doi{10.1007/JHEP01(2012)118}}.
\href{http://arxiv.org/abs/1112.4655}{{\tt arXiv:1112.4655}}.

\bibtype{Article}%
\bibitem[Aarts and Splittorff(2010)]{Aarts:2010gr}
\bibinfo{author}{Aarts G} and  \bibinfo{author}{Splittorff K}
  (\bibinfo{year}{2010}).
\bibinfo{title}{{Degenerate distributions in complex Langevin dynamics:
  one-dimensional QCD at finite chemical potential}}.
\bibinfo{journal}{{\em JHEP}} \bibinfo{volume}{08}: \bibinfo{pages}{017}.
  \bibinfo{doi}{\doi{10.1007/JHEP08(2010)017}}.
\href{http://arxiv.org/abs/1006.0332}{{\tt arXiv:1006.0332}}.

\bibtype{Article}%
\bibitem[Aarts and Stamatescu(2008)]{Aarts:2008rr}
\bibinfo{author}{Aarts G} and  \bibinfo{author}{Stamatescu IO}
  (\bibinfo{year}{2008}).
\bibinfo{title}{{Stochastic quantization at finite chemical potential}}.
\bibinfo{journal}{{\em JHEP}} \bibinfo{volume}{09}: \bibinfo{pages}{018}.
  \bibinfo{doi}{\doi{10.1088/1126-6708/2008/09/018}}.
\href{http://arxiv.org/abs/0807.1597}{{\tt arXiv:0807.1597}}.

\bibtype{Article}%
\bibitem[Aarts et al.(2010{\natexlab{a}})]{Aarts:2009dg}
\bibinfo{author}{Aarts G}, \bibinfo{author}{James FA}, \bibinfo{author}{Seiler
  E} and  \bibinfo{author}{Stamatescu IO} (\bibinfo{year}{2010}{\natexlab{a}}).
\bibinfo{title}{{Adaptive stepsize and instabilities in complex Langevin
  dynamics}}.
\bibinfo{journal}{{\em Phys. Lett. B}} \bibinfo{volume}{687}:
  \bibinfo{pages}{154--159}.
  \bibinfo{doi}{\doi{10.1016/j.physletb.2010.03.012}}.
\href{http://arxiv.org/abs/0912.0617}{{\tt arXiv:0912.0617}}.

\bibtype{Article}%
\bibitem[Aarts et al.(2010{\natexlab{b}})]{Aarts:2009uq}
\bibinfo{author}{Aarts G}, \bibinfo{author}{Seiler E} and
  \bibinfo{author}{Stamatescu IO} (\bibinfo{year}{2010}{\natexlab{b}}).
\bibinfo{title}{Complex {{Langevin}} method: {{When}} can it be trusted?}
\bibinfo{journal}{{\em Phys. Rev. D}} \bibinfo{volume}{81}
  (\bibinfo{number}{5}): \bibinfo{pages}{054508}.
  \bibinfo{doi}{\doi{10.1103/PhysRevD.81.054508}}.

\bibtype{Article}%
\bibitem[Aarts et al.(2011)]{Aarts:2011ax}
\bibinfo{author}{Aarts G}, \bibinfo{author}{James FA}, \bibinfo{author}{Seiler
  E} and  \bibinfo{author}{Stamatescu IO} (\bibinfo{year}{2011}).
\bibinfo{title}{{Complex Langevin: Etiology and Diagnostics of its Main
  Problem}}.
\bibinfo{journal}{{\em Eur. Phys. J. C}} \bibinfo{volume}{71}:
  \bibinfo{pages}{1756}. \bibinfo{doi}{\doi{10.1140/epjc/s10052-011-1756-5}}.
\href{http://arxiv.org/abs/1101.3270}{{\tt arXiv:1101.3270}}.

\bibtype{Article}%
\bibitem[Aarts et al.(2013{\natexlab{a}})]{Aarts:2013uxa}
\bibinfo{author}{Aarts G}, \bibinfo{author}{Bongiovanni L},
  \bibinfo{author}{Seiler E}, \bibinfo{author}{Sexty D} and
  \bibinfo{author}{Stamatescu IO} (\bibinfo{year}{2013}{\natexlab{a}}).
\bibinfo{title}{Controlling complex {{Langevin}} dynamics at finite density}.
\bibinfo{journal}{{\em Eur. Phys. J. A}} \bibinfo{volume}{49}
  (\bibinfo{number}{7}): \bibinfo{pages}{89}.
  \bibinfo{doi}{\doi{10.1140/epja/i2013-13089-4}}.

\bibtype{Article}%
\bibitem[Aarts et al.(2013{\natexlab{b}})]{Aarts:2012ft}
\bibinfo{author}{Aarts G}, \bibinfo{author}{James FA},
  \bibinfo{author}{Pawlowski JM}, \bibinfo{author}{Seiler E},
  \bibinfo{author}{Sexty D} and  \bibinfo{author}{Stamatescu IO}
  (\bibinfo{year}{2013}{\natexlab{b}}).
\bibinfo{title}{{Stability of complex Langevin dynamics in effective models}}.
\bibinfo{journal}{{\em JHEP}} \bibinfo{volume}{03}: \bibinfo{pages}{073}.
  \bibinfo{doi}{\doi{10.1007/JHEP03(2013)073}}.
\href{http://arxiv.org/abs/1212.5231}{{\tt arXiv:1212.5231}}.

\bibtype{Article}%
\bibitem[Aarts et al.(2014{\natexlab{a}})]{Aarts:2014nxa}
\bibinfo{author}{Aarts G}, \bibinfo{author}{Bongiovanni L},
  \bibinfo{author}{Seiler E} and  \bibinfo{author}{Sexty D}
  (\bibinfo{year}{2014}{\natexlab{a}}).
\bibinfo{title}{{Some remarks on Lefschetz thimbles and complex Langevin
  dynamics}}.
\bibinfo{journal}{{\em JHEP}} \bibinfo{volume}{10}: \bibinfo{pages}{159}.
  \bibinfo{doi}{\doi{10.1007/JHEP10(2014)159}}.
\href{http://arxiv.org/abs/1407.2090}{{\tt arXiv:1407.2090}}.

\bibtype{Article}%
\bibitem[Aarts et al.(2014{\natexlab{b}})]{Aarts:2014bwa}
\bibinfo{author}{Aarts G}, \bibinfo{author}{Seiler E}, \bibinfo{author}{Sexty
  D} and  \bibinfo{author}{Stamatescu IO} (\bibinfo{year}{2014}{\natexlab{b}}).
\bibinfo{title}{{Simulating QCD at nonzero baryon density to all orders in the
  hopping parameter expansion}}.
\bibinfo{journal}{{\em Phys. Rev. D}} \bibinfo{volume}{90}
  (\bibinfo{number}{11}): \bibinfo{pages}{114505}.
  \bibinfo{doi}{\doi{10.1103/PhysRevD.90.114505}}.
\href{http://arxiv.org/abs/1408.3770}{{\tt arXiv:1408.3770}}.

\bibtype{Article}%
\bibitem[Aarts et al.(2016)]{Aarts:2016qrv}
\bibinfo{author}{Aarts G}, \bibinfo{author}{Attanasio F},
  \bibinfo{author}{J\"ager B} and  \bibinfo{author}{Sexty D}
  (\bibinfo{year}{2016}).
\bibinfo{title}{{The QCD phase diagram in the limit of heavy quarks using
  complex Langevin dynamics}}.
\bibinfo{journal}{{\em JHEP}} \bibinfo{volume}{09}: \bibinfo{pages}{087}.
  \bibinfo{doi}{\doi{10.1007/JHEP09(2016)087}}.
\href{http://arxiv.org/abs/1606.05561}{{\tt arXiv:1606.05561}}.

\bibtype{Article}%
\bibitem[Aarts et al.(2017)]{Aarts:2017vrv}
\bibinfo{author}{Aarts G}, \bibinfo{author}{Seiler E}, \bibinfo{author}{Sexty
  D} and  \bibinfo{author}{Stamatescu IO} (\bibinfo{year}{2017}).
\bibinfo{title}{{Complex Langevin dynamics and zeroes of the fermion
  determinant}}.
\bibinfo{journal}{{\em JHEP}} \bibinfo{volume}{05}: \bibinfo{pages}{044}.
  \bibinfo{doi}{\doi{10.1007/JHEP05(2017)044}}.
\bibinfo{note}{[Erratum: JHEP 01, 128 (2018)]},
  \href{http://arxiv.org/abs/1701.02322}{{\tt arXiv:1701.02322}}.

\bibtype{Article}%
\bibitem[Aarts et al.(2023)]{Aarts:2023vsf}
\bibinfo{author}{Aarts G} and  et al. (\bibinfo{year}{2023}).
\bibinfo{title}{{Phase Transitions in Particle Physics}: {Results and
  Perspectives from Lattice Quantum Chromo-Dynamics}}.
\bibinfo{journal}{{\em Prog. Part. Nucl. Phys.}} \bibinfo{volume}{133}:
  \bibinfo{pages}{104070}. \bibinfo{doi}{\doi{10.1016/j.ppnp.2023.104070}}.
\href{http://arxiv.org/abs/2301.04382}{{\tt arXiv:2301.04382}}.

\bibtype{Article}%
\bibitem[Aarts et al.(2025{\natexlab{a}})]{Aarts:2025gyp}
\bibinfo{author}{Aarts G}, \bibinfo{author}{Fukushima K},
  \bibinfo{author}{Hatsuda T}, \bibinfo{author}{Ipp A}, \bibinfo{author}{Shi
  S}, \bibinfo{author}{Wang L} and  \bibinfo{author}{Zhou K}
  (\bibinfo{year}{2025}{\natexlab{a}}).
\bibinfo{title}{{Physics-driven learning for inverse problems in quantum
  chromodynamics}}.
\bibinfo{journal}{{\em Nature Rev. Phys.}} \bibinfo{volume}{7}
  (\bibinfo{number}{3}): \bibinfo{pages}{154--163}.
  \bibinfo{doi}{\doi{10.1038/s42254-024-00798-x}}.
\href{http://arxiv.org/abs/2501.05580}{{\tt arXiv:2501.05580}}.

\bibtype{Article}%
\bibitem[Aarts et al.(2025{\natexlab{b}})]{Aarts:2025lpi}
\bibinfo{author}{Aarts G}, \bibinfo{author}{Habibi DE}, \bibinfo{author}{Wang
  L} and  \bibinfo{author}{Zhou K} (\bibinfo{year}{2025}{\natexlab{b}}).
\bibinfo{title}{{Combining complex Langevin dynamics with score-based and
  energy-based diffusion models}}.
\bibinfo{journal}{{\em JHEP}} \bibinfo{volume}{12}: \bibinfo{pages}{160}.
  \bibinfo{doi}{\doi{10.1007/JHEP12(2025)160}}.
\href{http://arxiv.org/abs/2510.01328}{{\tt arXiv:2510.01328}}.

\bibtype{Article}%
\bibitem[Aarts et al.(2026)]{Aarts:2026zzr}
\bibinfo{author}{Aarts G}, \bibinfo{author}{Habibi DE}, \bibinfo{author}{Ipp
  A}, \bibinfo{author}{M{\"u}ller DI}, \bibinfo{author}{Ranner TR},
  \bibinfo{author}{Wang L}, \bibinfo{author}{Wang W} and  \bibinfo{author}{Zhu
  Q} (\bibinfo{year}{2026}).
\bibinfo{title}{{Generalizable Equivariant Diffusion Models for Non-Abelian
  Lattice Gauge Theory}} \href{http://arxiv.org/abs/2601.19552}{{\tt
  arXiv:2601.19552}}.

\bibtype{Inproceedings}%
\bibitem[Akiyama(2026)]{Akiyama:2026yhn}
\bibinfo{author}{Akiyama S} (\bibinfo{year}{2026}),
  \bibinfo{title}{{Renormalization group on tensor networks}},
  \bibinfo{booktitle}{{42th International Symposium on Lattice Field Theory}},
  \href{http://arxiv.org/abs/2603.02741}{{\tt arXiv:2603.02741}}.

\bibtype{Article}%
\bibitem[Akiyama and Kuramashi(2023)]{Akiyama:2023hvt}
\bibinfo{author}{Akiyama S} and  \bibinfo{author}{Kuramashi Y}
  (\bibinfo{year}{2023}).
\bibinfo{title}{{Critical endpoint of (3+1)-dimensional finite density
  {\ensuremath{\mathbb{Z}}}$_{3}$ gauge-Higgs model with tensor renormalization
  group}}.
\bibinfo{journal}{{\em JHEP}} \bibinfo{volume}{10}: \bibinfo{pages}{077}.
  \bibinfo{doi}{\doi{10.1007/JHEP10(2023)077}}.
\href{http://arxiv.org/abs/2304.07934}{{\tt arXiv:2304.07934}}.

\bibtype{Article}%
\bibitem[Akiyama and Kuramashi(2024)]{Akiyama:2024qer}
\bibinfo{author}{Akiyama S} and  \bibinfo{author}{Kuramashi Y}
  (\bibinfo{year}{2024}).
\bibinfo{title}{{Tensor renormalization group study of (1 + 1)-dimensional U(1)
  gauge-Higgs model at {\ensuremath{\theta}} = {\ensuremath{\pi}} with
  L{\"u}scher{\textquoteright}s admissibility condition}}.
\bibinfo{journal}{{\em JHEP}} \bibinfo{volume}{09}: \bibinfo{pages}{086}.
  \bibinfo{doi}{\doi{10.1007/JHEP09(2024)086}}.
\href{http://arxiv.org/abs/2407.10409}{{\tt arXiv:2407.10409}}.

\bibtype{Article}%
\bibitem[Akiyama et al.(2020)]{Akiyama:2020ntf}
\bibinfo{author}{Akiyama S}, \bibinfo{author}{Kadoh D},
  \bibinfo{author}{Kuramashi Y}, \bibinfo{author}{Yamashita T} and
  \bibinfo{author}{Yoshimura Y} (\bibinfo{year}{2020}).
\bibinfo{title}{{Tensor renormalization group approach to four-dimensional
  complex $\phi^4$ theory at finite density}}.
\bibinfo{journal}{{\em JHEP}} \bibinfo{volume}{09}: \bibinfo{pages}{177}.
  \bibinfo{doi}{\doi{10.1007/JHEP09(2020)177}}.
\href{http://arxiv.org/abs/2005.04645}{{\tt arXiv:2005.04645}}.

\bibtype{Article}%
\bibitem[Akiyama et al.(2021)]{Akiyama:2020soe}
\bibinfo{author}{Akiyama S}, \bibinfo{author}{Kuramashi Y},
  \bibinfo{author}{Yamashita T} and  \bibinfo{author}{Yoshimura Y}
  (\bibinfo{year}{2021}).
\bibinfo{title}{{Restoration of chiral symmetry in cold and dense
  Nambu--Jona-Lasinio model with tensor renormalization group}}.
\bibinfo{journal}{{\em JHEP}} \bibinfo{volume}{01}: \bibinfo{pages}{121}.
  \bibinfo{doi}{\doi{10.1007/JHEP01(2021)121}}.
\href{http://arxiv.org/abs/2009.11583}{{\tt arXiv:2009.11583}}.

\bibtype{Article}%
\bibitem[Albergo et al.(2019)]{Albergo:2019eim}
\bibinfo{author}{Albergo MS}, \bibinfo{author}{Kanwar G} and
  \bibinfo{author}{Shanahan PE} (\bibinfo{year}{2019}).
\bibinfo{title}{Flow-based generative models for {{Markov}} chain {{Monte
  Carlo}} in lattice field theory}.
\bibinfo{journal}{{\em Phys. Rev. D}} \bibinfo{volume}{100}
  (\bibinfo{number}{3}): \bibinfo{pages}{034515}.
  \bibinfo{doi}{\doi{10.1103/PhysRevD.100.034515}}.
\href{http://arxiv.org/abs/1904.12072}{{\tt arXiv:1904.12072}}.

\bibtype{Article}%
\bibitem[Alexandru et al.(2016{\natexlab{a}})]{Alexandru:2016san}
\bibinfo{author}{Alexandru A}, \bibinfo{author}{Basar G},
  \bibinfo{author}{Bedaque P}, \bibinfo{author}{Ridgway GW} and
  \bibinfo{author}{Warrington NC} (\bibinfo{year}{2016}{\natexlab{a}}).
\bibinfo{title}{{Study of symmetry breaking in a relativistic Bose gas using
  the contraction algorithm}}.
\bibinfo{journal}{{\em Phys. Rev. D}} \bibinfo{volume}{94}
  (\bibinfo{number}{4}): \bibinfo{pages}{045017}.
  \bibinfo{doi}{\doi{10.1103/PhysRevD.94.045017}}.
\href{http://arxiv.org/abs/1606.02742}{{\tt arXiv:1606.02742}}.

\bibtype{Article}%
\bibitem[Alexandru et al.(2016{\natexlab{b}})]{Alexandru:2015sua}
\bibinfo{author}{Alexandru A}, \bibinfo{author}{Basar G},
  \bibinfo{author}{Bedaque PF}, \bibinfo{author}{Ridgway GW} and
  \bibinfo{author}{Warrington NC} (\bibinfo{year}{2016}{\natexlab{b}}).
\bibinfo{title}{{Sign problem and Monte Carlo calculations beyond Lefschetz
  thimbles}}.
\bibinfo{journal}{{\em JHEP}} \bibinfo{volume}{05}: \bibinfo{pages}{053}.
  \bibinfo{doi}{\doi{10.1007/JHEP05(2016)053}}.
\href{http://arxiv.org/abs/1512.08764}{{\tt arXiv:1512.08764}}.

\bibtype{Article}%
\bibitem[Alexandru et al.(2017{\natexlab{a}})]{Alexandru:2017lqr}
\bibinfo{author}{Alexandru A}, \bibinfo{author}{Basar G},
  \bibinfo{author}{Bedaque PF} and  \bibinfo{author}{Ridgway GW}
  (\bibinfo{year}{2017}{\natexlab{a}}).
\bibinfo{title}{{Schwinger-Keldysh formalism on the lattice: A faster algorithm
  and its application to field theory}}.
\bibinfo{journal}{{\em Phys. Rev. D}} \bibinfo{volume}{95}
  (\bibinfo{number}{11}): \bibinfo{pages}{114501}.
  \bibinfo{doi}{\doi{10.1103/PhysRevD.95.114501}}.
\href{http://arxiv.org/abs/1704.06404}{{\tt arXiv:1704.06404}}.

\bibtype{Article}%
\bibitem[Alexandru et al.(2017{\natexlab{b}})]{Alexandru:2016ejd}
\bibinfo{author}{Alexandru A}, \bibinfo{author}{Basar G},
  \bibinfo{author}{Bedaque PF}, \bibinfo{author}{Ridgway GW} and
  \bibinfo{author}{Warrington NC} (\bibinfo{year}{2017}{\natexlab{b}}).
\bibinfo{title}{{Monte Carlo calculations of the finite density Thirring
  model}}.
\bibinfo{journal}{{\em Phys. Rev. D}} \bibinfo{volume}{95}
  (\bibinfo{number}{1}): \bibinfo{pages}{014502}.
  \bibinfo{doi}{\doi{10.1103/PhysRevD.95.014502}}.
\href{http://arxiv.org/abs/1609.01730}{{\tt arXiv:1609.01730}}.

\bibtype{Article}%
\bibitem[Alexandru et al.(2017{\natexlab{c}})]{Alexandru:2017oyw}
\bibinfo{author}{Alexandru A}, \bibinfo{author}{Basar G},
  \bibinfo{author}{Bedaque PF} and  \bibinfo{author}{Warrington NC}
  (\bibinfo{year}{2017}{\natexlab{c}}).
\bibinfo{title}{{Tempered transitions between thimbles}}.
\bibinfo{journal}{{\em Phys. Rev. D}} \bibinfo{volume}{96}
  (\bibinfo{number}{3}): \bibinfo{pages}{034513}.
  \bibinfo{doi}{\doi{10.1103/PhysRevD.96.034513}}.
\href{http://arxiv.org/abs/1703.02414}{{\tt arXiv:1703.02414}}.

\bibtype{Article}%
\bibitem[Alexandru et al.(2017{\natexlab{d}})]{Alexandru:2017czx}
\bibinfo{author}{Alexandru A}, \bibinfo{author}{Bedaque PF},
  \bibinfo{author}{Lamm H} and  \bibinfo{author}{Lawrence S}
  (\bibinfo{year}{2017}{\natexlab{d}}).
\bibinfo{title}{{Deep Learning Beyond Lefschetz Thimbles}}.
\bibinfo{journal}{{\em Phys. Rev. D}} \bibinfo{volume}{96}
  (\bibinfo{number}{9}): \bibinfo{pages}{094505}.
  \bibinfo{doi}{\doi{10.1103/PhysRevD.96.094505}}.
\href{http://arxiv.org/abs/1709.01971}{{\tt arXiv:1709.01971}}.

\bibtype{Article}%
\bibitem[Alexandru et al.(2018{\natexlab{a}})]{Alexandru:2018ngw}
\bibinfo{author}{Alexandru A}, \bibinfo{author}{Ba{\c{s}}ar G},
  \bibinfo{author}{Bedaque PF}, \bibinfo{author}{Lamm H} and
  \bibinfo{author}{Lawrence S} (\bibinfo{year}{2018}{\natexlab{a}}).
\bibinfo{title}{{Finite Density $QED_{1+1}$ Near Lefschetz Thimbles}}.
\bibinfo{journal}{{\em Phys. Rev. D}} \bibinfo{volume}{98}
  (\bibinfo{number}{3}): \bibinfo{pages}{034506}.
  \bibinfo{doi}{\doi{10.1103/PhysRevD.98.034506}}.
\href{http://arxiv.org/abs/1807.02027}{{\tt arXiv:1807.02027}}.

\bibtype{Article}%
\bibitem[Alexandru et al.(2018{\natexlab{b}})]{Alexandru:2018ddf}
\bibinfo{author}{Alexandru A}, \bibinfo{author}{Bedaque PF},
  \bibinfo{author}{Lamm H}, \bibinfo{author}{Lawrence S} and
  \bibinfo{author}{Warrington NC} (\bibinfo{year}{2018}{\natexlab{b}}).
\bibinfo{title}{{Fermions at Finite Density in 2+1 Dimensions with
  Sign-Optimized Manifolds}}.
\bibinfo{journal}{{\em Phys. Rev. Lett.}} \bibinfo{volume}{121}
  (\bibinfo{number}{19}): \bibinfo{pages}{191602}.
  \bibinfo{doi}{\doi{10.1103/PhysRevLett.121.191602}}.
\href{http://arxiv.org/abs/1808.09799}{{\tt arXiv:1808.09799}}.

\bibtype{Article}%
\bibitem[Alexandru et al.(2022)]{Alexandru:2020wrj}
\bibinfo{author}{Alexandru A}, \bibinfo{author}{Basar G},
  \bibinfo{author}{Bedaque PF} and  \bibinfo{author}{Warrington NC}
  (\bibinfo{year}{2022}).
\bibinfo{title}{{Complex paths around the sign problem}}.
\bibinfo{journal}{{\em Rev. Mod. Phys.}} \bibinfo{volume}{94}
  (\bibinfo{number}{1}): \bibinfo{pages}{015006}.
  \bibinfo{doi}{\doi{10.1103/RevModPhys.94.015006}}.
\href{http://arxiv.org/abs/2007.05436}{{\tt arXiv:2007.05436}}.

\bibtype{Article}%
\bibitem[Alharazin et al.(2026)]{Alharazin:2026lcb}
\bibinfo{author}{Alharazin H}, \bibinfo{author}{Panteleeva JY} and
  \bibinfo{author}{Sun BD} (\bibinfo{year}{2026}).
\bibinfo{title}{{Diffusion Models for SU(2) Lattice Gauge Theory in Two
  Dimensions}} \href{http://arxiv.org/abs/2602.09045}{{\tt arXiv:2602.09045}}.

\bibtype{Article}%
\bibitem[Alvestad et al.(2021)]{Alvestad:2021hsi}
\bibinfo{author}{Alvestad D}, \bibinfo{author}{Larsen R} and
  \bibinfo{author}{Rothkopf A} (\bibinfo{year}{2021}).
\bibinfo{title}{{Stable solvers for real-time Complex Langevin}}.
\bibinfo{journal}{{\em JHEP}} \bibinfo{volume}{08}: \bibinfo{pages}{138}.
  \bibinfo{doi}{\doi{10.1007/JHEP08(2021)138}}.
\href{http://arxiv.org/abs/2105.02735}{{\tt arXiv:2105.02735}}.

\bibtype{Article}%
\bibitem[Alvestad et al.(2023)]{Alvestad:2022abf}
\bibinfo{author}{Alvestad D}, \bibinfo{author}{Larsen R} and
  \bibinfo{author}{Rothkopf A} (\bibinfo{year}{2023}).
\bibinfo{title}{{Towards learning optimized kernels for complex Langevin}}.
\bibinfo{journal}{{\em JHEP}} \bibinfo{volume}{04}: \bibinfo{pages}{057}.
  \bibinfo{doi}{\doi{10.1007/JHEP04(2023)057}}.
\href{http://arxiv.org/abs/2211.15625}{{\tt arXiv:2211.15625}}.

\bibtype{Article}%
\bibitem[Alvestad et al.(2024)]{Alvestad:2023jgl}
\bibinfo{author}{Alvestad D}, \bibinfo{author}{Rothkopf A} and
  \bibinfo{author}{Sexty D} (\bibinfo{year}{2024}).
\bibinfo{title}{{Lattice real-time simulations with learned optimal kernels}}.
\bibinfo{journal}{{\em Phys. Rev. D}} \bibinfo{volume}{109}
  (\bibinfo{number}{3}): \bibinfo{pages}{L031502}.
  \bibinfo{doi}{\doi{10.1103/PhysRevD.109.L031502}}.
\href{http://arxiv.org/abs/2310.08053}{{\tt arXiv:2310.08053}}.

\bibtype{Article}%
\bibitem[Ambjorn and Yang(1985)]{Ambjorn:1985iw}
\bibinfo{author}{Ambjorn J} and  \bibinfo{author}{Yang SK}
  (\bibinfo{year}{1985}).
\bibinfo{title}{{Numerical Problems in Applying the Langevin Equation to
  Complex Effective Actions}}.
\bibinfo{journal}{{\em Phys. Lett. B}} \bibinfo{volume}{165}:
  \bibinfo{pages}{140}. \bibinfo{doi}{\doi{10.1016/0370-2693(85)90708-7}}.

\bibtype{Article}%
\bibitem[Ambjorn et al.(1986)]{Ambjorn:1986fz}
\bibinfo{author}{Ambjorn J}, \bibinfo{author}{Flensburg M} and
  \bibinfo{author}{Peterson C} (\bibinfo{year}{1986}).
\bibinfo{title}{{The Complex Langevin Equation and Monte Carlo Simulations of
  Actions With Static Charges}}.
\bibinfo{journal}{{\em Nucl. Phys. B}} \bibinfo{volume}{275}:
  \bibinfo{pages}{375--397}. \bibinfo{doi}{\doi{10.1016/0550-3213(86)90605-X}}.

\bibtype{Article}%
\bibitem[Asakawa et al.(2001)]{Asakawa:2000tr}
\bibinfo{author}{Asakawa M}, \bibinfo{author}{Hatsuda T} and
  \bibinfo{author}{Nakahara Y} (\bibinfo{year}{2001}).
\bibinfo{title}{{Maximum entropy analysis of the spectral functions in lattice
  QCD}}.
\bibinfo{journal}{{\em Prog. Part. Nucl. Phys.}} \bibinfo{volume}{46}:
  \bibinfo{pages}{459--508}.
  \bibinfo{doi}{\doi{10.1016/S0146-6410(01)00150-8}}.
\href{https://arxiv.org/abs/hep-lat/0011040}{{\tt hep-lat/0011040}}.

\bibtype{Article}%
\bibitem[Attanasio and J{\"a}ger(2019)]{Attanasio:2018rtq}
\bibinfo{author}{Attanasio F} and  \bibinfo{author}{J{\"a}ger B}
  (\bibinfo{year}{2019}).
\bibinfo{title}{{Dynamical stabilisation of complex Langevin simulations of
  QCD}}.
\bibinfo{journal}{{\em Eur. Phys. J. C}} \bibinfo{volume}{79}
  (\bibinfo{number}{1}): \bibinfo{pages}{16}.
  \bibinfo{doi}{\doi{10.1140/epjc/s10052-018-6512-7}}.
\href{http://arxiv.org/abs/1808.04400}{{\tt arXiv:1808.04400}}.

\bibtype{Article}%
\bibitem[Attanasio et al.(2022{\natexlab{a}})]{Attanasio:2022ogk}
\bibinfo{author}{Attanasio F}, \bibinfo{author}{J{\"a}ger B} and
  \bibinfo{author}{Ziegler FPG} (\bibinfo{year}{2022}{\natexlab{a}}).
\bibinfo{title}{{Equation of state from complex Langevin simulations}}.
\bibinfo{journal}{{\em EPJ Web Conf.}} \bibinfo{volume}{274}:
  \bibinfo{pages}{05012}. \bibinfo{doi}{\doi{10.1051/epjconf/202227405012}}.
\href{http://arxiv.org/abs/2211.10261}{{\tt arXiv:2211.10261}}.

\bibtype{Article}%
\bibitem[Attanasio et al.(2022{\natexlab{b}})]{Attanasio:2022mjd}
\bibinfo{author}{Attanasio F}, \bibinfo{author}{J{\"a}ger B} and
  \bibinfo{author}{Ziegler FPG} (\bibinfo{year}{2022}{\natexlab{b}}).
\bibinfo{title}{{QCD equation of state via the complex Langevin method}}
  \href{http://arxiv.org/abs/2203.13144}{{\tt arXiv:2203.13144}}.

\bibtype{Article}%
\bibitem[Batrouni et al.(1985)]{Batrouni:1985jn}
\bibinfo{author}{Batrouni GG}, \bibinfo{author}{Katz GR},
  \bibinfo{author}{Kronfeld AS}, \bibinfo{author}{Lepage GP},
  \bibinfo{author}{Svetitsky B} and  \bibinfo{author}{Wilson KG}
  (\bibinfo{year}{1985}).
\bibinfo{title}{{Langevin Simulations of Lattice Field Theories}}.
\bibinfo{journal}{{\em Phys. Rev. D}} \bibinfo{volume}{32}:
  \bibinfo{pages}{2736}. \bibinfo{doi}{\doi{10.1103/PhysRevD.32.2736}}.

\bibtype{Article}%
\bibitem[Berger et al.(2021)]{Berger:2019odf}
\bibinfo{author}{Berger CE}, \bibinfo{author}{Rammelm{\"u}ller L},
  \bibinfo{author}{Loheac AC}, \bibinfo{author}{Ehmann F},
  \bibinfo{author}{Braun J} and  \bibinfo{author}{Drut JE}
  (\bibinfo{year}{2021}).
\bibinfo{title}{{Complex Langevin and other approaches to the sign problem in
  quantum many-body physics}}.
\bibinfo{journal}{{\em Phys. Rept.}} \bibinfo{volume}{892}:
  \bibinfo{pages}{1--54}. \bibinfo{doi}{\doi{10.1016/j.physrep.2020.09.002}}.
\href{http://arxiv.org/abs/1907.10183}{{\tt arXiv:1907.10183}}.

\bibtype{Article}%
\bibitem[Berges and Sexty(2008)]{Berges:2007nr}
\bibinfo{author}{Berges J} and  \bibinfo{author}{Sexty D}
  (\bibinfo{year}{2008}).
\bibinfo{title}{{Real-time gauge theory simulations from stochastic
  quantization with optimized updating}}.
\bibinfo{journal}{{\em Nucl. Phys. B}} \bibinfo{volume}{799}:
  \bibinfo{pages}{306--329}.
  \bibinfo{doi}{\doi{10.1016/j.nuclphysb.2008.01.018}}.
\href{http://arxiv.org/abs/0708.0779}{{\tt arXiv:0708.0779}}.

\bibtype{Article}%
\bibitem[Berges and Stamatescu(2005)]{Berges:2005yt}
\bibinfo{author}{Berges J} and  \bibinfo{author}{Stamatescu IO}
  (\bibinfo{year}{2005}).
\bibinfo{title}{{Simulating nonequilibrium quantum fields with stochastic
  quantization techniques}}.
\bibinfo{journal}{{\em Phys. Rev. Lett.}} \bibinfo{volume}{95}:
  \bibinfo{pages}{202003}. \bibinfo{doi}{\doi{10.1103/PhysRevLett.95.202003}}.
\href{https://arxiv.org/abs/hep-lat/0508030}{{\tt hep-lat/0508030}}.

\bibtype{Article}%
\bibitem[Berges et al.(2007)]{Berges:2006xc}
\bibinfo{author}{Berges J}, \bibinfo{author}{Borsanyi S},
  \bibinfo{author}{Sexty D} and  \bibinfo{author}{Stamatescu IO}
  (\bibinfo{year}{2007}).
\bibinfo{title}{{Lattice simulations of real-time quantum fields}}.
\bibinfo{journal}{{\em Phys. Rev. D}} \bibinfo{volume}{75}:
  \bibinfo{pages}{045007}. \bibinfo{doi}{\doi{10.1103/PhysRevD.75.045007}}.
\href{https://arxiv.org/abs/hep-lat/0609058}{{\tt hep-lat/0609058}}.

\bibtype{Article}%
\bibitem[Berges et al.(2021)]{Berges:2020fwq}
\bibinfo{author}{Berges J}, \bibinfo{author}{Heller MP},
  \bibinfo{author}{Mazeliauskas A} and  \bibinfo{author}{Venugopalan R}
  (\bibinfo{year}{2021}).
\bibinfo{title}{{QCD thermalization: Ab initio approaches and interdisciplinary
  connections}}.
\bibinfo{journal}{{\em Rev. Mod. Phys.}} \bibinfo{volume}{93}
  (\bibinfo{number}{3}): \bibinfo{pages}{035003}.
  \bibinfo{doi}{\doi{10.1103/RevModPhys.93.035003}}.
\href{http://arxiv.org/abs/2005.12299}{{\tt arXiv:2005.12299}}.

\bibtype{Article}%
\bibitem[Boguslavski et al.(2023)]{Boguslavski:2022dee}
\bibinfo{author}{Boguslavski K}, \bibinfo{author}{Hotzy P} and
  \bibinfo{author}{M{\"u}ller DI} (\bibinfo{year}{2023}).
\bibinfo{title}{{Stabilizing complex Langevin for real-time gauge theories with
  an anisotropic kernel}}.
\bibinfo{journal}{{\em JHEP}} \bibinfo{volume}{06}: \bibinfo{pages}{011}.
  \bibinfo{doi}{\doi{10.1007/JHEP06(2023)011}}.
\href{http://arxiv.org/abs/2212.08602}{{\tt arXiv:2212.08602}}.

\bibtype{Article}%
\bibitem[Boguslavski et al.(2024)]{Boguslavski:2023unu}
\bibinfo{author}{Boguslavski K}, \bibinfo{author}{Hotzy P} and
  \bibinfo{author}{M{\"u}ller DI} (\bibinfo{year}{2024}).
\bibinfo{title}{{Real-time correlators in 3+1D thermal lattice gauge theory}}.
\bibinfo{journal}{{\em Phys. Rev. D}} \bibinfo{volume}{109}
  (\bibinfo{number}{9}): \bibinfo{pages}{094518}.
  \bibinfo{doi}{\doi{10.1103/PhysRevD.109.094518}}.
\href{http://arxiv.org/abs/2312.03063}{{\tt arXiv:2312.03063}}.

\bibtype{Article}%
\bibitem[Boguslavski et al.(2025)]{Boguslavski:2024yto}
\bibinfo{author}{Boguslavski K}, \bibinfo{author}{Hotzy P} and
  \bibinfo{author}{M{\"u}ller DI} (\bibinfo{year}{2025}).
\bibinfo{title}{{Lefschetz thimble-inspired weight regularizations for complex
  Langevin simulations}}.
\bibinfo{journal}{{\em SciPost Phys.}} \bibinfo{volume}{18}:
  \bibinfo{pages}{092}. \bibinfo{doi}{\doi{10.21468/SciPostPhys.18.3.092}}.
\href{http://arxiv.org/abs/2412.02396}{{\tt arXiv:2412.02396}}.

\bibtype{Article}%
\bibitem[Borsanyi and Parotto(2025)]{Borsanyi:2025ttb}
\bibinfo{author}{Borsanyi S} and  \bibinfo{author}{Parotto P}
  (\bibinfo{year}{2025}).
\bibinfo{title}{{The QCD phase diagram}}
  \href{http://arxiv.org/abs/2512.08843}{{\tt arXiv:2512.08843}}.

\bibtype{Article}%
\bibitem[Brandt et al.(2018)]{Brandt:2017oyy}
\bibinfo{author}{Brandt BB}, \bibinfo{author}{Endrodi G} and
  \bibinfo{author}{Schmalzbauer S} (\bibinfo{year}{2018}).
\bibinfo{title}{{QCD phase diagram for nonzero isospin-asymmetry}}.
\bibinfo{journal}{{\em Phys. Rev. D}} \bibinfo{volume}{97}
  (\bibinfo{number}{5}): \bibinfo{pages}{054514}.
  \bibinfo{doi}{\doi{10.1103/PhysRevD.97.054514}}.
\href{http://arxiv.org/abs/1712.08190}{{\tt arXiv:1712.08190}}.

\bibtype{Article}%
\bibitem[Cai et al.(2020)]{Cai:2019vmt}
\bibinfo{author}{Cai Z}, \bibinfo{author}{Di Y} and  \bibinfo{author}{Dong X}
  (\bibinfo{year}{2020}).
\bibinfo{title}{{How does Gauge Cooling Stabilize Complex Langevin?}}
\bibinfo{journal}{{\em Commun. Comput. Phys.}} \bibinfo{volume}{27}
  (\bibinfo{number}{5}): \bibinfo{pages}{1344--1377}.
  \bibinfo{doi}{\doi{10.4208/cicp.OA-2019-0126}}.
\href{http://arxiv.org/abs/1905.11683}{{\tt arXiv:1905.11683}}.

\bibtype{Article}%
\bibitem[Cai et al.(2021)]{Cai:2020tgd}
\bibinfo{author}{Cai Z}, \bibinfo{author}{Dong X} and  \bibinfo{author}{Kuang
  Y} (\bibinfo{year}{2021}).
\bibinfo{title}{{On the Validity of Complex Langevin Method for Path Integral
  Computations}}.
\bibinfo{journal}{{\em SIAM J. Sci. Comput.}} \bibinfo{volume}{43}
  (\bibinfo{number}{1}): \bibinfo{pages}{A685--A719}.
  \bibinfo{doi}{\doi{10.1137/20M1363224}}.
\href{http://arxiv.org/abs/2007.10198}{{\tt arXiv:2007.10198}}.

\bibtype{Article}%
\bibitem[Callaway et al.(1985)]{Callaway:1985vz}
\bibinfo{author}{Callaway DJE}, \bibinfo{author}{Cooper F},
  \bibinfo{author}{Klauder JR} and  \bibinfo{author}{Rose H}
  (\bibinfo{year}{1985}).
\bibinfo{title}{{Langevin Simulations in Minkowski Space}}.
\bibinfo{journal}{{\em Nucl. Phys. B}} \bibinfo{volume}{262}:
  \bibinfo{pages}{19--32}. \bibinfo{doi}{\doi{10.1016/0550-3213(85)90061-6}}.

\bibtype{Article}%
\bibitem[Chandrasekharan(2013)]{Chandrasekharan:2013rpa}
\bibinfo{author}{Chandrasekharan S} (\bibinfo{year}{2013}).
\bibinfo{title}{{Fermion Bag Approach to Fermion Sign Problems}}.
\bibinfo{journal}{{\em Eur. Phys. J. A}} \bibinfo{volume}{49}:
  \bibinfo{pages}{90}. \bibinfo{doi}{\doi{10.1140/epja/i2013-13090-y}}.
\href{http://arxiv.org/abs/1304.4900}{{\tt arXiv:1304.4900}}.

\bibtype{Article}%
\bibitem[Cohen(2003)]{Cohen:2003kd}
\bibinfo{author}{Cohen TD} (\bibinfo{year}{2003}).
\bibinfo{title}{{Functional integrals for QCD at nonzero chemical potential and
  zero density}}.
\bibinfo{journal}{{\em Phys. Rev. Lett.}} \bibinfo{volume}{91}:
  \bibinfo{pages}{222001}. \bibinfo{doi}{\doi{10.1103/PhysRevLett.91.222001}}.
\href{https://arxiv.org/abs/hep-ph/0307089}{{\tt hep-ph/0307089}}.

\bibtype{Article}%
\bibitem[Cohen(2026)]{Cohen:2026pzh}
\bibinfo{author}{Cohen TD} (\bibinfo{year}{2026}).
\bibinfo{title}{{The Silver Blaze Problem in QCD}}
  \href{http://arxiv.org/abs/2601.21053}{{\tt arXiv:2601.21053}}.

\bibtype{Article}%
\bibitem[Cranmer et al.(2023)]{Cranmer:2023xbe}
\bibinfo{author}{Cranmer K}, \bibinfo{author}{Kanwar G},
  \bibinfo{author}{Racani\`ere S}, \bibinfo{author}{Rezende DJ} and
  \bibinfo{author}{Shanahan PE} (\bibinfo{year}{2023}).
\bibinfo{title}{{Advances in machine-learning-based sampling motivated by
  lattice quantum chromodynamics}}.
\bibinfo{journal}{{\em Nature Rev. Phys.}} \bibinfo{volume}{5}
  (\bibinfo{number}{9}): \bibinfo{pages}{526--535}.
  \bibinfo{doi}{\doi{10.1038/s42254-023-00616-w}}.
\href{http://arxiv.org/abs/2309.01156}{{\tt arXiv:2309.01156}}.

\bibtype{Article}%
\bibitem[Cristoforetti et al.(2012)]{Cristoforetti:2012su}
\bibinfo{author}{Cristoforetti M}, \bibinfo{author}{Di~Renzo F} and
  \bibinfo{author}{Scorzato L} (\bibinfo{collaboration}{AuroraScience})
  (\bibinfo{year}{2012}).
\bibinfo{title}{{New approach to the sign problem in quantum field theories:
  High density QCD on a Lefschetz thimble}}.
\bibinfo{journal}{{\em Phys. Rev. D}} \bibinfo{volume}{86}:
  \bibinfo{pages}{074506}. \bibinfo{doi}{\doi{10.1103/PhysRevD.86.074506}}.
\href{http://arxiv.org/abs/1205.3996}{{\tt arXiv:1205.3996}}.

\bibtype{Article}%
\bibitem[Cristoforetti et al.(2013)]{Cristoforetti:2013wha}
\bibinfo{author}{Cristoforetti M}, \bibinfo{author}{Di~Renzo F},
  \bibinfo{author}{Mukherjee A} and  \bibinfo{author}{Scorzato L}
  (\bibinfo{year}{2013}).
\bibinfo{title}{{Monte Carlo simulations on the Lefschetz thimble: Taming the
  sign problem}}.
\bibinfo{journal}{{\em Phys. Rev. D}} \bibinfo{volume}{88}
  (\bibinfo{number}{5}): \bibinfo{pages}{051501}.
  \bibinfo{doi}{\doi{10.1103/PhysRevD.88.051501}}.
\href{http://arxiv.org/abs/1303.7204}{{\tt arXiv:1303.7204}}.

\bibtype{Article}%
\bibitem[Damgaard and H{\"u}ffel(1987)]{Damgaard:1987rr}
\bibinfo{author}{Damgaard PH} and  \bibinfo{author}{H{\"u}ffel H}
  (\bibinfo{year}{1987}).
\bibinfo{title}{Stochastic quantization}.
\bibinfo{journal}{{\em Phys. Rept.}} \bibinfo{volume}{152}
  (\bibinfo{number}{5}): \bibinfo{pages}{227--398}.
  \bibinfo{doi}{\doi{10.1016/0370-1573(87)90144-X}}.

\bibtype{Article}%
\bibitem[de~Forcrand(2009)]{deForcrand:2009zkb}
\bibinfo{author}{de~Forcrand P} (\bibinfo{year}{2009}).
\bibinfo{title}{{Simulating QCD at finite density}}.
\bibinfo{journal}{{\em PoS}} \bibinfo{volume}{LAT2009}: \bibinfo{pages}{010}.
  \bibinfo{doi}{\doi{10.22323/1.091.0010}}.
\href{http://arxiv.org/abs/1005.0539}{{\tt arXiv:1005.0539}}.

\bibtype{Article}%
\bibitem[Delgado~Mercado and Gattringer(2012)]{DelgadoMercado:2012gta}
\bibinfo{author}{Delgado~Mercado Y} and  \bibinfo{author}{Gattringer C}
  (\bibinfo{year}{2012}).
\bibinfo{title}{{Monte Carlo simulation of the SU(3) spin model with chemical
  potential in a flux representation}}.
\bibinfo{journal}{{\em Nucl. Phys. B}} \bibinfo{volume}{862}:
  \bibinfo{pages}{737--750}.
  \bibinfo{doi}{\doi{10.1016/j.nuclphysb.2012.05.009}}.
\href{http://arxiv.org/abs/1204.6074}{{\tt arXiv:1204.6074}}.

\bibtype{Article}%
\bibitem[Delgado~Mercado et al.(2013{\natexlab{a}})]{DelgadoMercado:2013ybm}
\bibinfo{author}{Delgado~Mercado Y}, \bibinfo{author}{Gattringer C} and
  \bibinfo{author}{Schmidt A} (\bibinfo{year}{2013}{\natexlab{a}}).
\bibinfo{title}{{Dual Lattice Simulation of the Abelian Gauge-Higgs Model at
  Finite Density: An Exploratory Proof of Concept Study}}.
\bibinfo{journal}{{\em Phys. Rev. Lett.}} \bibinfo{volume}{111}
  (\bibinfo{number}{14}): \bibinfo{pages}{141601}.
  \bibinfo{doi}{\doi{10.1103/PhysRevLett.111.141601}}.
\href{http://arxiv.org/abs/1307.6120}{{\tt arXiv:1307.6120}}.

\bibtype{Article}%
\bibitem[Delgado~Mercado et al.(2013{\natexlab{b}})]{DelgadoMercado:2012tte}
\bibinfo{author}{Delgado~Mercado Y}, \bibinfo{author}{Gattringer C} and
  \bibinfo{author}{Schmidt A} (\bibinfo{year}{2013}{\natexlab{b}}).
\bibinfo{title}{{Surface worm algorithm for abelian Gauge-Higgs systems on the
  lattice}}.
\bibinfo{journal}{{\em Comput. Phys. Commun.}} \bibinfo{volume}{184}:
  \bibinfo{pages}{1535--1546}. \bibinfo{doi}{\doi{10.1016/j.cpc.2013.02.001}}.
\href{http://arxiv.org/abs/1211.3436}{{\tt arXiv:1211.3436}}.

\bibtype{Article}%
\bibitem[D'Elia(2019)]{DElia:2018fjp}
\bibinfo{author}{D'Elia M} (\bibinfo{year}{2019}).
\bibinfo{title}{{High-Temperature QCD: theory overview}}.
\bibinfo{journal}{{\em Nucl. Phys. A}} \bibinfo{volume}{982}:
  \bibinfo{pages}{99--105}.
  \bibinfo{doi}{\doi{10.1016/j.nuclphysa.2018.10.042}}.
\href{http://arxiv.org/abs/1809.10660}{{\tt arXiv:1809.10660}}.

\bibtype{Article}%
\bibitem[D'Elia and Lombardo(2003)]{DElia:2002tig}
\bibinfo{author}{D'Elia M} and  \bibinfo{author}{Lombardo MP}
  (\bibinfo{year}{2003}).
\bibinfo{title}{{Finite density QCD via imaginary chemical potential}}.
\bibinfo{journal}{{\em Phys. Rev. D}} \bibinfo{volume}{67}:
  \bibinfo{pages}{014505}. \bibinfo{doi}{\doi{10.1103/PhysRevD.67.014505}}.
\href{https://arxiv.org/abs/hep-lat/0209146}{{\tt hep-lat/0209146}}.

\bibtype{Article}%
\bibitem[Detmold et al.(2020)]{Detmold:2020ncp}
\bibinfo{author}{Detmold W}, \bibinfo{author}{Kanwar G},
  \bibinfo{author}{Wagman ML} and  \bibinfo{author}{Warrington NC}
  (\bibinfo{year}{2020}).
\bibinfo{title}{{Path integral contour deformations for noisy observables}}.
\bibinfo{journal}{{\em Phys. Rev. D}} \bibinfo{volume}{102}
  (\bibinfo{number}{1}): \bibinfo{pages}{014514}.
  \bibinfo{doi}{\doi{10.1103/PhysRevD.102.014514}}.
\href{http://arxiv.org/abs/2003.05914}{{\tt arXiv:2003.05914}}.

\bibtype{Article}%
\bibitem[Detmold et al.(2021)]{Detmold:2021ulb}
\bibinfo{author}{Detmold W}, \bibinfo{author}{Kanwar G}, \bibinfo{author}{Lamm
  H}, \bibinfo{author}{Wagman ML} and  \bibinfo{author}{Warrington NC}
  (\bibinfo{year}{2021}).
\bibinfo{title}{{Path integral contour deformations for observables in $SU(N)$
  gauge theory}}.
\bibinfo{journal}{{\em Phys. Rev. D}} \bibinfo{volume}{103}
  (\bibinfo{number}{9}): \bibinfo{pages}{094517}.
  \bibinfo{doi}{\doi{10.1103/PhysRevD.103.094517}}.
\href{http://arxiv.org/abs/2101.12668}{{\tt arXiv:2101.12668}}.

\bibtype{Article}%
\bibitem[Dhindsa et al.(2025)]{Dhindsa:2025xfv}
\bibinfo{author}{Dhindsa NS}, \bibinfo{author}{Joseph A} and
  \bibinfo{author}{Longia V} (\bibinfo{year}{2025}).
\bibinfo{title}{{Gradient and Hessian-Based temperature estimator in lattice
  gauge theories: a diagnostic tool for stability and consistency in numerical
  simulations}}.
\bibinfo{journal}{{\em JHEP}} \bibinfo{volume}{10}: \bibinfo{pages}{015}.
  \bibinfo{doi}{\doi{10.1007/JHEP10(2025)015}}.
\href{http://arxiv.org/abs/2508.05595}{{\tt arXiv:2508.05595}}.

\bibtype{Article}%
\bibitem[Di~Renzo and Eruzzi(2018)]{DiRenzo:2017igr}
\bibinfo{author}{Di~Renzo F} and  \bibinfo{author}{Eruzzi G}
  (\bibinfo{year}{2018}).
\bibinfo{title}{{One-dimensional QCD in thimble regularization}}.
\bibinfo{journal}{{\em Phys. Rev. D}} \bibinfo{volume}{97}
  (\bibinfo{number}{1}): \bibinfo{pages}{014503}.
  \bibinfo{doi}{\doi{10.1103/PhysRevD.97.014503}}.
\href{http://arxiv.org/abs/1709.10468}{{\tt arXiv:1709.10468}}.

\bibtype{Article}%
\bibitem[Di~Renzo and Zambello(2022)]{DiRenzo:2021kcw}
\bibinfo{author}{Di~Renzo F} and  \bibinfo{author}{Zambello K}
  (\bibinfo{year}{2022}).
\bibinfo{title}{{Solution of the Thirring model in thimble regularization}}.
\bibinfo{journal}{{\em Phys. Rev. D}} \bibinfo{volume}{105}
  (\bibinfo{number}{5}): \bibinfo{pages}{054501}.
  \bibinfo{doi}{\doi{10.1103/PhysRevD.105.054501}}.
\href{http://arxiv.org/abs/2109.02511}{{\tt arXiv:2109.02511}}.

\bibtype{Inproceedings}%
\bibitem[Ding(2026)]{Ding:2026gao}
\bibinfo{author}{Ding HT} (\bibinfo{year}{2026}), \bibinfo{title}{{Lattice QCD
  at finite temperature and density}}, \bibinfo{booktitle}{{42th International
  Symposium on Lattice Field Theory}},
  \href{http://arxiv.org/abs/2603.16230}{{\tt arXiv:2603.16230}}.

\bibtype{Article}%
\bibitem[Ding et al.(2015)]{Ding:2015ona}
\bibinfo{author}{Ding HT}, \bibinfo{author}{Karsch F} and
  \bibinfo{author}{Mukherjee S} (\bibinfo{year}{2015}).
\bibinfo{title}{{Thermodynamics of strong-interaction matter from Lattice
  QCD}}.
\bibinfo{journal}{{\em Int. J. Mod. Phys. E}} \bibinfo{volume}{24}
  (\bibinfo{number}{10}): \bibinfo{pages}{1530007}.
  \bibinfo{doi}{\doi{10.1142/S0218301315300076}}.
\href{http://arxiv.org/abs/1504.05274}{{\tt arXiv:1504.05274}}.

\bibtype{Article}%
\bibitem[Dong et al.(2020)]{Dong:2020mtk}
\bibinfo{author}{Dong X}, \bibinfo{author}{Cai Z} and  \bibinfo{author}{Di Y}
  (\bibinfo{year}{2020}).
\bibinfo{title}{{Alternating Descent Method for Gauge Cooling of Complex
  Langevin Simulations}}.
\bibinfo{journal}{{\em Phys. Rev. D}} \bibinfo{volume}{102}
  (\bibinfo{number}{5}): \bibinfo{pages}{054518}.
  \bibinfo{doi}{\doi{10.1103/PhysRevD.102.054518}}.
\href{http://arxiv.org/abs/2008.06654}{{\tt arXiv:2008.06654}}.

\bibtype{Article}%
\bibitem[Endres(2007)]{Endres:2006xu}
\bibinfo{author}{Endres MG} (\bibinfo{year}{2007}).
\bibinfo{title}{{Method for simulating O(N) lattice models at finite density}}.
\bibinfo{journal}{{\em Phys. Rev. D}} \bibinfo{volume}{75}:
  \bibinfo{pages}{065012}. \bibinfo{doi}{\doi{10.1103/PhysRevD.75.065012}}.
\href{https://arxiv.org/abs/hep-lat/0610029}{{\tt hep-lat/0610029}}.

\bibtype{Article}%
\bibitem[Fodor et al.(2007)]{Fodor:2007vv}
\bibinfo{author}{Fodor Z}, \bibinfo{author}{Katz SD} and
  \bibinfo{author}{Schmidt C} (\bibinfo{year}{2007}).
\bibinfo{title}{{The Density of states method at non-zero chemical potential}}.
\bibinfo{journal}{{\em JHEP}} \bibinfo{volume}{03}: \bibinfo{pages}{121}.
  \bibinfo{doi}{\doi{10.1088/1126-6708/2007/03/121}}.
\href{https://arxiv.org/abs/hep-lat/0701022}{{\tt hep-lat/0701022}}.

\bibtype{Article}%
\bibitem[Fodor et al.(2015)]{Fodor:2015doa}
\bibinfo{author}{Fodor Z}, \bibinfo{author}{Katz SD}, \bibinfo{author}{Sexty D}
  and  \bibinfo{author}{T{\"o}r{\"o}k C} (\bibinfo{year}{2015}).
\bibinfo{title}{{Complex Langevin dynamics for dynamical QCD at nonzero
  chemical potential: A comparison with multiparameter reweighting}}.
\bibinfo{journal}{{\em Phys. Rev. D}} \bibinfo{volume}{92}
  (\bibinfo{number}{9}): \bibinfo{pages}{094516}.
  \bibinfo{doi}{\doi{10.1103/PhysRevD.92.094516}}.
\href{http://arxiv.org/abs/1508.05260}{{\tt arXiv:1508.05260}}.

\bibtype{Article}%
\bibitem[Fujii et al.(2013)]{Fujii:2013sra}
\bibinfo{author}{Fujii H}, \bibinfo{author}{Honda D}, \bibinfo{author}{Kato M},
  \bibinfo{author}{Kikukawa Y}, \bibinfo{author}{Komatsu S} and
  \bibinfo{author}{Sano T} (\bibinfo{year}{2013}).
\bibinfo{title}{{Hybrid Monte Carlo on Lefschetz thimbles - A study of the
  residual sign problem}}.
\bibinfo{journal}{{\em JHEP}} \bibinfo{volume}{10}: \bibinfo{pages}{147}.
  \bibinfo{doi}{\doi{10.1007/JHEP10(2013)147}}.
\href{http://arxiv.org/abs/1309.4371}{{\tt arXiv:1309.4371}}.

\bibtype{Article}%
\bibitem[Fujisawa et al.(2022)]{Fujisawa:2021hxh}
\bibinfo{author}{Fujisawa G}, \bibinfo{author}{Nishimura J},
  \bibinfo{author}{Sakai K} and  \bibinfo{author}{Yosprakob A}
  (\bibinfo{year}{2022}).
\bibinfo{title}{{Backpropagating Hybrid Monte Carlo algorithm for fast
  Lefschetz thimble calculations}}.
\bibinfo{journal}{{\em JHEP}} \bibinfo{volume}{04}: \bibinfo{pages}{179}.
  \bibinfo{doi}{\doi{10.1007/JHEP04(2022)179}}.
\href{http://arxiv.org/abs/2112.10519}{{\tt arXiv:2112.10519}}.

\bibtype{Article}%
\bibitem[Fukuma(2025)]{Fukuma:2025gya}
\bibinfo{author}{Fukuma M} (\bibinfo{year}{2025}).
\bibinfo{title}{{Worldvolume Hybrid Monte Carlo algorithm for group manifolds}}
  \href{http://arxiv.org/abs/2506.12002}{{\tt arXiv:2506.12002}}.

\bibtype{Article}%
\bibitem[Fukuma and Matsumoto(2021)]{Fukuma:2020fez}
\bibinfo{author}{Fukuma M} and  \bibinfo{author}{Matsumoto N}
  (\bibinfo{year}{2021}).
\bibinfo{title}{{Worldvolume approach to the tempered Lefschetz thimble
  method}}.
\bibinfo{journal}{{\em PTEP}} \bibinfo{volume}{2021} (\bibinfo{number}{2}):
  \bibinfo{pages}{023B08}. \bibinfo{doi}{\doi{10.1093/ptep/ptab010}}.
\href{http://arxiv.org/abs/2012.08468}{{\tt arXiv:2012.08468}}.

\bibtype{Article}%
\bibitem[Fukuma and Namekawa(2025)]{Fukuma:2025cxg}
\bibinfo{author}{Fukuma M} and  \bibinfo{author}{Namekawa Y}
  (\bibinfo{year}{2025}).
\bibinfo{title}{{Enhancing the ergodicity of Worldvolume HMC via embedding
  Generalized-thimble HMC}} \href{http://arxiv.org/abs/2508.02659}{{\tt
  arXiv:2508.02659}}.

\bibtype{Article}%
\bibitem[Fukuma and Umeda(2017)]{Fukuma:2017fjq}
\bibinfo{author}{Fukuma M} and  \bibinfo{author}{Umeda N}
  (\bibinfo{year}{2017}).
\bibinfo{title}{{Parallel tempering algorithm for integration over Lefschetz
  thimbles}}.
\bibinfo{journal}{{\em PTEP}} \bibinfo{volume}{2017} (\bibinfo{number}{7}):
  \bibinfo{pages}{073B01}. \bibinfo{doi}{\doi{10.1093/ptep/ptx081}}.
\href{http://arxiv.org/abs/1703.00861}{{\tt arXiv:1703.00861}}.

\bibtype{Article}%
\bibitem[Fukuma et al.(2021)]{Fukuma:2021aoo}
\bibinfo{author}{Fukuma M}, \bibinfo{author}{Matsumoto N} and
  \bibinfo{author}{Namekawa Y} (\bibinfo{year}{2021}).
\bibinfo{title}{{Statistical analysis method for the worldvolume hybrid Monte
  Carlo algorithm}}.
\bibinfo{journal}{{\em PTEP}} \bibinfo{volume}{2021} (\bibinfo{number}{12}):
  \bibinfo{pages}{123B02}. \bibinfo{doi}{\doi{10.1093/ptep/ptab133}}.
\href{http://arxiv.org/abs/2107.06858}{{\tt arXiv:2107.06858}}.

\bibtype{Article}%
\bibitem[Funcke et al.(2023)]{Funcke:2023jbq}
\bibinfo{author}{Funcke L}, \bibinfo{author}{Hartung T},
  \bibinfo{author}{Jansen K} and  \bibinfo{author}{K{\"u}hn S}
  (\bibinfo{year}{2023}).
\bibinfo{title}{{Review on Quantum Computing for Lattice Field Theory}}.
\bibinfo{journal}{{\em PoS}} \bibinfo{volume}{LATTICE2022}:
  \bibinfo{pages}{228}. \bibinfo{doi}{\doi{10.22323/1.430.0228}}.
\href{http://arxiv.org/abs/2302.00467}{{\tt arXiv:2302.00467}}.

\bibtype{Article}%
\bibitem[G{\"a}ntgen et al.(2024)]{Gantgen:2023byf}
\bibinfo{author}{G{\"a}ntgen C}, \bibinfo{author}{Berkowitz E},
  \bibinfo{author}{Luu T}, \bibinfo{author}{Ostmeyer J} and
  \bibinfo{author}{Rodekamp M} (\bibinfo{year}{2024}).
\bibinfo{title}{{Fermionic sign problem minimization by constant path integral
  contour shifts}}.
\bibinfo{journal}{{\em Phys. Rev. B}} \bibinfo{volume}{109}
  (\bibinfo{number}{19}): \bibinfo{pages}{195158}.
  \bibinfo{doi}{\doi{10.1103/PhysRevB.109.195158}}.
\href{http://arxiv.org/abs/2307.06785}{{\tt arXiv:2307.06785}}.

\bibtype{Article}%
\bibitem[Gattringer(2011)]{Gattringer:2011gq}
\bibinfo{author}{Gattringer C} (\bibinfo{year}{2011}).
\bibinfo{title}{{Flux representation of an effective Polyakov loop model for
  QCD thermodynamics}}.
\bibinfo{journal}{{\em Nucl. Phys. B}} \bibinfo{volume}{850}:
  \bibinfo{pages}{242--252}.
  \bibinfo{doi}{\doi{10.1016/j.nuclphysb.2011.04.018}}.
\href{http://arxiv.org/abs/1104.2503}{{\tt arXiv:1104.2503}}.

\bibtype{Article}%
\bibitem[Gattringer(2014)]{Gattringer:2014nxa}
\bibinfo{author}{Gattringer C} (\bibinfo{year}{2014}).
\bibinfo{title}{{New developments for dual methods in lattice field theory at
  non-zero density}}.
\bibinfo{journal}{{\em PoS}} \bibinfo{volume}{LATTICE2013}:
  \bibinfo{pages}{002}. \bibinfo{doi}{\doi{10.22323/1.187.0002}}.
\href{http://arxiv.org/abs/1401.7788}{{\tt arXiv:1401.7788}}.

\bibtype{Article}%
\bibitem[Gattringer and Kloiber(2013)]{Gattringer:2012df}
\bibinfo{author}{Gattringer C} and  \bibinfo{author}{Kloiber T}
  (\bibinfo{year}{2013}).
\bibinfo{title}{{Lattice study of the Silver Blaze phenomenon for a charged
  scalar $\phi^4$ field}}.
\bibinfo{journal}{{\em Nucl. Phys. B}} \bibinfo{volume}{869}:
  \bibinfo{pages}{56--73}.
  \bibinfo{doi}{\doi{10.1016/j.nuclphysb.2012.12.005}}.
\href{http://arxiv.org/abs/1206.2954}{{\tt arXiv:1206.2954}}.

\bibtype{Book}%
\bibitem[Gattringer and Lang(2010)]{Gattringer:2010zz}
\bibinfo{author}{Gattringer C} and  \bibinfo{author}{Lang CB}
  (\bibinfo{year}{2010}).
\bibinfo{title}{{Quantum chromodynamics on the lattice}},
  \bibinfo{volume}{(788)}, \bibinfo{publisher}{Springer},
  \bibinfo{address}{Berlin}.
\bibinfo{comment}{ISBN} \bibinfo{isbn}{978-3-642-01849-7, 978-3-642-01850-3}.
\bibinfo{doi}{\doi{10.1007/978-3-642-01850-3}}.

\bibtype{Article}%
\bibitem[Gattringer and Langfeld(2016)]{Gattringer:2016kco}
\bibinfo{author}{Gattringer C} and  \bibinfo{author}{Langfeld K}
  (\bibinfo{year}{2016}).
\bibinfo{title}{{Approaches to the sign problem in lattice field theory}}.
\bibinfo{journal}{{\em Int. J. Mod. Phys. A}} \bibinfo{volume}{31}
  (\bibinfo{number}{22}): \bibinfo{pages}{1643007}.
  \bibinfo{doi}{\doi{10.1142/S0217751X16430077}}.
\href{http://arxiv.org/abs/1603.09517}{{\tt arXiv:1603.09517}}.

\bibtype{Article}%
\bibitem[Gattringer et al.(2015)]{Gattringer:2015baa}
\bibinfo{author}{Gattringer C}, \bibinfo{author}{Kloiber T} and
  \bibinfo{author}{M{\"u}ller-Preussker M} (\bibinfo{year}{2015}).
\bibinfo{title}{{Dual simulation of the two-dimensional lattice U(1)
  gauge-Higgs model with a topological term}}.
\bibinfo{journal}{{\em Phys. Rev. D}} \bibinfo{volume}{92}
  (\bibinfo{number}{11}): \bibinfo{pages}{114508}.
  \bibinfo{doi}{\doi{10.1103/PhysRevD.92.114508}}.
\href{http://arxiv.org/abs/1508.00681}{{\tt arXiv:1508.00681}}.

\bibtype{Article}%
\bibitem[Giordano et al.(2023)]{Giordano:2023ppk}
\bibinfo{author}{Giordano M}, \bibinfo{author}{Pasztor A},
  \bibinfo{author}{Pesznyak D} and  \bibinfo{author}{Tulipant Z}
  (\bibinfo{year}{2023}).
\bibinfo{title}{{Alleviating the sign problem in a chiral random matrix model
  with contour deformations}}.
\bibinfo{journal}{{\em Phys. Rev. D}} \bibinfo{volume}{108}
  (\bibinfo{number}{9}): \bibinfo{pages}{094507}.
  \bibinfo{doi}{\doi{10.1103/PhysRevD.108.094507}}.
\href{http://arxiv.org/abs/2301.12947}{{\tt arXiv:2301.12947}}.

\bibtype{Article}%
\bibitem[Halimeh et al.(2025)]{Halimeh:2025vvp}
\bibinfo{author}{Halimeh JC}, \bibinfo{author}{Mueller N},
  \bibinfo{author}{Knolle J}, \bibinfo{author}{Papi{\'c} Z} and
  \bibinfo{author}{Davoudi Z} (\bibinfo{year}{2025}).
\bibinfo{title}{{Quantum simulation of out-of-equilibrium dynamics in gauge
  theories}} \href{http://arxiv.org/abs/2509.03586}{{\tt arXiv:2509.03586}}.

\bibtype{Article}%
\bibitem[Hansen and Sexty(2025)]{Hansen:2024lkn}
\bibinfo{author}{Hansen MW} and  \bibinfo{author}{Sexty D}
  (\bibinfo{year}{2025}).
\bibinfo{title}{{Testing dynamical stabilization of complex Langevin
  simulations of QCD}}.
\bibinfo{journal}{{\em Phys. Rev. D}} \bibinfo{volume}{111}
  (\bibinfo{number}{5}): \bibinfo{pages}{054508}.
  \bibinfo{doi}{\doi{10.1103/PhysRevD.111.054508}}.
\href{http://arxiv.org/abs/2405.20709}{{\tt arXiv:2405.20709}}.

\bibtype{Article}%
\bibitem[Hansen et al.(2025)]{Hansen:2024kjm}
\bibinfo{author}{Hansen MW}, \bibinfo{author}{Mandl M}, \bibinfo{author}{Seiler
  E} and  \bibinfo{author}{Sexty D} (\bibinfo{year}{2025}).
\bibinfo{title}{{Role of integration cycles in complex Langevin simulations}}.
\bibinfo{journal}{{\em Phys. Rev. D}} \bibinfo{volume}{111}
  (\bibinfo{number}{7}): \bibinfo{pages}{074502}.
  \bibinfo{doi}{\doi{10.1103/PhysRevD.111.074502}}.
\href{http://arxiv.org/abs/2412.17137}{{\tt arXiv:2412.17137}}.

\bibtype{Article}%
\bibitem[Hasenfratz and Karsch(1983)]{Hasenfratz:1983ba}
\bibinfo{author}{Hasenfratz P} and  \bibinfo{author}{Karsch F}
  (\bibinfo{year}{1983}).
\bibinfo{title}{{Chemical Potential on the Lattice}}.
\bibinfo{journal}{{\em Phys. Lett. B}} \bibinfo{volume}{125}:
  \bibinfo{pages}{308--310}. \bibinfo{doi}{\doi{10.1016/0370-2693(83)91290-X}}.

\bibtype{Article}%
\bibitem[Hisayoshi et al.(2025)]{Hisayoshi:2025adi}
\bibinfo{author}{Hisayoshi K}, \bibinfo{author}{Kashiwa K},
  \bibinfo{author}{Namekawa Y} and  \bibinfo{author}{Takase H}
  (\bibinfo{year}{2025}).
\bibinfo{title}{{Path optimization method for the sign problem caused by the
  fermion determinant}}.
\bibinfo{journal}{{\em Phys. Rev. D}} \bibinfo{volume}{111}
  (\bibinfo{number}{9}): \bibinfo{pages}{094503}.
  \bibinfo{doi}{\doi{10.1103/PhysRevD.111.094503}}.
\href{http://arxiv.org/abs/2502.02804}{{\tt arXiv:2502.02804}}.

\bibtype{Article}%
\bibitem[Ihssen et al.(2026)]{Ihssen:2026njd}
\bibinfo{author}{Ihssen F}, \bibinfo{author}{Kapust R} and
  \bibinfo{author}{Pawlowski JM} (\bibinfo{year}{2026}).
\bibinfo{title}{{Solving sign problems with physics-informed kernels}}
  \href{http://arxiv.org/abs/2603.03159}{{\tt arXiv:2603.03159}}.

\bibtype{Article}%
\bibitem[Ito et al.(2020)]{Ito:2020mys}
\bibinfo{author}{Ito Y}, \bibinfo{author}{Matsufuru H},
  \bibinfo{author}{Namekawa Y}, \bibinfo{author}{Nishimura J},
  \bibinfo{author}{Shimasaki S}, \bibinfo{author}{Tsuchiya A} and
  \bibinfo{author}{Tsutsui S} (\bibinfo{year}{2020}).
\bibinfo{title}{{Complex Langevin calculations in QCD at finite density}}.
\bibinfo{journal}{{\em JHEP}} \bibinfo{volume}{10}: \bibinfo{pages}{144}.
  \bibinfo{doi}{\doi{10.1007/JHEP10(2020)144}}.
\href{http://arxiv.org/abs/2007.08778}{{\tt arXiv:2007.08778}}.

\bibtype{Article}%
\bibitem[Jeon and Yaffe(1996)]{Jeon:1995zm}
\bibinfo{author}{Jeon S} and  \bibinfo{author}{Yaffe LG}
  (\bibinfo{year}{1996}).
\bibinfo{title}{{From quantum field theory to hydrodynamics: Transport
  coefficients and effective kinetic theory}}.
\bibinfo{journal}{{\em Phys. Rev. D}} \bibinfo{volume}{53}:
  \bibinfo{pages}{5799--5809}. \bibinfo{doi}{\doi{10.1103/PhysRevD.53.5799}}.
\href{https://arxiv.org/abs/hep-ph/9512263}{{\tt hep-ph/9512263}}.

\bibtype{Article}%
\bibitem[Joseph and Kumar(2025{\natexlab{a}})]{Joseph:2025xbn}
\bibinfo{author}{Joseph A} and  \bibinfo{author}{Kumar A}
  (\bibinfo{year}{2025}{\natexlab{a}}).
\bibinfo{title}{{Configurational Temperature as a Diagnostic for Complex
  Langevin Dynamics in the 3D XY Model}}
  \href{http://arxiv.org/abs/2509.13314}{{\tt arXiv:2509.13314}}.

\bibtype{Article}%
\bibitem[Joseph and Kumar(2025{\natexlab{b}})]{Joseph:2025fcd}
\bibinfo{author}{Joseph A} and  \bibinfo{author}{Kumar A}
  (\bibinfo{year}{2025}{\natexlab{b}}).
\bibinfo{title}{{Thermodynamic Diagnostics for Complex Langevin Simulations:
  The Role of Configurational Temperature}}
  \href{http://arxiv.org/abs/2509.08287}{{\tt arXiv:2509.08287}}.

\bibtype{Article}%
\bibitem[Kadoh et al.(2020)]{Kadoh:2019ube}
\bibinfo{author}{Kadoh D}, \bibinfo{author}{Kuramashi Y},
  \bibinfo{author}{Nakamura Y}, \bibinfo{author}{Sakai R},
  \bibinfo{author}{Takeda S} and  \bibinfo{author}{Yoshimura Y}
  (\bibinfo{year}{2020}).
\bibinfo{title}{{Investigation of complex $\phi^{4}$ theory at finite density
  in two dimensions using TRG}}.
\bibinfo{journal}{{\em JHEP}} \bibinfo{volume}{02}: \bibinfo{pages}{161}.
  \bibinfo{doi}{\doi{10.1007/JHEP02(2020)161}}.
\href{http://arxiv.org/abs/1912.13092}{{\tt arXiv:1912.13092}}.

\bibtype{Inproceedings}%
\bibitem[Kanwar(2024)]{Kanwar:2024ujc}
\bibinfo{author}{Kanwar G} (\bibinfo{year}{2024}), \bibinfo{title}{{Flow-based
  sampling for lattice field theories}}, \bibinfo{booktitle}{{40th
  International Symposium on Lattice Field Theory}},
  \href{http://arxiv.org/abs/2401.01297}{{\tt arXiv:2401.01297}}.

\bibtype{Article}%
\bibitem[Kanwar et al.(2020)]{Kanwar:2020xzo}
\bibinfo{author}{Kanwar G}, \bibinfo{author}{Albergo MS},
  \bibinfo{author}{Boyda D}, \bibinfo{author}{Cranmer K},
  \bibinfo{author}{Hackett DC}, \bibinfo{author}{Racani{\`e}re S},
  \bibinfo{author}{Rezende DJ} and  \bibinfo{author}{Shanahan PE}
  (\bibinfo{year}{2020}).
\bibinfo{title}{Equivariant {{Flow-Based Sampling}} for {{Lattice Gauge
  Theory}}}.
\bibinfo{journal}{{\em Phys. Rev. Lett.}} \bibinfo{volume}{125}
  (\bibinfo{number}{12}): \bibinfo{pages}{121601}.
  \bibinfo{doi}{\doi{10.1103/PhysRevLett.125.121601}}.
\href{http://arxiv.org/abs/2003.06413}{{\tt arXiv:2003.06413}}.

\bibtype{Article}%
\bibitem[Karsch and Wyld(1985)]{Karsch:1985cb}
\bibinfo{author}{Karsch F} and  \bibinfo{author}{Wyld HW}
  (\bibinfo{year}{1985}).
\bibinfo{title}{{Complex Langevin Simulation of the SU(3) Spin Model With
  Nonzero Chemical Potential}}.
\bibinfo{journal}{{\em Phys. Rev. Lett.}} \bibinfo{volume}{55}:
  \bibinfo{pages}{2242}. \bibinfo{doi}{\doi{10.1103/PhysRevLett.55.2242}}.

\bibtype{Article}%
\bibitem[Kashiwa and Mori(2020)]{Kashiwa:2020brj}
\bibinfo{author}{Kashiwa K} and  \bibinfo{author}{Mori Y}
  (\bibinfo{year}{2020}).
\bibinfo{title}{{Path optimization for $U(1)$ gauge theory with complexified
  parameters}}.
\bibinfo{journal}{{\em Phys. Rev. D}} \bibinfo{volume}{102}
  (\bibinfo{number}{5}): \bibinfo{pages}{054519}.
  \bibinfo{doi}{\doi{10.1103/PhysRevD.102.054519}}.
\href{http://arxiv.org/abs/2007.04167}{{\tt arXiv:2007.04167}}.

\bibtype{Article}%
\bibitem[Kashiwa et al.(2019)]{Kashiwa:2019lkv}
\bibinfo{author}{Kashiwa K}, \bibinfo{author}{Mori Y} and
  \bibinfo{author}{Ohnishi A} (\bibinfo{year}{2019}).
\bibinfo{title}{{Application of the path optimization method to the sign
  problem in an effective model of QCD with a repulsive vector-type
  interaction}}.
\bibinfo{journal}{{\em Phys. Rev. D}} \bibinfo{volume}{99}
  (\bibinfo{number}{11}): \bibinfo{pages}{114005}.
  \bibinfo{doi}{\doi{10.1103/PhysRevD.99.114005}}.
\href{http://arxiv.org/abs/1903.03679}{{\tt arXiv:1903.03679}}.

\bibtype{Article}%
\bibitem[Keldysh(1965)]{Keldysh:1964ud}
\bibinfo{author}{Keldysh LV} (\bibinfo{year}{1965}).
\bibinfo{title}{{Diagram Technique for Nonequilibrium Processes}}.
\bibinfo{journal}{{\em Sov. Phys. JETP}} \bibinfo{volume}{20}:
  \bibinfo{pages}{1018--1026}. \bibinfo{doi}{\doi{10.1142/9789811279461_0007}}.

\bibtype{Inproceedings}%
\bibitem[Klauder(1983)]{klauder}
\bibinfo{author}{Klauder JR} (\bibinfo{year}{1983}), \bibinfo{title}{Stochastic
  quantization}, \bibinfo{editor}{Mitter H} and  \bibinfo{editor}{Lang CB},
  (Eds.), \bibinfo{booktitle}{Recent Developments in High-Energy Physics},
  \bibinfo{publisher}{Springer Vienna}, \bibinfo{address}{Vienna},
  \bibinfo{pages}{251--281}.

\bibtype{Article}%
\bibitem[Kogut and Sinclair(2019)]{Kogut:2019qmi}
\bibinfo{author}{Kogut JB} and  \bibinfo{author}{Sinclair DK}
  (\bibinfo{year}{2019}).
\bibinfo{title}{{Applying Complex Langevin Simulations to Lattice QCD at Finite
  Density}}.
\bibinfo{journal}{{\em Phys. Rev. D}} \bibinfo{volume}{100}
  (\bibinfo{number}{5}): \bibinfo{pages}{054512}.
  \bibinfo{doi}{\doi{10.1103/PhysRevD.100.054512}}.
\href{http://arxiv.org/abs/1903.02622}{{\tt arXiv:1903.02622}}.

\bibtype{Article}%
\bibitem[Kuramashi and Yoshimura(2020)]{Kuramashi:2019cgs}
\bibinfo{author}{Kuramashi Y} and  \bibinfo{author}{Yoshimura Y}
  (\bibinfo{year}{2020}).
\bibinfo{title}{{Tensor renormalization group study of two-dimensional U(1)
  lattice gauge theory with a $\theta$ term}}.
\bibinfo{journal}{{\em JHEP}} \bibinfo{volume}{04}: \bibinfo{pages}{089}.
  \bibinfo{doi}{\doi{10.1007/JHEP04(2020)089}}.
\href{http://arxiv.org/abs/1911.06480}{{\tt arXiv:1911.06480}}.

\bibtype{Article}%
\bibitem[Lai et al.(2025)]{2510.21890}
\bibinfo{author}{Lai CH}, \bibinfo{author}{Song Y}, \bibinfo{author}{Kim D},
  \bibinfo{author}{Mitsufuji Y} and  \bibinfo{author}{Ermon S}
  (\bibinfo{year}{2025}).
\bibinfo{title}{{The Principles of Diffusion Models}}
  \href{http://arxiv.org/abs/2510.21890}{{\tt arXiv:2510.21890}}.

\bibtype{Article}%
\bibitem[Lampl and Sexty(2025)]{Lampl:2023xpb}
\bibinfo{author}{Lampl NM} and  \bibinfo{author}{Sexty D}
  (\bibinfo{year}{2025}).
\bibinfo{title}{{Real time simulations of scalar fields with kernelled complex
  Langevin equation}}.
\bibinfo{journal}{{\em PoS}} \bibinfo{volume}{LATTICE2024}:
  \bibinfo{pages}{073}. \bibinfo{doi}{\doi{10.22323/1.466.0073}}.
\href{http://arxiv.org/abs/2309.06103}{{\tt arXiv:2309.06103}}.

\bibtype{Article}%
\bibitem[Langfeld et al.(2012)]{Langfeld:2012ah}
\bibinfo{author}{Langfeld K}, \bibinfo{author}{Lucini B} and
  \bibinfo{author}{Rago A} (\bibinfo{year}{2012}).
\bibinfo{title}{{The density of states in gauge theories}}.
\bibinfo{journal}{{\em Phys. Rev. Lett.}} \bibinfo{volume}{109}:
  \bibinfo{pages}{111601}. \bibinfo{doi}{\doi{10.1103/PhysRevLett.109.111601}}.
\href{http://arxiv.org/abs/1204.3243}{{\tt arXiv:1204.3243}}.

\bibtype{Article}%
\bibitem[Lawrence(2025)]{Lawrence:2025rnk}
\bibinfo{author}{Lawrence S} (\bibinfo{year}{2025}).
\bibinfo{title}{{Machine-learning approaches to accelerating lattice
  simulations}}.
\bibinfo{journal}{{\em PoS}} \bibinfo{volume}{LATTICE2024}:
  \bibinfo{pages}{010}. \bibinfo{doi}{\doi{10.22323/1.466.0010}}.
\href{http://arxiv.org/abs/2502.02670}{{\tt arXiv:2502.02670}}.

\bibtype{Article}%
\bibitem[Lawrence and Yamauchi(2021)]{Lawrence:2021izu}
\bibinfo{author}{Lawrence S} and  \bibinfo{author}{Yamauchi Y}
  (\bibinfo{year}{2021}).
\bibinfo{title}{{Normalizing Flows and the Real-Time Sign Problem}}.
\bibinfo{journal}{{\em Phys. Rev. D}} \bibinfo{volume}{103}
  (\bibinfo{number}{11}): \bibinfo{pages}{114509}.
  \bibinfo{doi}{\doi{10.1103/PhysRevD.103.114509}}.
\href{http://arxiv.org/abs/2101.05755}{{\tt arXiv:2101.05755}}.

\bibtype{Article}%
\bibitem[Lawrence and Yamauchi(2023)]{Lawrence:2022dba}
\bibinfo{author}{Lawrence S} and  \bibinfo{author}{Yamauchi Y}
  (\bibinfo{year}{2023}).
\bibinfo{title}{{Deep learning of fermion sign fluctuations}}.
\bibinfo{journal}{{\em Phys. Rev. D}} \bibinfo{volume}{107}
  (\bibinfo{number}{11}): \bibinfo{pages}{114505}.
  \bibinfo{doi}{\doi{10.1103/PhysRevD.107.114505}}.
\href{http://arxiv.org/abs/2212.14606}{{\tt arXiv:2212.14606}}.

\bibtype{Article}%
\bibitem[Lawrence and Yamauchi(2024)]{Lawrence:2023sfc}
\bibinfo{author}{Lawrence S} and  \bibinfo{author}{Yamauchi Y}
  (\bibinfo{year}{2024}).
\bibinfo{title}{{Convex optimization of contour deformations}}.
\bibinfo{journal}{{\em Phys. Rev. D}} \bibinfo{volume}{110}
  (\bibinfo{number}{1}): \bibinfo{pages}{014508}.
  \bibinfo{doi}{\doi{10.1103/PhysRevD.110.014508}}.
\href{http://arxiv.org/abs/2311.13002}{{\tt arXiv:2311.13002}}.

\bibtype{Article}%
\bibitem[Levin and Nave(2007)]{Levin:2006jai}
\bibinfo{author}{Levin M} and  \bibinfo{author}{Nave CP}
  (\bibinfo{year}{2007}).
\bibinfo{title}{{Tensor renormalization group approach to 2D classical lattice
  models}}.
\bibinfo{journal}{{\em Phys. Rev. Lett.}} \bibinfo{volume}{99}:
  \bibinfo{pages}{120601}. \bibinfo{doi}{\doi{10.1103/PhysRevLett.99.120601}}.
\href{https://arxiv.org/abs/cond-mat/0611687}{{\tt cond-mat/0611687}}.

\bibtype{Article}%
\bibitem[Lin et al.(2024)]{Lin:2023svo}
\bibinfo{author}{Lin Y}, \bibinfo{author}{Detmold W}, \bibinfo{author}{Kanwar
  G}, \bibinfo{author}{Shanahan PE} and  \bibinfo{author}{Wagman ML}
  (\bibinfo{year}{2024}).
\bibinfo{title}{{Signal-to-noise improvement through neural network contour
  deformations for 3D
  {\ensuremath{\boldsymbol{\mathit{S}}}}{\ensuremath{\boldsymbol{\mathit{U}}}}(2)
  lattice gauge theory}}.
\bibinfo{journal}{{\em PoS}} \bibinfo{volume}{LATTICE2023}:
  \bibinfo{pages}{043}. \bibinfo{doi}{\doi{10.22323/1.453.0043}}.
\href{http://arxiv.org/abs/2309.00600}{{\tt arXiv:2309.00600}}.

\bibtype{Article}%
\bibitem[Luo and Kuramashi(2023)]{Luo:2022eje}
\bibinfo{author}{Luo X} and  \bibinfo{author}{Kuramashi Y}
  (\bibinfo{year}{2023}).
\bibinfo{title}{{Tensor renormalization group approach to (1+1)-dimensional
  SU(2) principal chiral model at finite density}}.
\bibinfo{journal}{{\em Phys. Rev. D}} \bibinfo{volume}{107}
  (\bibinfo{number}{9}): \bibinfo{pages}{094509}.
  \bibinfo{doi}{\doi{10.1103/PhysRevD.107.094509}}.
\href{http://arxiv.org/abs/2208.13991}{{\tt arXiv:2208.13991}}.

\bibtype{Article}%
\bibitem[Luo and Kuramashi(2024)]{Luo:2024lbh}
\bibinfo{author}{Luo X} and  \bibinfo{author}{Kuramashi Y}
  (\bibinfo{year}{2024}).
\bibinfo{title}{{Quantum phase transition of (1+1)-dimensional O(3) nonlinear
  sigma model at finite density with tensor renormalization group}}.
\bibinfo{journal}{{\em JHEP}} \bibinfo{volume}{11}: \bibinfo{pages}{144}.
  \bibinfo{doi}{\doi{10.1007/JHEP11(2024)144}}.
\href{http://arxiv.org/abs/2406.08865}{{\tt arXiv:2406.08865}}.

\bibtype{Article}%
\bibitem[Luo and Kuramashi(2025)]{Luo:2025qtv}
\bibinfo{author}{Luo X} and  \bibinfo{author}{Kuramashi Y}
  (\bibinfo{year}{2025}).
\bibinfo{title}{{Critical endpoints of three-dimensional finite density SU(3)
  spin model with tensor renormalization group}}.
\bibinfo{journal}{{\em JHEP}} \bibinfo{volume}{07}: \bibinfo{pages}{036}.
  \bibinfo{doi}{\doi{10.1007/JHEP07(2025)036}}.
\href{http://arxiv.org/abs/2503.05144}{{\tt arXiv:2503.05144}}.

\bibtype{Article}%
\bibitem[Mandl(2026)]{Mandl:2026vdc}
\bibinfo{author}{Mandl M} (\bibinfo{year}{2026}).
\bibinfo{title}{{Correctness criteria for complex Langevin}}
  \href{http://arxiv.org/abs/2604.12388}{{\tt arXiv:2604.12388}}.

\bibtype{Inproceedings}%
\bibitem[Mandl et al.(2025{\natexlab{a}})]{Mandl:2025ins}
\bibinfo{author}{Mandl M}, \bibinfo{author}{Seiler E} and
  \bibinfo{author}{Sexty D} (\bibinfo{year}{2025}{\natexlab{a}}),
  \bibinfo{title}{{Complex Langevin simulations with a kernel}},
  \bibinfo{booktitle}{{42th International Symposium on Lattice Field Theory}},
  \href{http://arxiv.org/abs/2512.14153}{{\tt arXiv:2512.14153}}.

\bibtype{Article}%
\bibitem[Mandl et al.(2025{\natexlab{b}})]{Mandl:2025mav}
\bibinfo{author}{Mandl M}, \bibinfo{author}{Seiler E} and
  \bibinfo{author}{Sexty D} (\bibinfo{year}{2025}{\natexlab{b}}).
\bibinfo{title}{{Necessary and sufficient conditions for correctness of complex
  Langevin}}.
\bibinfo{journal}{{\em J. Phys. A}} \bibinfo{volume}{58}
  (\bibinfo{number}{49}): \bibinfo{pages}{495202}.
  \bibinfo{doi}{\doi{10.1088/1751-8121/ae2245}}.
\href{http://arxiv.org/abs/2508.14512}{{\tt arXiv:2508.14512}}.

\bibtype{Article}%
\bibitem[Mandl et al.(2026)]{Mandl:2026ngb}
\bibinfo{author}{Mandl M}, \bibinfo{author}{Sexty D} and
  \bibinfo{author}{Unterhuber D} (\bibinfo{year}{2026}).
\bibinfo{title}{{Finite-density equation of state of hot QCD using the complex
  Langevin equation}} \href{http://arxiv.org/abs/2604.19649}{{\tt
  arXiv:2604.19649}}.

\bibtype{Article}%
\bibitem[Matsen et al.(2026)]{Matsen2026}
\bibinfo{author}{Matsen MW}, \bibinfo{author}{Ramírez J},
  \bibinfo{author}{Willis JD} and  \bibinfo{author}{Pina PD}
  (\bibinfo{year}{2026}).
\bibinfo{title}{{Accessing the universal phase behavior of block copolymer
  melts with complex-Langevin field-theoretic simulations}}.
\bibinfo{journal}{{\em The Journal of Chemical Physics}} \bibinfo{volume}{164}
  (\bibinfo{number}{1}): \bibinfo{pages}{014905}.
ISSN \bibinfo{issn}{0021-9606}. \bibinfo{doi}{\doi{10.1063/5.0311060}}.
\bibinfo{url}{\url{https://doi.org/10.1063/5.0311060}}.

\bibtype{Article}%
\bibitem[Mollgaard and Splittorff(2015)]{Mollgaard:2014mga}
\bibinfo{author}{Mollgaard A} and  \bibinfo{author}{Splittorff K}
  (\bibinfo{year}{2015}).
\bibinfo{title}{{Full simulation of chiral random matrix theory at nonzero
  chemical potential by complex Langevin}}.
\bibinfo{journal}{{\em Phys. Rev. D}} \bibinfo{volume}{91}
  (\bibinfo{number}{3}): \bibinfo{pages}{036007}.
  \bibinfo{doi}{\doi{10.1103/PhysRevD.91.036007}}.
\href{http://arxiv.org/abs/1412.2729}{{\tt arXiv:1412.2729}}.

\bibtype{Article}%
\bibitem[Mori et al.(2017)]{Mori:2017pne}
\bibinfo{author}{Mori Y}, \bibinfo{author}{Kashiwa K} and
  \bibinfo{author}{Ohnishi A} (\bibinfo{year}{2017}).
\bibinfo{title}{{Toward solving the sign problem with path optimization
  method}}.
\bibinfo{journal}{{\em Phys. Rev. D}} \bibinfo{volume}{96}
  (\bibinfo{number}{11}): \bibinfo{pages}{111501}.
  \bibinfo{doi}{\doi{10.1103/PhysRevD.96.111501}}.
\href{http://arxiv.org/abs/1705.05605}{{\tt arXiv:1705.05605}}.

\bibtype{Article}%
\bibitem[Mori et al.(2018)]{Mori:2017nwj}
\bibinfo{author}{Mori Y}, \bibinfo{author}{Kashiwa K} and
  \bibinfo{author}{Ohnishi A} (\bibinfo{year}{2018}).
\bibinfo{title}{{Application of a neural network to the sign problem via the
  path optimization method}}.
\bibinfo{journal}{{\em PTEP}} \bibinfo{volume}{2018} (\bibinfo{number}{2}):
  \bibinfo{pages}{023B04}. \bibinfo{doi}{\doi{10.1093/ptep/ptx191}}.
\href{http://arxiv.org/abs/1709.03208}{{\tt arXiv:1709.03208}}.

\bibtype{Article}%
\bibitem[Nagata(2022)]{Nagata:2021ugx}
\bibinfo{author}{Nagata K} (\bibinfo{year}{2022}).
\bibinfo{title}{{Finite-density lattice QCD and sign problem: Current status
  and open problems}}.
\bibinfo{journal}{{\em Prog. Part. Nucl. Phys.}} \bibinfo{volume}{127}:
  \bibinfo{pages}{103991}. \bibinfo{doi}{\doi{10.1016/j.ppnp.2022.103991}}.
\href{http://arxiv.org/abs/2108.12423}{{\tt arXiv:2108.12423}}.

\bibtype{Article}%
\bibitem[Nagata et al.(2016{\natexlab{a}})]{Nagata:2016vkn}
\bibinfo{author}{Nagata K}, \bibinfo{author}{Nishimura J} and
  \bibinfo{author}{Shimasaki S} (\bibinfo{year}{2016}{\natexlab{a}}).
\bibinfo{title}{{Argument for justification of the complex Langevin method and
  the condition for correct convergence}}.
\bibinfo{journal}{{\em Phys. Rev. D}} \bibinfo{volume}{94}
  (\bibinfo{number}{11}): \bibinfo{pages}{114515}.
  \bibinfo{doi}{\doi{10.1103/PhysRevD.94.114515}}.
\href{http://arxiv.org/abs/1606.07627}{{\tt arXiv:1606.07627}}.

\bibtype{Article}%
\bibitem[Nagata et al.(2016{\natexlab{b}})]{Nagata:2015uga}
\bibinfo{author}{Nagata K}, \bibinfo{author}{Nishimura J} and
  \bibinfo{author}{Shimasaki S} (\bibinfo{year}{2016}{\natexlab{b}}).
\bibinfo{title}{{Justification of the complex Langevin method with the gauge
  cooling procedure}}.
\bibinfo{journal}{{\em PTEP}} \bibinfo{volume}{2016} (\bibinfo{number}{1}):
  \bibinfo{pages}{013B01}. \bibinfo{doi}{\doi{10.1093/ptep/ptv173}}.
\href{http://arxiv.org/abs/1508.02377}{{\tt arXiv:1508.02377}}.

\bibtype{Article}%
\bibitem[Nagata et al.(2018)]{Nagata:2018net}
\bibinfo{author}{Nagata K}, \bibinfo{author}{Nishimura J} and
  \bibinfo{author}{Shimasaki S} (\bibinfo{year}{2018}).
\bibinfo{title}{{Testing the criterion for correct convergence in the complex
  Langevin method}}.
\bibinfo{journal}{{\em JHEP}} \bibinfo{volume}{05}: \bibinfo{pages}{004}.
  \bibinfo{doi}{\doi{10.1007/JHEP05(2018)004}}.
\href{http://arxiv.org/abs/1802.01876}{{\tt arXiv:1802.01876}}.

\bibtype{Article}%
\bibitem[Namekawa and Fukuma(2024)]{Namekawa:2024ert}
\bibinfo{author}{Namekawa Y} and  \bibinfo{author}{Fukuma M}
  (\bibinfo{year}{2024}).
\bibinfo{title}{{Applying the Worldvolume Hybrid Monte Carlo method to the
  finite-density complex $\phi^4$ model and the Hubbard model}}.
\bibinfo{journal}{{\em PoS}} \bibinfo{volume}{LATTICE2023}:
  \bibinfo{pages}{178}. \bibinfo{doi}{\doi{10.22323/1.453.0178}}.

\bibtype{Article}%
\bibitem[Nishimura and Shimasaki(2015)]{Nishimura:2015pba}
\bibinfo{author}{Nishimura J} and  \bibinfo{author}{Shimasaki S}
  (\bibinfo{year}{2015}).
\bibinfo{title}{{New Insights into the Problem with a Singular Drift Term in
  the Complex Langevin Method}}.
\bibinfo{journal}{{\em Phys. Rev. D}} \bibinfo{volume}{92}
  (\bibinfo{number}{1}): \bibinfo{pages}{011501}.
  \bibinfo{doi}{\doi{10.1103/PhysRevD.92.011501}}.
\href{http://arxiv.org/abs/1504.08359}{{\tt arXiv:1504.08359}}.

\bibtype{Article}%
\bibitem[Nishimura and Shimasaki(2017)]{Nishimura:2017vav}
\bibinfo{author}{Nishimura J} and  \bibinfo{author}{Shimasaki S}
  (\bibinfo{year}{2017}).
\bibinfo{title}{{Combining the complex Langevin method and the generalized
  Lefschetz-thimble method}}.
\bibinfo{journal}{{\em JHEP}} \bibinfo{volume}{06}: \bibinfo{pages}{023}.
  \bibinfo{doi}{\doi{10.1007/JHEP06(2017)023}}.
\href{http://arxiv.org/abs/1703.09409}{{\tt arXiv:1703.09409}}.

\bibtype{Article}%
\bibitem[Nishimura et al.(2023)]{Nishimura:2023dky}
\bibinfo{author}{Nishimura J}, \bibinfo{author}{Sakai K} and
  \bibinfo{author}{Yosprakob A} (\bibinfo{year}{2023}).
\bibinfo{title}{{A new picture of quantum tunneling in the real-time path
  integral from Lefschetz thimble calculations}}.
\bibinfo{journal}{{\em JHEP}} \bibinfo{volume}{09}: \bibinfo{pages}{110}.
  \bibinfo{doi}{\doi{10.1007/JHEP09(2023)110}}.
\href{http://arxiv.org/abs/2307.11199}{{\tt arXiv:2307.11199}}.

\bibtype{Article}%
\bibitem[Nishimura et al.(2024)]{Nishimura:2024bou}
\bibinfo{author}{Nishimura J}, \bibinfo{author}{Sakai K} and
  \bibinfo{author}{Yosprakob A} (\bibinfo{year}{2024}).
\bibinfo{title}{{Preconditioned flow as a solution to the hierarchical growth
  problem in the generalized Lefschetz thimble method}}.
\bibinfo{journal}{{\em JHEP}} \bibinfo{volume}{07}: \bibinfo{pages}{174}.
  \bibinfo{doi}{\doi{10.1007/JHEP07(2024)174}}.
\href{http://arxiv.org/abs/2404.16589}{{\tt arXiv:2404.16589}}.

\bibtype{Article}%
\bibitem[Okamoto et al.(1989)]{Okamoto:1988ru}
\bibinfo{author}{Okamoto H}, \bibinfo{author}{Okano K},
  \bibinfo{author}{Schulke L} and  \bibinfo{author}{Tanaka S}
  (\bibinfo{year}{1989}).
\bibinfo{title}{{The Role of a Kernel in Complex Langevin Systems}}.
\bibinfo{journal}{{\em Nucl. Phys. B}} \bibinfo{volume}{324}:
  \bibinfo{pages}{684--714}. \bibinfo{doi}{\doi{10.1016/0550-3213(89)90526-9}}.

\bibtype{Article}%
\bibitem[Okano et al.(1991)]{Okano:1991tz}
\bibinfo{author}{Okano K}, \bibinfo{author}{Schulke L} and
  \bibinfo{author}{Zheng B} (\bibinfo{year}{1991}).
\bibinfo{title}{{Kernel controlled complex Langevin simulation: Field dependent
  kernel}}.
\bibinfo{journal}{{\em Phys. Lett. B}} \bibinfo{volume}{258}:
  \bibinfo{pages}{421--426}. \bibinfo{doi}{\doi{10.1016/0370-2693(91)91111-8}}.

\bibtype{Article}%
\bibitem[Parisi(1983)]{Parisi:1983mgm}
\bibinfo{author}{Parisi G} (\bibinfo{year}{1983}).
\bibinfo{title}{On complex probabilities}.
\bibinfo{journal}{{\em Physics Letters B}} \bibinfo{volume}{131}
  (\bibinfo{number}{4}): \bibinfo{pages}{393--395}.
  \bibinfo{doi}{\doi{10.1016/0370-2693(83)90525-7}}.

\bibtype{Article}%
\bibitem[Parisi and Wu(1980)]{Parisi:1980ys}
\bibinfo{author}{Parisi G} and  \bibinfo{author}{Wu YS} (\bibinfo{year}{1980}).
\bibinfo{title}{{Perturbation theory without gauge fixing}}.
\bibinfo{journal}{{\em Sci. China, A}} \bibinfo{volume}{24}
  (\bibinfo{number}{ASITP-80-004}): \bibinfo{pages}{483}.

\bibtype{Article}%
\bibitem[Pasztor and Pesznyak(2026)]{Pasztor:2025wsj}
\bibinfo{author}{Pasztor A} and  \bibinfo{author}{Pesznyak D}
  (\bibinfo{year}{2026}).
\bibinfo{title}{{Is it worth the effort to find Lefschetz thimbles? Integration
  contours with numerically optimal signal-to-noise ratios in simple fermionic
  toy models}}.
\bibinfo{journal}{{\em Phys. Rev. D}} \bibinfo{volume}{113}
  (\bibinfo{number}{1}): \bibinfo{pages}{014506}.
  \bibinfo{doi}{\doi{10.1103/fywf-77b4}}.
\href{http://arxiv.org/abs/2509.07881}{{\tt arXiv:2509.07881}}.

\bibtype{Article}%
\bibitem[Pawlowski and Urban(2023)]{Pawlowski:2022rdn}
\bibinfo{author}{Pawlowski JM} and  \bibinfo{author}{Urban JM}
  (\bibinfo{year}{2023}).
\bibinfo{title}{{Flow-based density of states for complex actions}}.
\bibinfo{journal}{{\em Phys. Rev. D}} \bibinfo{volume}{108}
  (\bibinfo{number}{5}): \bibinfo{pages}{054511}.
  \bibinfo{doi}{\doi{10.1103/PhysRevD.108.054511}}.
\href{http://arxiv.org/abs/2203.01243}{{\tt arXiv:2203.01243}}.

\bibtype{Article}%
\bibitem[Pawlowski et al.(2021)]{Pawlowski:2021bbu}
\bibinfo{author}{Pawlowski JM}, \bibinfo{author}{Scherzer M},
  \bibinfo{author}{Schmidt C}, \bibinfo{author}{Ziegler FPG} and
  \bibinfo{author}{Ziesch{\'e} F} (\bibinfo{year}{2021}).
\bibinfo{title}{{Simulating Yang-Mills theories with a complex coupling}}.
\bibinfo{journal}{{\em Phys. Rev. D}} \bibinfo{volume}{103}
  (\bibinfo{number}{9}): \bibinfo{pages}{094505}.
  \bibinfo{doi}{\doi{10.1103/PhysRevD.103.094505}}.
\href{http://arxiv.org/abs/2101.03938}{{\tt arXiv:2101.03938}}.

\bibtype{Article}%
\bibitem[Pehlevan and Guralnik(2009)]{Pehlevan:2007eq}
\bibinfo{author}{Pehlevan C} and  \bibinfo{author}{Guralnik G}
  (\bibinfo{year}{2009}).
\bibinfo{title}{{Complex Langevin Equations and Schwinger-Dyson Equations}}.
\bibinfo{journal}{{\em Nucl. Phys. B}} \bibinfo{volume}{811}:
  \bibinfo{pages}{519--536}.
  \bibinfo{doi}{\doi{10.1016/j.nuclphysb.2008.11.034}}.
\href{http://arxiv.org/abs/0710.3756}{{\tt arXiv:0710.3756}}.

\bibtype{Article}%
\bibitem[Prokof’ev and Svistunov(2001)]{Prokof_ev_2001}
\bibinfo{author}{Prokof’ev N} and  \bibinfo{author}{Svistunov B}
  (\bibinfo{year}{2001}).
\bibinfo{title}{{Worm Algorithms for Classical Statistical Models}}.
\bibinfo{journal}{{\em Physical Review Letters}} \bibinfo{volume}{87}
  (\bibinfo{number}{16}).
ISSN \bibinfo{issn}{1079-7114}.
  \bibinfo{doi}{\doi{10.1103/physrevlett.87.160601}}.
\href{https://arxiv.org/abs/0103146}{{\tt 0103146}},
  \bibinfo{url}{\url{http://dx.doi.org/10.1103/PhysRevLett.87.160601}}.

\bibtype{Article}%
\bibitem[Rodekamp et al.(2022)]{Rodekamp:2022xpf}
\bibinfo{author}{Rodekamp M}, \bibinfo{author}{Berkowitz E},
  \bibinfo{author}{G{\"a}ntgen C}, \bibinfo{author}{Krieg S},
  \bibinfo{author}{Luu T} and  \bibinfo{author}{Ostmeyer J}
  (\bibinfo{year}{2022}).
\bibinfo{title}{{Mitigating the Hubbard sign problem with complex-valued neural
  networks}}.
\bibinfo{journal}{{\em Phys. Rev. B}} \bibinfo{volume}{106}
  (\bibinfo{number}{12}): \bibinfo{pages}{125139}.
  \bibinfo{doi}{\doi{10.1103/PhysRevB.106.125139}}.
\href{http://arxiv.org/abs/2203.00390}{{\tt arXiv:2203.00390}}.

\bibtype{Article}%
\bibitem[Rothkopf(2020)]{Rothkopf:2019ipj}
\bibinfo{author}{Rothkopf A} (\bibinfo{year}{2020}).
\bibinfo{title}{{Heavy Quarkonium in Extreme Conditions}}.
\bibinfo{journal}{{\em Phys. Rept.}} \bibinfo{volume}{858}:
  \bibinfo{pages}{1--117}. \bibinfo{doi}{\doi{10.1016/j.physrep.2020.02.006}}.
\href{http://arxiv.org/abs/1912.02253}{{\tt arXiv:1912.02253}}.

\bibtype{Article}%
\bibitem[Salcedo and Seiler(2019)]{Salcedo:2018fvt}
\bibinfo{author}{Salcedo LL} and  \bibinfo{author}{Seiler E}
  (\bibinfo{year}{2019}).
\bibinfo{title}{{Schwinger--Dyson equations and line integrals}}.
\bibinfo{journal}{{\em J. Phys. A}} \bibinfo{volume}{52} (\bibinfo{number}{3}):
  \bibinfo{pages}{035201}. \bibinfo{doi}{\doi{10.1088/1751-8121/aaefca}}.
\href{http://arxiv.org/abs/1809.06888}{{\tt arXiv:1809.06888}}.

\bibtype{Article}%
\bibitem[Samberger et al.(2026)]{Samberger:2025hsr}
\bibinfo{author}{Samberger T}, \bibinfo{author}{Bloch J},
  \bibinfo{author}{Lohmayer R} and  \bibinfo{author}{Wettig T}
  (\bibinfo{year}{2026}).
\bibinfo{title}{{Tensor-network formulation of QCD in the strong-coupling
  expansion}}.
\bibinfo{journal}{{\em Nucl. Phys. B}} \bibinfo{volume}{1024}:
  \bibinfo{pages}{117267}.
  \bibinfo{doi}{\doi{10.1016/j.nuclphysb.2025.117267}}.
\href{http://arxiv.org/abs/2508.09891}{{\tt arXiv:2508.09891}}.

\bibtype{Article}%
\bibitem[Scherzer et al.(2019)]{Scherzer:2018hid}
\bibinfo{author}{Scherzer M}, \bibinfo{author}{Seiler E},
  \bibinfo{author}{Sexty D} and  \bibinfo{author}{Stamatescu IO}
  (\bibinfo{year}{2019}).
\bibinfo{title}{{Complex Langevin and boundary terms}}.
\bibinfo{journal}{{\em Phys. Rev. D}} \bibinfo{volume}{99}
  (\bibinfo{number}{1}): \bibinfo{pages}{014512}.
  \bibinfo{doi}{\doi{10.1103/PhysRevD.99.014512}}.
\href{http://arxiv.org/abs/1808.05187}{{\tt arXiv:1808.05187}}.

\bibtype{Article}%
\bibitem[Scherzer et al.(2020{\natexlab{a}})]{Scherzer:2019lrh}
\bibinfo{author}{Scherzer M}, \bibinfo{author}{Seiler E},
  \bibinfo{author}{Sexty D} and  \bibinfo{author}{Stamatescu IO}
  (\bibinfo{year}{2020}{\natexlab{a}}).
\bibinfo{title}{{Controlling Complex Langevin simulations of lattice models by
  boundary term analysis}}.
\bibinfo{journal}{{\em Phys. Rev. D}} \bibinfo{volume}{101}
  (\bibinfo{number}{1}): \bibinfo{pages}{014501}.
  \bibinfo{doi}{\doi{10.1103/PhysRevD.101.014501}}.
\href{http://arxiv.org/abs/1910.09427}{{\tt arXiv:1910.09427}}.

\bibtype{Article}%
\bibitem[Scherzer et al.(2020{\natexlab{b}})]{Scherzer:2020kiu}
\bibinfo{author}{Scherzer M}, \bibinfo{author}{Sexty D} and
  \bibinfo{author}{Stamatescu IO} (\bibinfo{year}{2020}{\natexlab{b}}).
\bibinfo{title}{{Deconfinement transition line with the complex Langevin
  equation up to $\mu /T \sim 5$}}.
\bibinfo{journal}{{\em Phys. Rev. D}} \bibinfo{volume}{102}
  (\bibinfo{number}{1}): \bibinfo{pages}{014515}.
  \bibinfo{doi}{\doi{10.1103/PhysRevD.102.014515}}.
\href{http://arxiv.org/abs/2004.05372}{{\tt arXiv:2004.05372}}.

\bibtype{Article}%
\bibitem[Schmidt(2025)]{Schmidt:2025ppy}
\bibinfo{author}{Schmidt C} (\bibinfo{year}{2025}).
\bibinfo{title}{{Selected topics on the QCD phase diagram at finite temperature
  and density}}.
\bibinfo{journal}{{\em PoS}} \bibinfo{volume}{LATTICE2024}:
  \bibinfo{pages}{005}. \bibinfo{doi}{\doi{10.22323/1.466.0005}}.
\href{http://arxiv.org/abs/2504.00629}{{\tt arXiv:2504.00629}}.

\bibtype{Article}%
\bibitem[Schwinger(1961)]{Schwinger:1960qe}
\bibinfo{author}{Schwinger JS} (\bibinfo{year}{1961}).
\bibinfo{title}{{Brownian motion of a quantum oscillator}}.
\bibinfo{journal}{{\em J. Math. Phys.}} \bibinfo{volume}{2}:
  \bibinfo{pages}{407--432}. \bibinfo{doi}{\doi{10.1063/1.1703727}}.

\bibtype{Article}%
\bibitem[Seiler(2020)]{Seiler:2020mkh}
\bibinfo{author}{Seiler E} (\bibinfo{year}{2020}).
\bibinfo{title}{{Complex Langevin: Boundary terms at poles}}.
\bibinfo{journal}{{\em Phys. Rev. D}} \bibinfo{volume}{102}
  (\bibinfo{number}{9}): \bibinfo{pages}{094507}.
  \bibinfo{doi}{\doi{10.1103/PhysRevD.102.094507}}.
\href{http://arxiv.org/abs/2006.04714}{{\tt arXiv:2006.04714}}.

\bibtype{Article}%
\bibitem[Seiler and Wosiek(2017)]{Seiler:2017vwj}
\bibinfo{author}{Seiler E} and  \bibinfo{author}{Wosiek J}
  (\bibinfo{year}{2017}).
\bibinfo{title}{{Positive Representations of a Class of Complex Measures}}.
\bibinfo{journal}{{\em J. Phys. A}} \bibinfo{volume}{50}
  (\bibinfo{number}{49}): \bibinfo{pages}{495403}.
  \bibinfo{doi}{\doi{10.1088/1751-8121/aa9310}}.
\href{http://arxiv.org/abs/1702.06012}{{\tt arXiv:1702.06012}}.

\bibtype{Article}%
\bibitem[Seiler et al.(2013)]{Seiler:2012wz}
\bibinfo{author}{Seiler E}, \bibinfo{author}{Sexty D} and
  \bibinfo{author}{Stamatescu IO} (\bibinfo{year}{2013}).
\bibinfo{title}{{Gauge cooling in complex Langevin for QCD with heavy quarks}}.
\bibinfo{journal}{{\em Phys. Lett. B}} \bibinfo{volume}{723}:
  \bibinfo{pages}{213--216}.
  \bibinfo{doi}{\doi{10.1016/j.physletb.2013.04.062}}.
\href{http://arxiv.org/abs/1211.3709}{{\tt arXiv:1211.3709}}.

\bibtype{Article}%
\bibitem[Seiler et al.(2024)]{Seiler:2023kes}
\bibinfo{author}{Seiler E}, \bibinfo{author}{Sexty D} and
  \bibinfo{author}{Stamatescu IO} (\bibinfo{year}{2024}).
\bibinfo{title}{{Complex Langevin: Correctness criteria, boundary terms, and
  spectrum}}.
\bibinfo{journal}{{\em Phys. Rev. D}} \bibinfo{volume}{109}
  (\bibinfo{number}{1}): \bibinfo{pages}{014509}.
  \bibinfo{doi}{\doi{10.1103/PhysRevD.109.014509}}.
\href{http://arxiv.org/abs/2304.00563}{{\tt arXiv:2304.00563}}.

\bibtype{Article}%
\bibitem[Sexty(2014{\natexlab{a}})]{Sexty:2014dxa}
\bibinfo{author}{Sexty D} (\bibinfo{year}{2014}{\natexlab{a}}).
\bibinfo{title}{{New algorithms for finite density QCD}}.
\bibinfo{journal}{{\em PoS}} \bibinfo{volume}{LATTICE2014}:
  \bibinfo{pages}{016}. \bibinfo{doi}{\doi{10.22323/1.214.0016}}.
\href{http://arxiv.org/abs/1410.8813}{{\tt arXiv:1410.8813}}.

\bibtype{Article}%
\bibitem[Sexty(2014{\natexlab{b}})]{Sexty:2013ica}
\bibinfo{author}{Sexty D} (\bibinfo{year}{2014}{\natexlab{b}}).
\bibinfo{title}{{Simulating full QCD at nonzero density using the complex
  Langevin equation}}.
\bibinfo{journal}{{\em Phys. Lett. B}} \bibinfo{volume}{729}:
  \bibinfo{pages}{108--111}.
  \bibinfo{doi}{\doi{10.1016/j.physletb.2014.01.019}}.
\href{http://arxiv.org/abs/1307.7748}{{\tt arXiv:1307.7748}}.

\bibtype{Article}%
\bibitem[Sexty(2019)]{Sexty:2019vqx}
\bibinfo{author}{Sexty D} (\bibinfo{year}{2019}).
\bibinfo{title}{{Calculating the equation of state of dense quark-gluon plasma
  using the complex Langevin equation}}.
\bibinfo{journal}{{\em Phys. Rev. D}} \bibinfo{volume}{100}
  (\bibinfo{number}{7}): \bibinfo{pages}{074503}.
  \bibinfo{doi}{\doi{10.1103/PhysRevD.100.074503}}.
\href{http://arxiv.org/abs/1907.08712}{{\tt arXiv:1907.08712}}.

\bibtype{Article}%
\bibitem[Shimizu and Kuramashi(2014{\natexlab{a}})]{Shimizu:2014fsa}
\bibinfo{author}{Shimizu Y} and  \bibinfo{author}{Kuramashi Y}
  (\bibinfo{year}{2014}{\natexlab{a}}).
\bibinfo{title}{{Critical behavior of the lattice Schwinger model with a
  topological term at $\theta=\pi$ using the Grassmann tensor renormalization
  group}}.
\bibinfo{journal}{{\em Phys. Rev. D}} \bibinfo{volume}{90}
  (\bibinfo{number}{7}): \bibinfo{pages}{074503}.
  \bibinfo{doi}{\doi{10.1103/PhysRevD.90.074503}}.
\href{http://arxiv.org/abs/1408.0897}{{\tt arXiv:1408.0897}}.

\bibtype{Article}%
\bibitem[Shimizu and Kuramashi(2014{\natexlab{b}})]{Shimizu:2014uva}
\bibinfo{author}{Shimizu Y} and  \bibinfo{author}{Kuramashi Y}
  (\bibinfo{year}{2014}{\natexlab{b}}).
\bibinfo{title}{{Grassmann tensor renormalization group approach to one-flavor
  lattice Schwinger model}}.
\bibinfo{journal}{{\em Phys. Rev. D}} \bibinfo{volume}{90}
  (\bibinfo{number}{1}): \bibinfo{pages}{014508}.
  \bibinfo{doi}{\doi{10.1103/PhysRevD.90.014508}}.
\href{http://arxiv.org/abs/1403.0642}{{\tt arXiv:1403.0642}}.

\bibtype{Book}%
\bibitem[Smit(2002)]{Smit:2002ug}
\bibinfo{author}{Smit J} (\bibinfo{year}{2002}).
\bibinfo{title}{{Introduction to quantum fields on a lattice: A robust mate}},
  \bibinfo{volume}{(15)}, \bibinfo{publisher}{Cambridge University Press}.
\bibinfo{comment}{ISBN} \bibinfo{isbn}{978-0-511-89373-5, 978-0-521-89051-9}.

\bibtype{Article}%
\bibitem[Son and Stephanov(2001)]{Son:2000xc}
\bibinfo{author}{Son DT} and  \bibinfo{author}{Stephanov MA}
  (\bibinfo{year}{2001}).
\bibinfo{title}{{QCD at finite isospin density}}.
\bibinfo{journal}{{\em Phys. Rev. Lett.}} \bibinfo{volume}{86}:
  \bibinfo{pages}{592--595}. \bibinfo{doi}{\doi{10.1103/PhysRevLett.86.592}}.
\href{https://arxiv.org/abs/hep-ph/0005225}{{\tt hep-ph/0005225}}.

\bibtype{Article}%
\bibitem[Sugimoto et al.(2026)]{Sugimoto:2026wnw}
\bibinfo{author}{Sugimoto Y}, \bibinfo{author}{Akiyama S} and
  \bibinfo{author}{Kuramashi Y} (\bibinfo{year}{2026}).
\bibinfo{title}{{Tensor renormalization group study of cold and dense QCD in
  the strong coupling limit}} \href{http://arxiv.org/abs/2601.20690}{{\tt
  arXiv:2601.20690}}.

\bibtype{Article}%
\bibitem[Tanizaki et al.(2016)]{Tanizaki:2015rda}
\bibinfo{author}{Tanizaki Y}, \bibinfo{author}{Hidaka Y} and
  \bibinfo{author}{Hayata T} (\bibinfo{year}{2016}).
\bibinfo{title}{{Lefschetz-thimble analysis of the sign problem in one-site
  fermion model}}.
\bibinfo{journal}{{\em New J. Phys.}} \bibinfo{volume}{18}
  (\bibinfo{number}{3}): \bibinfo{pages}{033002}.
  \bibinfo{doi}{\doi{10.1088/1367-2630/18/3/033002}}.
\href{http://arxiv.org/abs/1509.07146}{{\tt arXiv:1509.07146}}.

\bibtype{Article}%
\bibitem[Tomiya(2025)]{Tomiya:2025quf}
\bibinfo{author}{Tomiya A} (\bibinfo{year}{2025}).
\bibinfo{title}{{Machine Learning for Lattice QCD}}.
\bibinfo{journal}{{\em J. Phys. Soc. Jap.}} \bibinfo{volume}{94}
  (\bibinfo{number}{3}): \bibinfo{pages}{031006}.
  \bibinfo{doi}{\doi{10.7566/JPSJ.94.031006}}.

\bibtype{Article}%
\bibitem[Troyer and Wiese(2005)]{Troyer:2004ge}
\bibinfo{author}{Troyer M} and  \bibinfo{author}{Wiese UJ}
  (\bibinfo{year}{2005}).
\bibinfo{title}{{Computational complexity and fundamental limitations to
  fermionic quantum Monte Carlo simulations}}.
\bibinfo{journal}{{\em Phys. Rev. Lett.}} \bibinfo{volume}{94}:
  \bibinfo{pages}{170201}. \bibinfo{doi}{\doi{10.1103/PhysRevLett.94.170201}}.
\href{https://arxiv.org/abs/cond-mat/0408370}{{\tt cond-mat/0408370}}.

\bibtype{Article}%
\bibitem[Tsutsui et al.(2025)]{Tsutsui:2025jez}
\bibinfo{author}{Tsutsui S}, \bibinfo{author}{Asano Y}, \bibinfo{author}{Ito
  Y}, \bibinfo{author}{Matsufuru H}, \bibinfo{author}{Namekawa Y},
  \bibinfo{author}{Nishimura J}, \bibinfo{author}{Shimasaki S} and
  \bibinfo{author}{Tsuchiya A} (\bibinfo{year}{2025}).
\bibinfo{title}{{On the validity of the complex Langevin method near the
  deconfining phase transition in QCD at finite density}}.
\bibinfo{journal}{{\em JHEP}} \bibinfo{volume}{10}: \bibinfo{pages}{108}.
  \bibinfo{doi}{\doi{10.1007/JHEP10(2025)108}}.
\href{http://arxiv.org/abs/2505.06551}{{\tt arXiv:2505.06551}}.

\bibtype{Article}%
\bibitem[Ulybyshev et al.(2020)]{Ulybyshev:2019fte}
\bibinfo{author}{Ulybyshev M}, \bibinfo{author}{Winterowd C} and
  \bibinfo{author}{Zafeiropoulos S} (\bibinfo{year}{2020}).
\bibinfo{title}{{Lefschetz thimbles decomposition for the Hubbard model on the
  hexagonal lattice}}.
\bibinfo{journal}{{\em Phys. Rev. D}} \bibinfo{volume}{101}
  (\bibinfo{number}{1}): \bibinfo{pages}{014508}.
  \bibinfo{doi}{\doi{10.1103/PhysRevD.101.014508}}.
\href{http://arxiv.org/abs/1906.07678}{{\tt arXiv:1906.07678}}.

\bibtype{Article}%
\bibitem[Wang et al.(2024)]{Wang:2023exq}
\bibinfo{author}{Wang L}, \bibinfo{author}{Aarts G} and  \bibinfo{author}{Zhou
  K} (\bibinfo{year}{2024}).
\bibinfo{title}{{Diffusion models as stochastic quantization in lattice field
  theory}}.
\bibinfo{journal}{{\em JHEP}} \bibinfo{volume}{05}: \bibinfo{pages}{060}.
  \bibinfo{doi}{\doi{10.1007/JHEP05(2024)060}}.
\href{http://arxiv.org/abs/2309.17082}{{\tt arXiv:2309.17082}}.

\bibtype{Article}%
\bibitem[Weingarten(2002)]{Weingarten:2002xs}
\bibinfo{author}{Weingarten D} (\bibinfo{year}{2002}).
\bibinfo{title}{{Complex probabilities on $\mathbb{R}^N$ as real probabilities
  on $\mathbb{C}^N$ and an application to path integrals}}.
\bibinfo{journal}{{\em Phys. Rev. Lett.}} \bibinfo{volume}{89}:
  \bibinfo{pages}{240201}. \bibinfo{doi}{\doi{10.1103/PhysRevLett.89.240201}}.
\href{https://arxiv.org/abs/quant-ph/0210195}{{\tt quant-ph/0210195}}.

\bibtype{Article}%
\bibitem[Willis and Matsen(2024)]{Matsen2024}
\bibinfo{author}{Willis JD} and  \bibinfo{author}{Matsen MW}
  (\bibinfo{year}{2024}).
\bibinfo{title}{{Stabilizing complex-Langevin field-theoretic simulations for
  block copolymer melts}}.
\bibinfo{journal}{{\em The Journal of Chemical Physics}} \bibinfo{volume}{161}
  (\bibinfo{number}{24}): \bibinfo{pages}{244903}.
ISSN \bibinfo{issn}{0021-9606}. \bibinfo{doi}{\doi{10.1063/5.0245363}}.
\bibinfo{url}{\url{https://doi.org/10.1063/5.0245363}}.

\bibtype{Article}%
\bibitem[Wilson(1974)]{PhysRevD.10.2445}
\bibinfo{author}{Wilson KG} (\bibinfo{year}{1974}).
\bibinfo{title}{Confinement of quarks}.
\bibinfo{journal}{{\em Phys. Rev. D}} \bibinfo{volume}{10}:
  \bibinfo{pages}{2445--2459}. \bibinfo{doi}{\doi{10.1103/PhysRevD.10.2445}}.
\bibinfo{url}{\url{https://link.aps.org/doi/10.1103/PhysRevD.10.2445}}.

\bibtype{Article}%
\bibitem[Witten(2011)]{Witten:2010cx}
\bibinfo{author}{Witten E} (\bibinfo{year}{2011}).
\bibinfo{title}{{Analytic Continuation Of Chern-Simons Theory}}.
\bibinfo{journal}{{\em AMS/IP Stud.~Adv.~Math.}} \bibinfo{volume}{50}:
  \bibinfo{pages}{347--446}.
\href{http://arxiv.org/abs/1001.2933}{{\tt arXiv:1001.2933}}.

\bibtype{Article}%
\bibitem[Yosprakob and Okunishi(2025)]{Yosprakob:2024sfd}
\bibinfo{author}{Yosprakob A} and  \bibinfo{author}{Okunishi K}
  (\bibinfo{year}{2025}).
\bibinfo{title}{{Tensor Renormalization Group Study of the 3D SU(2) and SU(3)
  Gauge Theories with the Reduced Tensor Network Formulation}}.
\bibinfo{journal}{{\em PTEP}} \bibinfo{volume}{2025} (\bibinfo{number}{3}):
  \bibinfo{pages}{033B06}. \bibinfo{doi}{\doi{10.1093/ptep/ptaf028}}.
\href{http://arxiv.org/abs/2406.16763}{{\tt arXiv:2406.16763}}.

\bibtype{Article}%
\bibitem[Zhu et al.(2026)]{Zhu:2025pmw}
\bibinfo{author}{Zhu Q}, \bibinfo{author}{Aarts G}, \bibinfo{author}{Wang W},
  \bibinfo{author}{Zhou K} and  \bibinfo{author}{Wang L}
  (\bibinfo{year}{2026}).
\bibinfo{title}{{Physics-conditioned diffusion models for lattice gauge
  theory}}.
\bibinfo{journal}{{\em JHEP}} \bibinfo{volume}{03}: \bibinfo{pages}{111}.
  \bibinfo{doi}{\doi{10.1007/JHEP03(2026)111}}.
\href{http://arxiv.org/abs/2502.05504}{{\tt arXiv:2502.05504}}.

\end{thebibliography*}

\end{document}